\documentclass[12pt]{iopart}

\usepackage[T1]{fontenc}
\usepackage[british]{babel}
\usepackage{amssymb}
\usepackage{cite}
\usepackage{iopams}
\usepackage{color}
\usepackage{graphicx}
\usepackage{mathrsfs}
\usepackage[colorlinks=true,linkcolor=blue,citecolor=blue]{hyperref}

\begin{document}

\title[Fractional Brownian motion in shallow external potentials]{Absence
of confinement and non-Boltzmann stationary states of fractional Brownian
motion in shallow external potentials}

\author{Tobias Guggenberger$^{\dagger}$, Aleksei Chechkin$^{\dagger,\S,
\ddagger}$ and Ralf Metzler$^{\dagger}$}
\address{$\dagger$ Institute of Physics \& Astronomy, University of Potsdam,
14476 Potsdam-Golm, Germany\\
$\S$Faculty of Pure and Applied Mathematics, Hugo Steinhaus Centre,
Wroc\l{}aw University of Science and Technology, Wyspianskiego 27,
50-370 Wroc{\l}aw, Poland\\
$\ddagger$ Akhiezer Institute for Theoretical Physics, Kharkov 61108, Ukraine}
\eads{\mailto{rmetzler@uni-potsdam.de}}

\begin{abstract}
We study the diffusive motion of a particle in a subharmonic potential of
the form $U(x)=|x|^c$ ($0<c<2$) driven by long-range correlated, stationary
fractional Gaussian noise $\xi_{\alpha}(t)$ with $0<\alpha\le2$. In the
absence of the potential the particle exhibits free fractional Brownian
motion with anomalous diffusion exponent $\alpha$. While for an harmonic
external potential the dynamics converges to a Gaussian stationary state,
from extensive numerical analysis we here demonstrate that stationary states
for shallower than harmonic potentials exist only as long as the relation
$c>2(1-1/\alpha)$ holds. We analyse the motion in terms of the mean squared
displacement and (when it exists) the stationary probability density function
(PDF). Moreover we discuss analogies of non-stationarity of L{\'e}vy
flights in shallow external potentials.
\end{abstract}

\section{Introduction}
\label{sec:introduction}

In his seminal PhD thesis published in 1931, Kappler presents the Gaussian
equilibrium distribution (Boltzmannian) for the angular co-ordinate of a
torsional balance driven by thermal noise \cite{kappler}. This result is
expected from equilibrium statistical physics \cite{landau}, as long as
the angle is sufficiently small and thus the restoring effect on the angular
motion, exerted by the suspending glass thread, can be approximated by a
Hookean force. On microscopic scales such an harmonic confinement and the
associated equilibrium fluctuations for a diffusing particle in water can
be effected by a polymeric tether \cite{schafer,simon}.

Harmonic confinement of micron-sized dielectric tracer particles in simple
liquids is now routinely achieved by optical tweezers \cite{franosch}. The
equilibration from a non-equilibrium initial condition of the tracer can be
derived from the associated Fokker-Planck-Smoluchowski or Langevin equations
and turns out to be exponentially fast \cite{landau1,vankampen,coffey}. In more
complex fluids such as viscoelastic liquids the relaxation to an equilibrium
situation of a tracer confined by an optical tweezers trap still occurs
albeit with more complex dynamics including transient non-ergodicity
\cite{lene1,pre12,jochen}. For ageing, weakly non-ergodic dynamics the
approach to the Boltzmannian state may be much slower \cite{mebakla,report}
and, when time-averaged observables are evaluated, obscured by a crossover
to a power-law instead of a plateau \cite{stas,staspccp}, as shown in optical
tweezers measurements of tracer particles \cite{lene} and for the relative
motion of subunits of single protein molecules \cite{jeremy,xie}.

What happens when the external potential deviates from the conventional
harmonic shape? Steeper than harmonic potentials occur, for instance, when
the harmonic approximation of the symmetric potential no longer holds and the
next order, quartic term needs to be considered. The Boltzmannian in such
potentials is flatter around the centre and decays more abruptly at
larger distances. For L{\'e}vy flights governed by power-law jump
length distributions $\simeq|x|^{-1-\mu}$ with $0<\mu<2$ such steeper
than harmonic potentials effect non-Boltzmannian, multimodal stationary
probability density functions (PDFs) \cite{chechkin,chechkin1,spagno,capala}.
For fractional Brownian motion driven by power-law correlated, fractional
Gaussian noise (FGN, see below for the definition) superharmonic external
potentials also lead to non-Boltzmannian PDFs, that in the superdiffusive
case may assume multimodal states \cite{tobias1}. Similar effects occur
on a finite interval with reflecting boundaries \cite{tobias}. Shallower
than harmonic
potentials may emerge as entropic forces, e.g., in specific geometries
of confining channels \cite{igorejp,haenggi}, and confining, symmetric
linear potentials are often analysed as prototype cases \cite{risken}.
Finally, logarithmic potentials are, e.g., known from laser traps
\cite{eli}. In potentials of the generic form $U(x)\simeq|x|^c$ with
$0<c<2$ L{\'e}vy flights were shown to be confined only when the
scaling exponent $c$ of the potential fulfils the inequality $c>2-\mu$
\cite{dybiec2010}.

Here we study the behaviour of a particle driven by FGN in shallower
than harmonic potentials. Despite the fact that FGN is a Gaussian process
we demonstrate that---similar to L{\'e}vy flights driven by white L{\'e}vy
noise with a diverging variance of the amplitude PDF---a stationary state only
exists as long
as the potential scaling exponent satisfies the relation $c>2(1-1/\alpha)$,
where $\alpha$ is the anomalous diffusion exponent of the free
FBM with the MSD $\langle X^2(t)\rangle\simeq t^{\alpha}$. For subdiffusive
and normal-diffusive FBM ($0<\alpha\le1$), that is, any positive value
of $c$ will induce confinement. While for L{\'e}vy flights non-stationarity
in shallow potentials emerges when for smaller $\mu$ the increased propensity
for long jumps outcompetes the confining tendency of the potential, for FBM
non-stationarity occurs when the driving FGN is sufficiently persistent
(positively correlated). In addition, we also report
details on the behaviour of the tails of the emerging stationary PDF such
as the dependence of the stationary MSD on the scaling exponent $c$ and
the anomalous diffusion exponent $\alpha$. The rich behaviour of FBM in
external confinement is an important further building block in the study
of this widely applied yet often surprising non-Markovian process.

The paper is structured as follows. We introduce our model and detail the
numerical implementation in section \ref{sec:theModel}. The results are
presented in section \ref{eq:results}, with a focus on the MSD as well as
the PDF of the process. We draw our Conclusions in section
\ref{sec:conclusion}.

\section{The model}
\label{sec:theModel}

We first define free FBM and introduce the governing overdamped stochastic
equation along with the associated discretisation scheme. We also state
our conjecture on the existence of stationary states in subharmonic
external potentials.

\subsection{\label{sec:fbmAndFgn}Fractional Brownian motion and fractional
Gaussian noise}

Free FBM is a zero-mean Gaussian process with two-time auto-covariance
function \cite{kolmo}
\begin{equation}
\label{eq:autocovFbm}
\langle B_{\alpha}(t_1)B_{\alpha}(t_2)\rangle=K\big[t_1^\alpha+t_2^\alpha
-\left|t_1-t_2\right|^\alpha\big],\qquad0<\alpha\leq2,
\end{equation}
whose limit is the MSD $\langle B_{\alpha}^2(t)\rangle=2Kt^\alpha$ for $t_1=
t_2=t$. The PDF of FBM for natural boundary conditions ($\lim_{|x|\to\infty}
P(x,t)=0$) is given by the Gaussian
\begin{equation}
\label{eq:pdfFbm}
P(x,t)=\frac{1}{\sqrt{4\pi Kt^\alpha}}\exp\left(-\frac{x^2}{4Kt^\alpha}\right).
\end{equation}
For $\alpha=1$ FBM reduces to a Brownian motion.

Since the sample paths of FBM are almost surely continuous but not
differentiable \cite{mandelbrot1968} we follow Mandelbrot and van Ness
and define FGN as the difference quotient
\cite{mandelbrot1968}
\begin{equation}
\label{eq:defFgn}
\xi_{\alpha}(t)=\frac{B_{\alpha}(t+\delta t)-B_{\alpha}(t)}{\delta t},
\end{equation}
where $\delta t>0$ is a small but finite time step. It follows that FGN is
a zero-mean stationary Gaussian process whose auto-covariance function is
readily obtained from \eref{eq:autocovFbm} and \eref{eq:defFgn},
\begin{equation}
\label{eq:autocovFgn}
\langle\xi_{\alpha}(t)\xi_{\alpha}(t+\tau)\rangle=K(\delta t)^{\alpha-2}
\left(\left|\frac{\tau}{\delta t}+1\right|^\alpha+\left|\frac{\tau}{\delta
t}-1 \right|^\alpha-2\left|\frac{\tau}{\delta t}\right|^\alpha\right).
\end{equation}
The variance of FGN is thus $\langle\xi_{\alpha}^2(t)\rangle=2K(\delta t)^{
\alpha-2}$. At times much longer than the time step, $\tau\gg\delta t$, one
has
\begin{equation}
\label{eq:asympAutocovFgn}
\langle\xi_{\alpha}(t)\xi_{\alpha}(t+\tau)\rangle\sim\alpha(\alpha-1)K
\tau^{\alpha-2},
\end{equation}
and hence the correlations are positive (negative) for $\alpha>1$ ($\alpha<
1$). We further mention that
\begin{equation}
\label{eq:autocovIntFgn}
\int_0^\infty\langle\xi_{\alpha}(t)\xi_{\alpha}(t+\tau)\rangle d\tau=\cases{
0,&$0<\alpha<1$\\
K,&$\alpha=1$\\
\infty,&$1<\alpha\le2$}.
\end{equation}
Equations \eref{eq:asympAutocovFgn} and \eref{eq:autocovIntFgn} demonstrate
the fundamental difference between persistent ($1<\alpha<2$) and anti-persistent
($0<\alpha<1$) FGN with their positive and negative autocorrelations,
respectively. In particular, we emphasise the vanishing integral over the
noise auto-covariance in the anti-persistent case.

Considering $\delta t$ to be "infinitesimally small", FGN can be taken as
the formal "derivative" of FBM so that $B_{\alpha}(t)=\int_0^t\xi_{\alpha}
(t') dt'$. In this case, the auto-covariance for $1\le\alpha\le2$
can formally be derived by writing $\xi_{\alpha}(t)=dB_{\alpha}(t)
/dt$, pulling the time derivatives out of the expectation value and using
the auto-covariance \eref{eq:autocovFbm} of FBM (see, e.g., \cite{qian2003}).

Finally, let us mention the ballistic limit $\alpha=2$ for which $\langle
\xi_{\alpha}(t)\xi_{\alpha}(t+\tau)\rangle=2K$ such that the FGN becomes
time-independent and hence perfectly correlated. More precisely, $\xi_{\alpha}
(t)=V$ is a Gaussian-distributed random variable with zero mean and variance
$2K$, and thus FBM reduces to a \emph{random line\/} $B_{\alpha}(t)=\int_0^t
\xi_{\alpha}(t')dt'=Vt$. In physical terms, in the ballistic limit FBM
describes a linear in time motion with a symmetric Gaussian random velocity.

\subsection{FBM in a subharmonic potential}
\label{sec:fbmSubharPot}

We investigate the diffusive motion of particles governed by the overdamped
(i.e., for dynamics neglecting inertial terms) Langevin equation
\begin{equation}
\label{eq:ovdamLang}
\frac{dX(t)}{dt}=-\frac{dU}{dx}(X(t))+\xi_{\alpha}(t)
\end{equation}
with the subharmonic potential
\begin{equation}
\label{eq:defSubharPot}
U(x)=|x|^c,\qquad0<c<2
\end{equation}
and the FGN $\xi_{\alpha}(t)$. The (deterministic) initial condition is $X(0)
=x_0\in\mathbb{R}$. The force acting on the particle reads $F(x)=-\frac{dU(x)
}{dx}=-c\mbox{ sign}(x)|x|^{c-1}$, where $\mathrm{sign}(x)$ denotes the
sign function.

For numerical simulations we used the Euler-Maruyama discretisation scheme
(see, for instance, \cite{kloeden2011}) to generate (approximate) sample
trajectories $\hat{X}_n=\hat{X}(t_n)\approx X(t_n)$ with equidistant time
points $t_n=\epsilon n$ ($\epsilon>0$, $n=0,1,\ldots,N$):
\begin{equation}
\label{eq:discScheme}
\hat{X}_0=x_0,\\
\hat{X}_{n+1}=\hat{X}_n-c|\hat{X}_n|^{c-1}\mathrm{sign}(\hat{X}_n)\epsilon
+\epsilon^{\alpha/2}\Delta B_{\alpha}(n).
\end{equation}
Here, $\Delta B_{\alpha}(n)$ is the unit increment of FBM, $\Delta
B_{\alpha}(n)=B_{\alpha}(n+1)-B_{\alpha}(n)$.\footnote{We first note that
since FBM is a self-similar process with self-similarity index $H=\alpha/2$,
one has $B_{\alpha}(t_n)=B_{\alpha}(\epsilon n)=\epsilon^{\alpha/2}
B_{\alpha}(n)$. We further note that in the ballistic limit ($\alpha=2$)
$\Delta B_{\alpha}(n)=V$ is a Gaussian distributed random variable with
zero mean and variance $2K$.} To generate sample trajectories of FBM we used
the Cholesky method \cite{dieker2004}.

\subsection{Conjecture about existence of stationary states}
\label{sec:conjecture}

An analogous situation as described by the overdamped Langevin equation
\eref{eq:ovdamLang} with a subharmonic potential \eref{eq:defSubharPot}
for a symmetric stable L{\'e}vy noise---instead of the
FGN studied here---was investigated in \cite{dybiec2010}. The authors
showed that a necessary condition for the existence of stationary states
is $c>2-\mu$, where $\mu$ denotes the stability index of the noise. For
sufficiently shallow potentials, that is, the particle is spreading
indefinitely, and thus the MSD is continuously increasing as function
of time \cite{dybiec2010}. When the condition $c>2-\mu$ is not satisfied
the competition with the external potential, tending to confine the
particle, is shifted in favour of the
long jumps of the L{\'e}vy flight. Indeed, the propensity for such long
jumps is due to the stable distribution of the noise amplitude with
tail $\simeq|x|^{-1-\mu}$. We also note that in an harmonic external
potential, the stationary state of a L{\'e}vy flight has the
same L{\'e}vy index $\mu$ as the driving L{\'e}vy stable noise \cite{sune}.
L{\'e}vy flights are Markovian. In external potentials, based on
their formulations in terms of a Langevin equation with L{\'e}vy stable noise
\cite{fogedby,fogedby1,sune,chechkinjsp} or Fokker-Planck equations with
space-fractional derivatives \cite{epl,report}, the asymptotic behaviour
can be derived analytically or from scaling arguments \cite{sune,chechkin,
chechkin1,chechkinjsp,dybiec2010}.

Due to the long-ranged autocorrelation property of FGN, FBM is a strongly
non-Markovian process \cite{qian2003,mandelbrot1968} and does not fulfil
the semi-martingale property \cite{weron}. FBM is thus not amenable to
many standard analysis techniques, for instance, to calculate first-passage
times (see the discussion in the Conclusion section). However, we here build
the following argument on the self-similarity property of FBM in comparison
to L{\'e}vy flights. Namely, the integral over stable L{\'e}vy noise
is a L{\'e}vy flight, which is self-similar with self-similarity
index $H=1/\mu$, so that the necessary condition for the existence of
stationary states for L{\'e}vy flights can be rewritten as $c>2-1/H$.
Analogously the integral over FGN is an FBM, which is self-similar with
self-similarity index $H=\alpha/2$ \cite{qian2003,mandelbrot1968}. Hence,
by analogy we arrive at the following conjecture: The dynamics given
by \eref{eq:ovdamLang}, driven by FGN, in the potential \eref{eq:defSubharPot}
has a long-time stationary solution if
\begin{equation}
\label{eq:statConj}
c>c_\mathrm{crit}=2\left(1-\frac{1}{\alpha}\right)\quad\iff\quad\alpha<
\alpha_\mathrm{crit}=\frac{2}{2-c}.
\end{equation}
Here we denoted the critical values for the scaling exponent of the external
potential and the corresponding critical value for the correlation exponent
of the FGN by $c_\mathrm{crit}$ and $\alpha_\mathrm{crit}$, respectively. 

Our main focus is to check this conjecture numerically using the MSD
$\langle(X(t)-x_0)^2\rangle$ as a measure of stationarity. Subsequently we
will examine the properties of the long-time stationary PDF $P(x)=\lim_{t
\to\infty}P(x,t)$ of the system (if it exists). Our detailed analysis
based on extensive simulations provides strong arguments for the validity
of the conjecture (\ref{eq:statConj}).

\section{Results}
\label{eq:results}

In all simulations we employ a normalised FGN (FBM), that is, we set the
diffusivity $K=1/2$. For all simulations with $c\geq1$ we set the initial
position to the origin, $x_0=0$. For $c<1$ we set $x_0=0.1$, to avoid
divergence of the force at the initial position. The discretisation time
step was chosen between $\epsilon
=0.05$ and $\epsilon=0.001$, and the ensemble size ranged from several ten
to several million trajectories.

Before we present our numerical results, let us briefly discuss two special
cases, that can be solved analytically.

\subsection{Brownian case}
\label{sec:brownianCase}

In the Brownian case ($\alpha=1$) the FGN reduces to a white Gaussian noise
with $\delta$-correlation, $\langle \xi_{\alpha}(t_1) \xi_{\alpha}(t_2)\rangle
=2K\delta(t_1-t_2)$, and hence the PDF of the process  $X(t)$ in the
Langevin equation \eref{eq:ovdamLang} satisfies a Fokker-Planck equation
whose long-time stationary solution for the potential (\ref{eq:defSubharPot})
is given by the Boltzmann PDF
\begin{equation}
\label{eq:statPdfBrown}
\fl P_{\mathrm{st}}(x)=\frac{1}{\mathscr{N}}\exp(-U(x)/K),\qquad\mathscr{N}
=\int_{-\infty}^{\infty}\exp(-U(x)/K)dx=\frac{2K^{1/c}}{c}\Gamma(1/c),
\end{equation}
where $\Gamma(z)$ denotes the complete gamma function. Thus
the first moment in the stationary state is zero,
$\langle X_\mathrm{st}\rangle=0$, and the second moment is
\begin{equation}
\label{eq:statEmsdBrown}
\langle X_\mathrm{st}^2\rangle=K^{2/c}\frac{\Gamma(3/c)}{\Gamma(1/c)}.
\end{equation}
Note that the second moment, although finite for all $c>0$, tends to infinity
for $c\to0$, which simply corresponds to the non-existence of a stationary
state in the unconfined case.\footnote{We note in passing that for $c\to
\infty$ the second moment converges to the value $1/3$, which equals the
value of the second moment for the uniform distribution on the interval $[-1,
1]$ and corresponds to the potential converging to the infinite box potential
on $[-1,1]$, i.e., reflecting walls at $x=\pm1$.}

\subsection{Harmonic case}
\label{sec:harmonicCase}

In the harmonic case ($c=2$) the time-dependent first and second moment
\cite{pre12,jochen} can be obtained directly from the Langevin equation
\begin{equation}
\label{eq:FirAndSecMomLeqHarPot}
\fl\eqalign{\langle X(t)\rangle=x_0e^{-2t},\\
\langle X^2(t)\rangle=x_0^2e^{-4t}+2Kt^\alpha e^{-2t}+\frac{K}{2^\alpha}
\gamma(\alpha+1,2t)-\frac{2K}{\alpha+1}t^{\alpha+1}e^{-4t}M(\alpha+1,
\alpha+2,2t),}
\end{equation}
where $\gamma(z,t)=\int_0^ts^{z-1}e^{-s}ds$ ($\mathrm{Re}(z)>0$, $t\geq0$)
is the incomplete gamma function of the upper bound, and
$M(a,b,z)$ is the Kummer function which for $b>a>0$ has
the integral representation \cite{abramowitz72}
\begin{equation}
\label{eq:kummer}
M(a,b,z)=\frac{\Gamma(b)}{\Gamma(b-a)\Gamma(a)}\int_0^1e^{zs}s^{a-1}(1-s)^{
b-a-1}ds\quad(z\in\mathbb{C}).
\end{equation}
In the long-time limit the first moment converges to zero, $\langle X_
\mathrm{st}\rangle=0$, and the second moment assumes the limiting value
\begin{equation}
\label{eq:statEmsdHarPot}
\langle X_\mathrm{st}^2\rangle=\frac{K}{2^\alpha}\Gamma(\alpha+1).
\end{equation}
The explicit dependence on the anomalous diffusion exponent $\alpha$
underlines the non-equilibrium nature of FBM \cite{oleksii}, that is
not subject to the fluctuation-dissipation theorem in contrast to the
generalised Langevin equation \cite{zwanzig}. FGN in the FBM dynamics
is therefore also often described as "external noise" \cite{klimo}.

Additionally one can show that the PDF defined by the Langevin equation
\eref{eq:ovdamLang} with an arbitrary stationary Gaussian noise $\eta(t)$
satisfies the following generalised Fokker-Planck equation
\cite{adelman1976,haenggi1994}\footnote{We emphatically note that this
partial differential equation formulation cannot be used to calculate
the behaviour of FBM close to absorbing or reflecting boundaries, see
the discussion in the Conclusions section.}
\begin{equation}
\label{eq:fpeHarmonic}
\frac{\partial}{\partial t}P(x,t)=\frac{\partial}{\partial x}\left[2xP
(x,t)\right]+D(t)\frac{\partial^2}{\partial x^2}P(x,t),
\end{equation}
with the time-dependent diffusion coefficient
\begin{equation}
\label{eq:fpeHarmonic2}
D(t)=\int_0^te^{-2\tau}\langle\eta(t)\eta(t+\tau)\rangle d\tau.
\end{equation}
For FGN, $\eta(t)=\xi_\alpha(t)$, we obtain
\begin{equation}
\label{eq:fpeHarmonic3}
D(t)=\alpha Kt^{\alpha-1}e^{-2t}+\frac{\alpha K}{2^{\alpha-1}}\gamma(
\alpha,2t)\stackrel{t\to\infty}{\longrightarrow}\frac{K}{2^{\alpha-1}}
\Gamma(\alpha+1)=2\langle X_\mathrm{st}^2\rangle.
\end{equation}
Thus, the long-time stationary Fokker-Planck equation reads
\begin{equation}
\label{eq:statFpeHarmonic}
0=2xP_{\mathrm{st}}(x)+2\langle X_\mathrm{st}^2\rangle\frac{d}{dx}P_{
\mathrm{st}}(x)
\end{equation}
and has the Gaussian solution
\begin{equation}
\label{eq:solStatFpeHarmonic}
P_{\mathrm{st}}(x)=\frac{1}{\sqrt{2\pi\sigma_{\mathrm{st}}^2}}\exp\left(
-\frac{x^2}{2\sigma_{\mathrm{st}}^2}\right),
\end{equation}
where $\sigma_{\mathrm{st}}^2=\langle X_\mathrm{st}^2\rangle$ is the
stationary variance. As can be checked by insertion, the solution of the
time-dependent Fokker-Planck equation \eref{eq:fpeHarmonic} is given by
the shifted Gaussian
\begin{equation}
\label{eq:solFpeHar}
P(x,t)=\frac{1}{\sqrt{2\pi\sigma^2(t)}}\exp\left(-\frac{(x-\mu(t))^2}{2
\sigma^2(t)}\right),
\end{equation}
with $\mu(t)=\langle X(t)\rangle$ and $\sigma^2(t)=\langle X^2(t)\rangle
-\mu^2(t)$ given by expressions \eref{eq:FirAndSecMomLeqHarPot}.

\subsection{The general case}
\label{sec:generalCase}

We first consider the MSD and determine for which parameter values of the
scaling exponent $c$ of the potential and the autocorrelation exponent
$\alpha$ of the driving FGN it converges to a plateau value thus indicating
confinement, or whether it continues to grow indefinitely. We then evaluate
the PDF of the process and quantify its non-Gaussianity for the stationary
cases.

\begin{figure}
\includegraphics[width=0.48\textwidth]{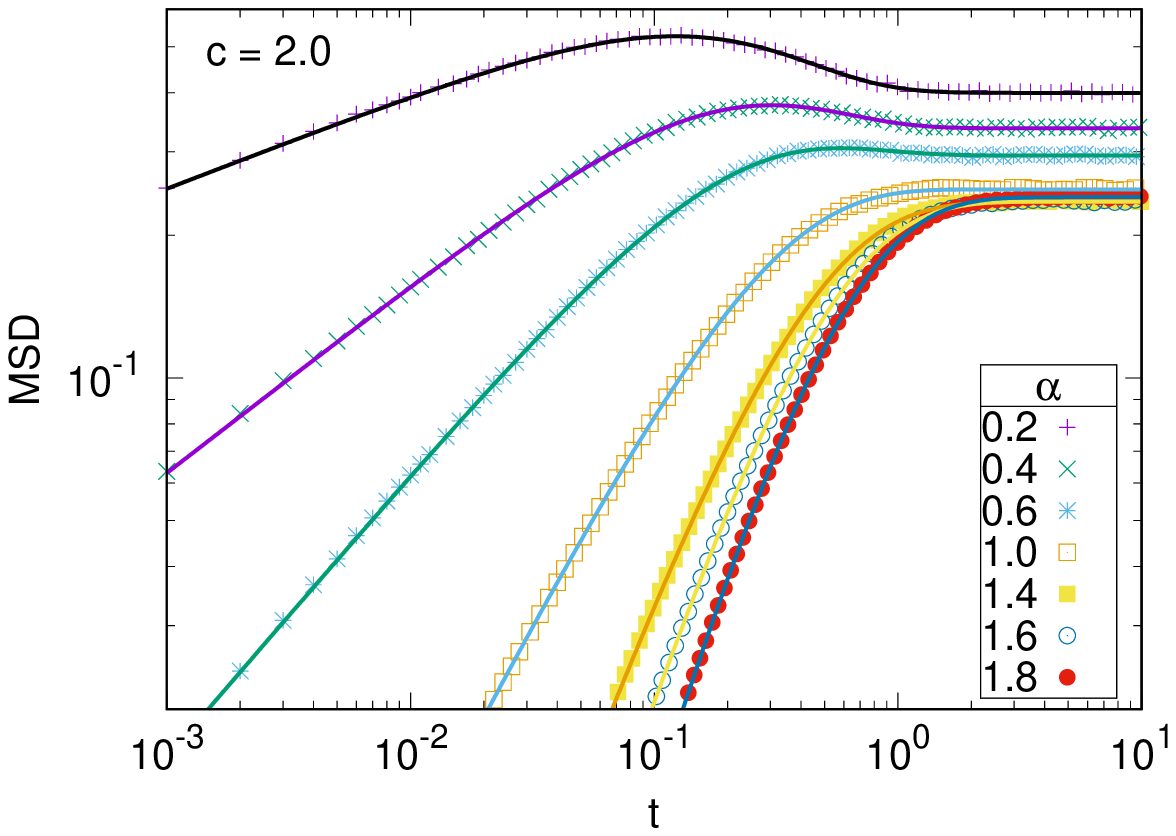}
\includegraphics[width=0.48\textwidth]{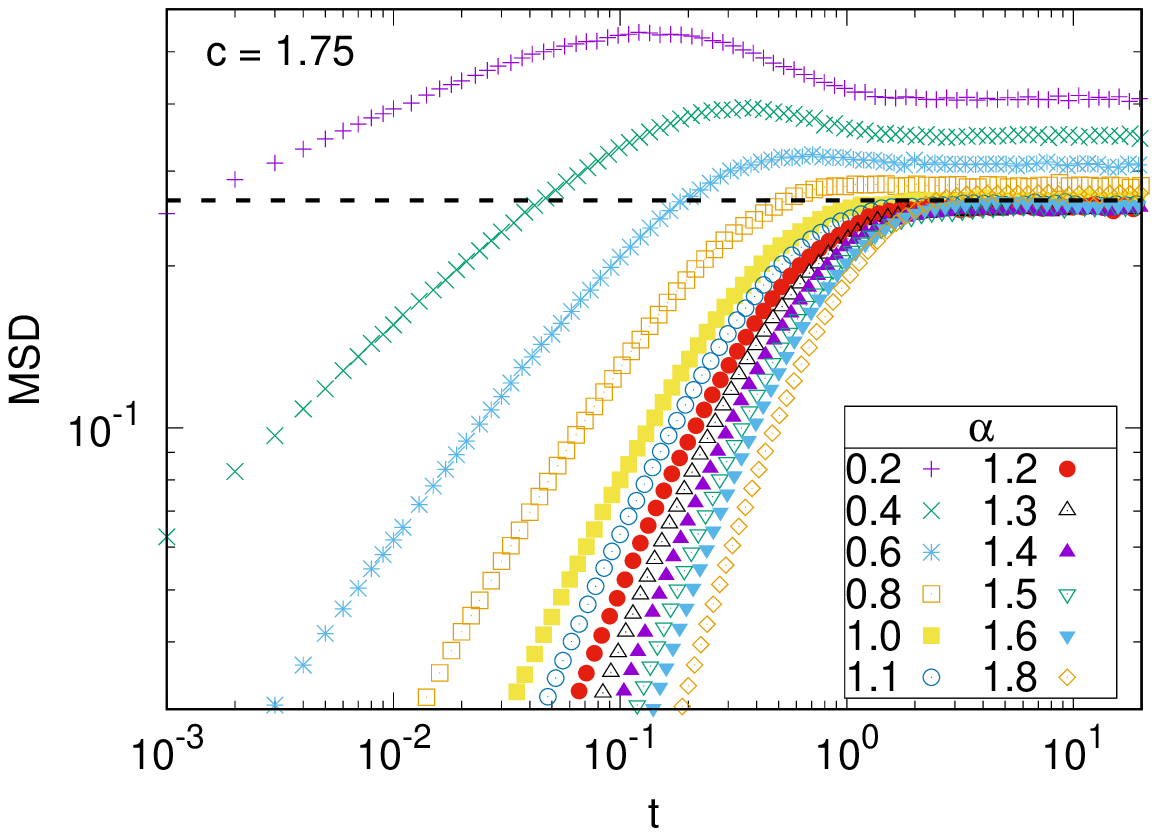}\\
\includegraphics[width=0.48\textwidth]{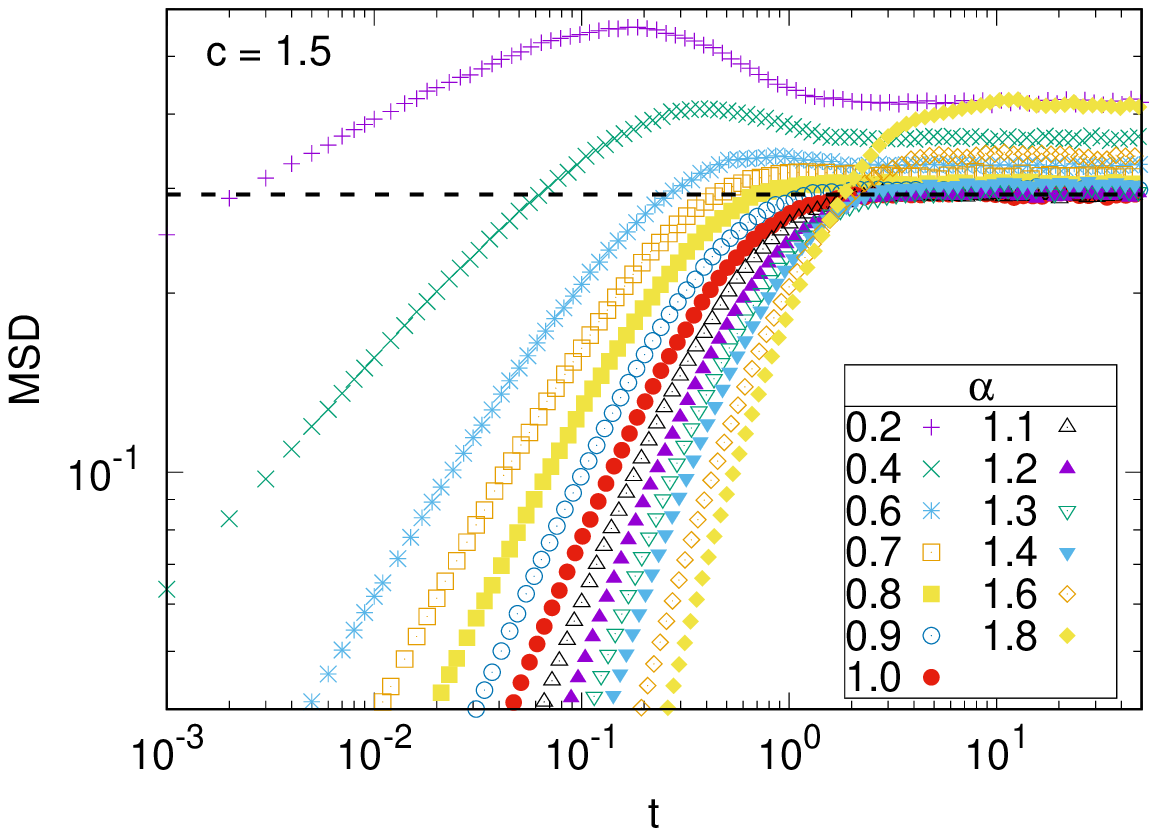}
\includegraphics[width=0.48\textwidth]{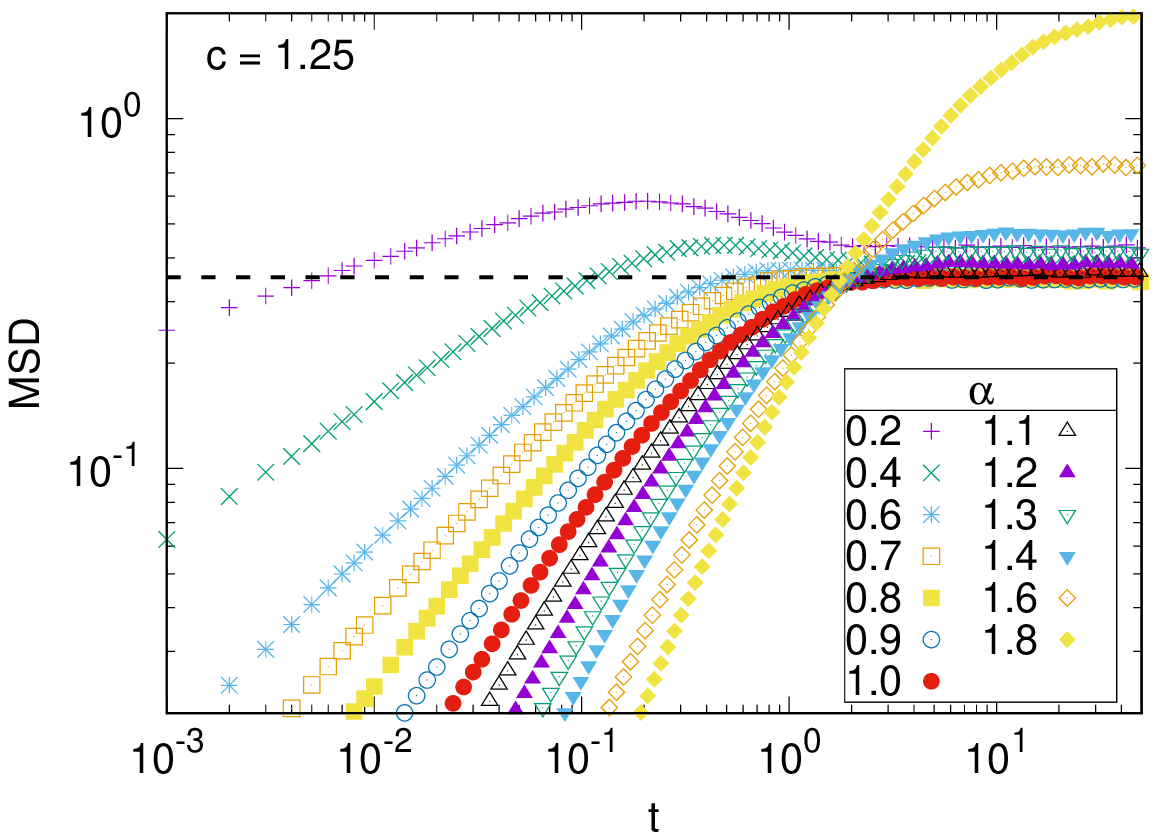}
\caption{MSD for the potential scaling exponents $c=2.0$, $1.75$, $1.5$, and
$1.25$, each shown for different anomalous diffusion exponents $\alpha$.
The solid lines in the top left panel show the theoretical MSD
\eref{eq:FirAndSecMomLeqHarPot} in the harmonic case. The horizontal dashed
lines show the theoretically predicted stationary MSD \eref{eq:statEmsdBrown}
in the Brownian case.}
\label{fig:emsd-c>1}
\end{figure}

\begin{figure}
\includegraphics[width=0.48\textwidth]{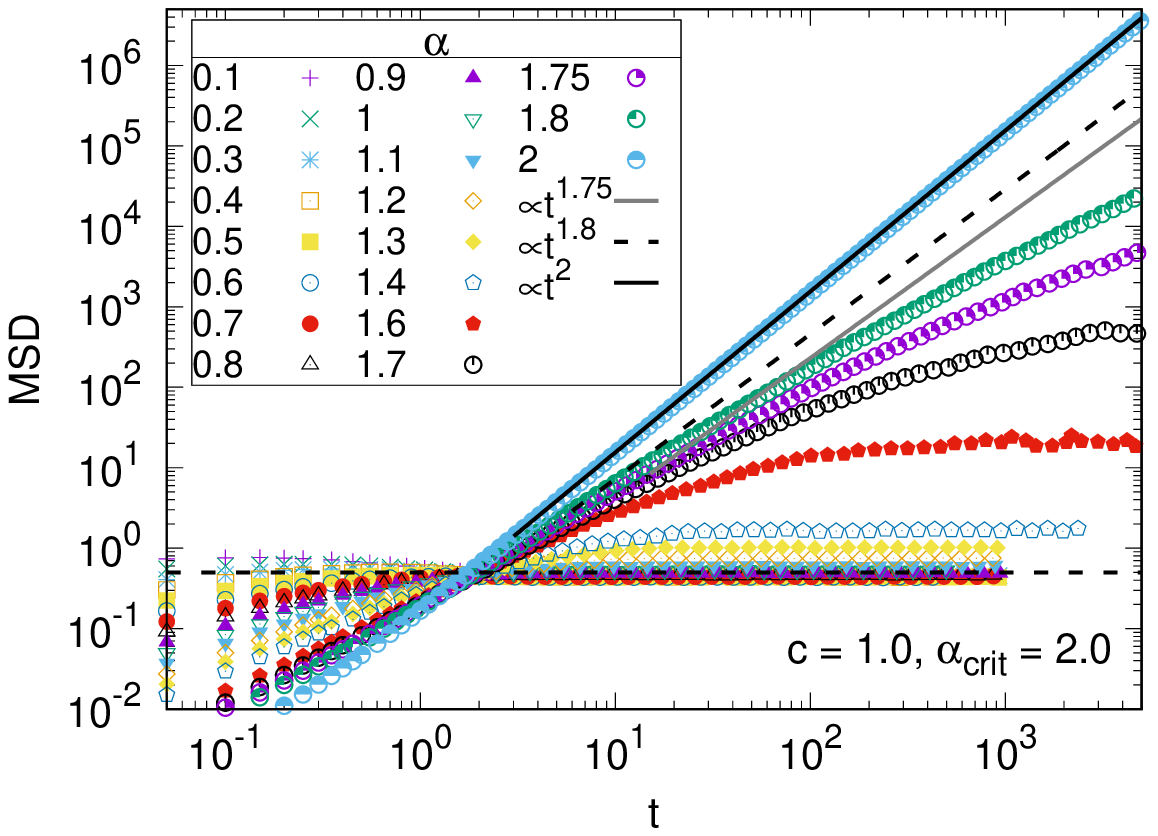}
\includegraphics[width=0.48\textwidth]{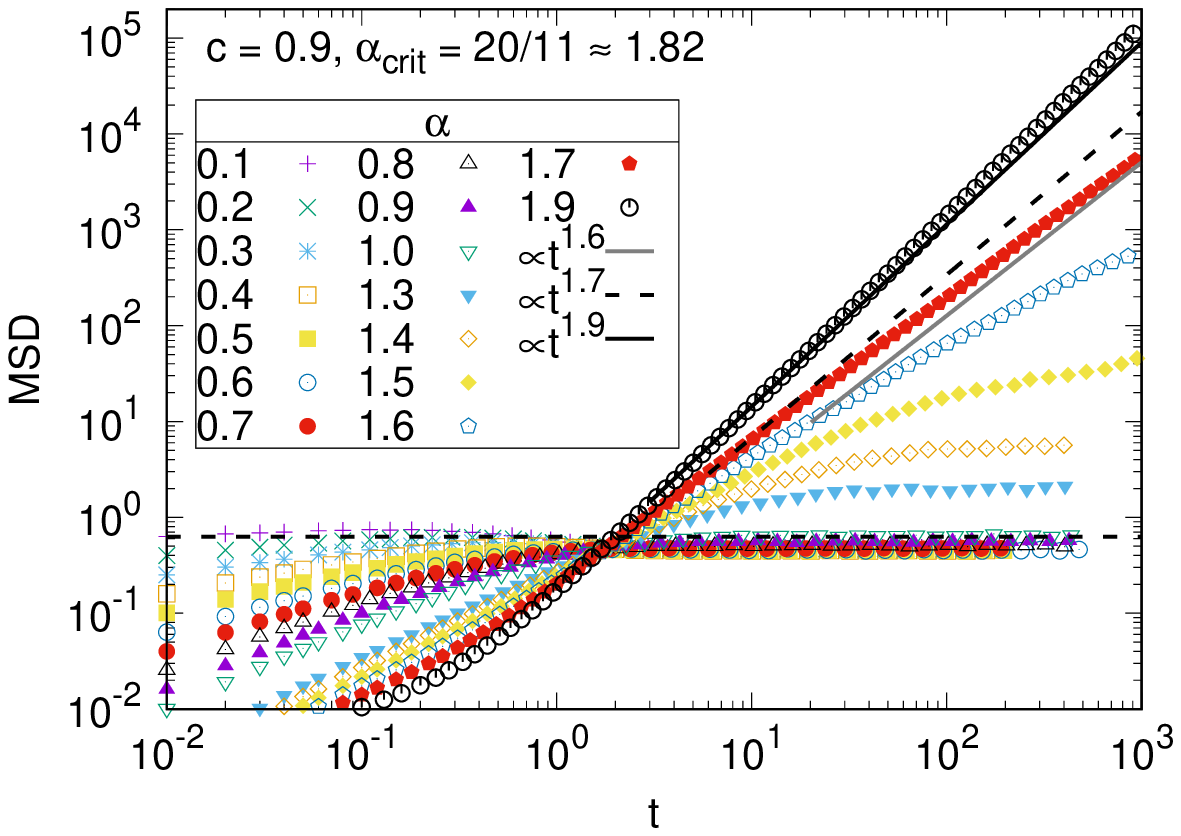}\\
\includegraphics[width=0.48\textwidth]{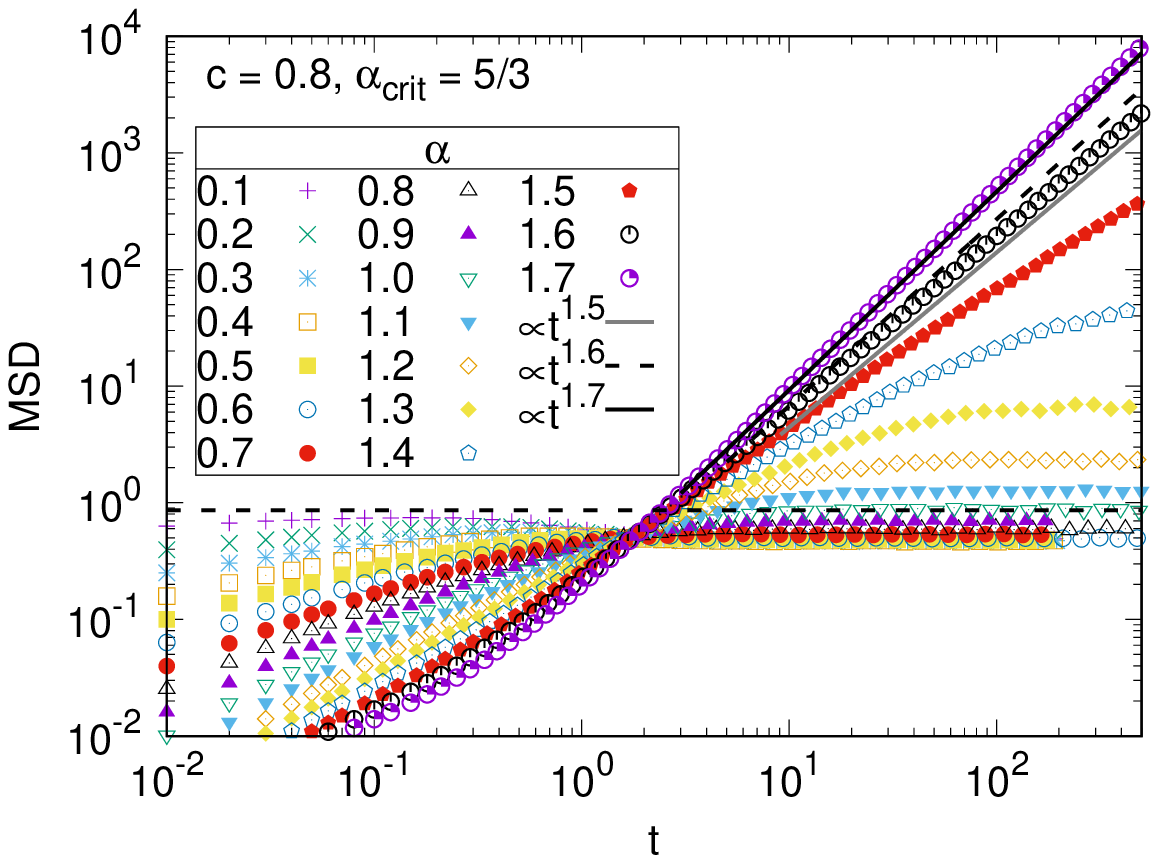}
\includegraphics[width=0.48\textwidth]{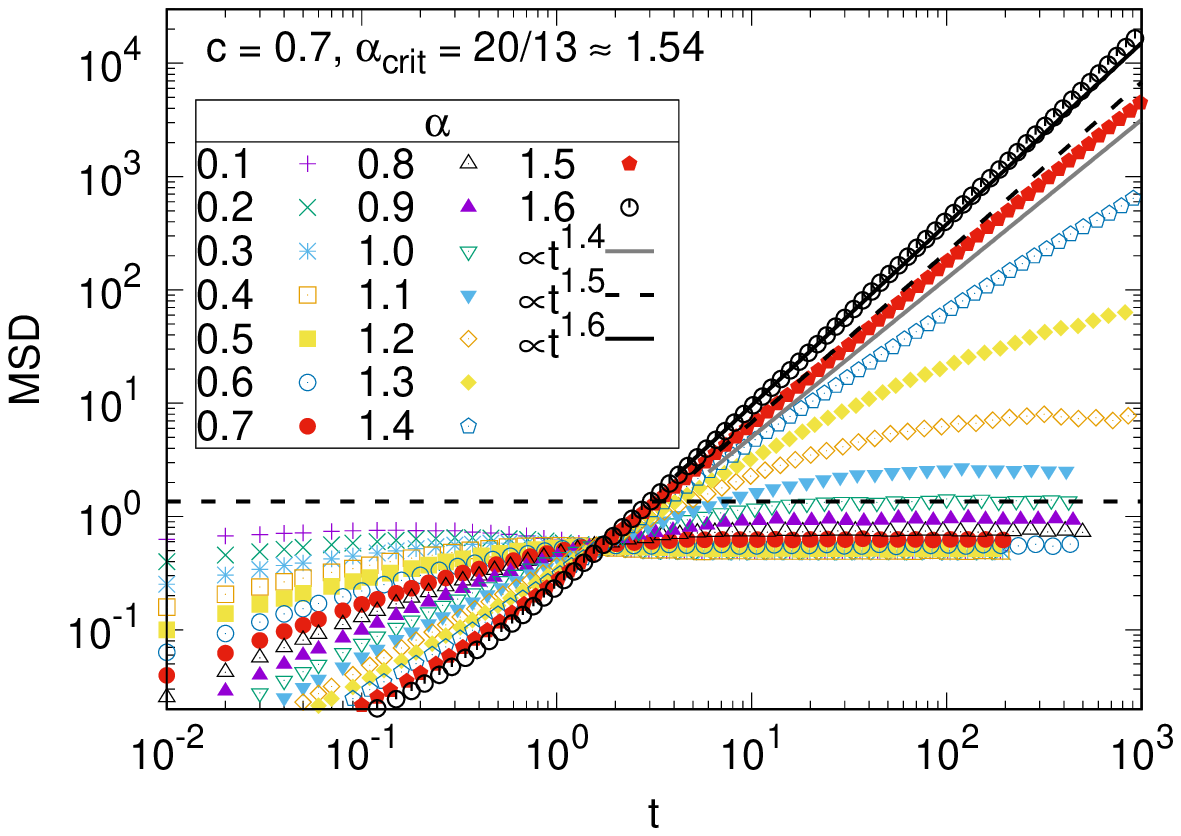}
\caption{MSD for $c=1.0$, $0.9$, $0.8$, and $0.7$, each for different
$\alpha$. The lines show the growth of the MSD of free FBM ($\propto
t^\alpha$ with arbitrary prefactors), see the keys. The horizontal
dashed lines show the theoretical stationary MSD \eref{eq:statEmsdBrown}
in the Brownian case.}
\label{fig:emsd-c<=1}
\end{figure}

\subsubsection{MSD.}
\label{sec:emsd}

Figures \ref{fig:emsd-c>1} and \ref{fig:emsd-c<=1} show the MSD for fixed
scaling exponent $c>1$ and $c\leq1$, respectively, each for different
values of the FGN-exponent $\alpha$. According to our conjecture
\eref{eq:statConj} as long as $c>1$ stationary states should exist for all
values of $\alpha\leq2$. As can be seen in figure \ref{fig:emsd-c>1} the MSD
indeed clearly converges to a stationary value for all $c$ and $\alpha$. We
also note that our simulation results agree well with the theory in the
Brownian and harmonic cases, given by expressions \eref{eq:statEmsdBrown} and
\eref{eq:FirAndSecMomLeqHarPot}.

For $c=1$ stationary states should exist for all $\alpha<\alpha_\mathrm{
crit}=2$, whereas in the ballistic limit $\alpha=2$, no stationary state
should exist. As demonstrated by the top left panel for $c=1$ in figure
\ref{fig:emsd-c<=1} the MSD reaches stationarity for FGN-exponents up to
$\alpha=1.7$. For $\alpha$-values in the range $1.7<\alpha<2$ stationarity
is not fully reached. We attribute this to an increasingly slower
convergence to stationarity for larger $\alpha$, as the comparison to the
growth of the MSD of the corresponding free FBM ($\propto t^\alpha$) clearly
shows a decelerating growth of the MSD when the external potential is present.
In contrast, in the ballistic limit, for which no stationary state should
exist, the MSD grows perfectly proportional to that of free ballistic motion
($\propto t^2$) without any slowing-down.

For $c<1$ stationary states should exist for all $\alpha<\alpha_\mathrm{crit}
=2/(2-c)$ and should not exist for $\alpha\geq\alpha_\mathrm{crit}$. Here, as
shown in figure~\ref{fig:emsd-c<=1} our observation on the existence of
stationary states is analogous to the case $c=1$. Namely, for smaller $\alpha$
values the MSD clearly reaches stationarity. For larger $\alpha$ values, that
still fulfil the criterion $\alpha<\alpha_\mathrm{crit}$ but get close to the
conjectured critical value $\alpha_\mathrm{crit}$ the convergence to
stationarity becomes increasingly slow and stationarity is not fully reached.
Again, the comparison to the growth of the MSD of the corresponding free FBM
($\propto t^\alpha$) clearly shows a decelerating growth of the MSD in those
cases, whereas for $\alpha\geq\alpha_\mathrm{crit}$, for which no stationary
states should exist, the growth of the MSD does not decelerate and is
proportional or even a bit faster than for the corresponding free FBM. The
effect that the observed motion in the presence of the potential accelerates
slightly and eventually catches up with the MSD of the corresponding free
FBM may be understood as follows: initially the particle strongly responds
to the confining potential. Later, when the particle moves away from the
origin and experiences a decreasing restoring force, it more and more moves
like a free particle.

\begin{figure}
\includegraphics[width=0.48\textwidth]{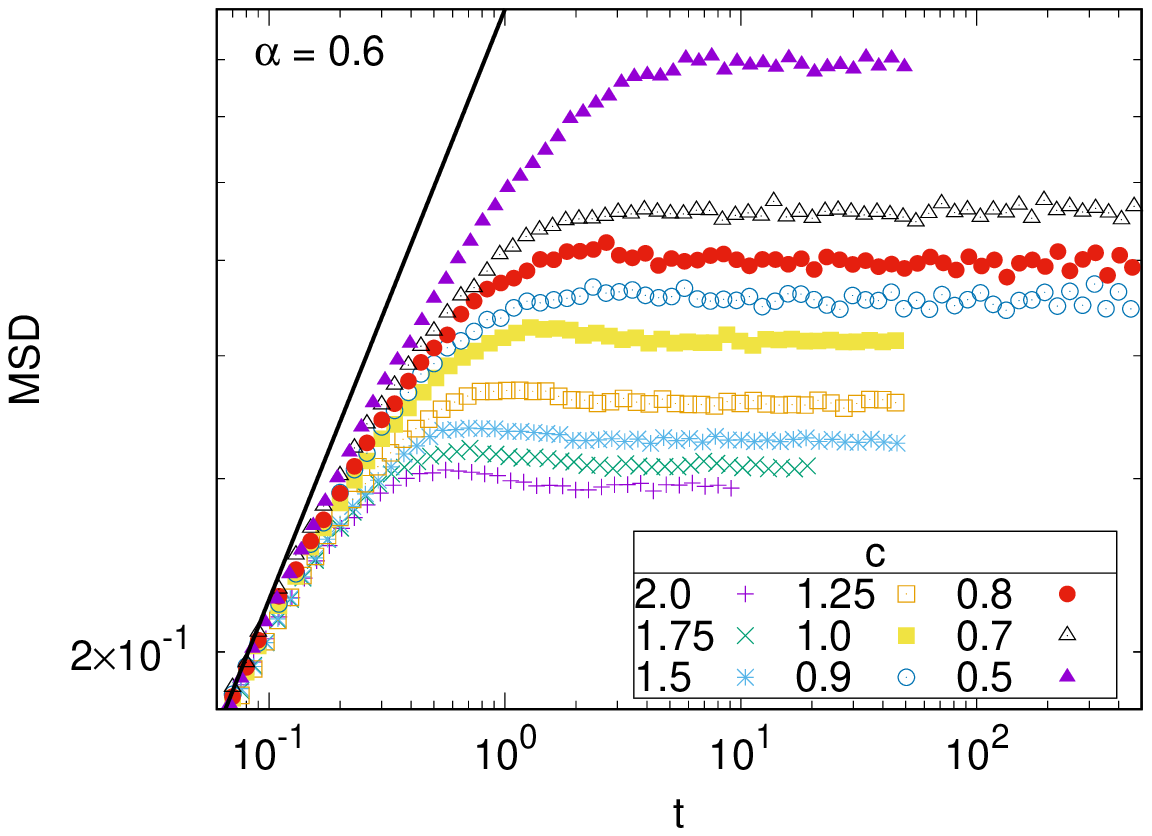}
\includegraphics[width=0.48\textwidth]{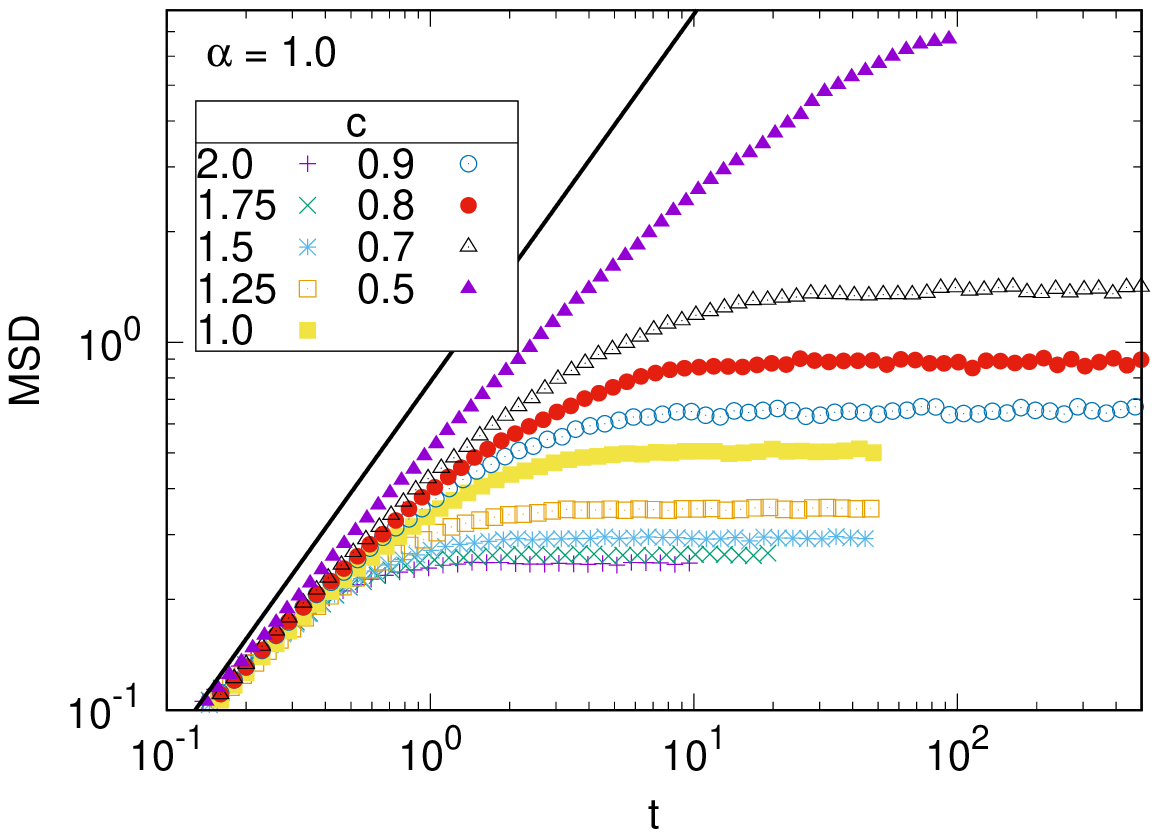}\\
\includegraphics[width=0.48\textwidth]{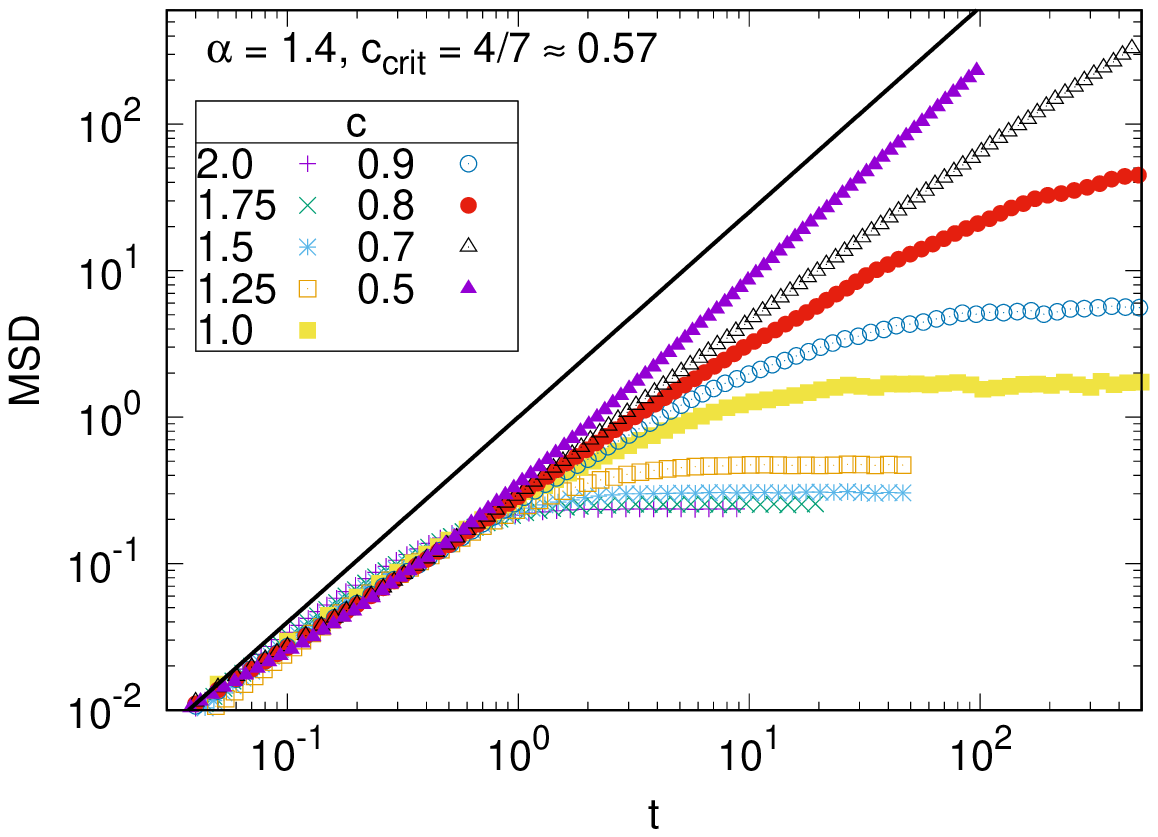}
\includegraphics[width=0.48\textwidth]{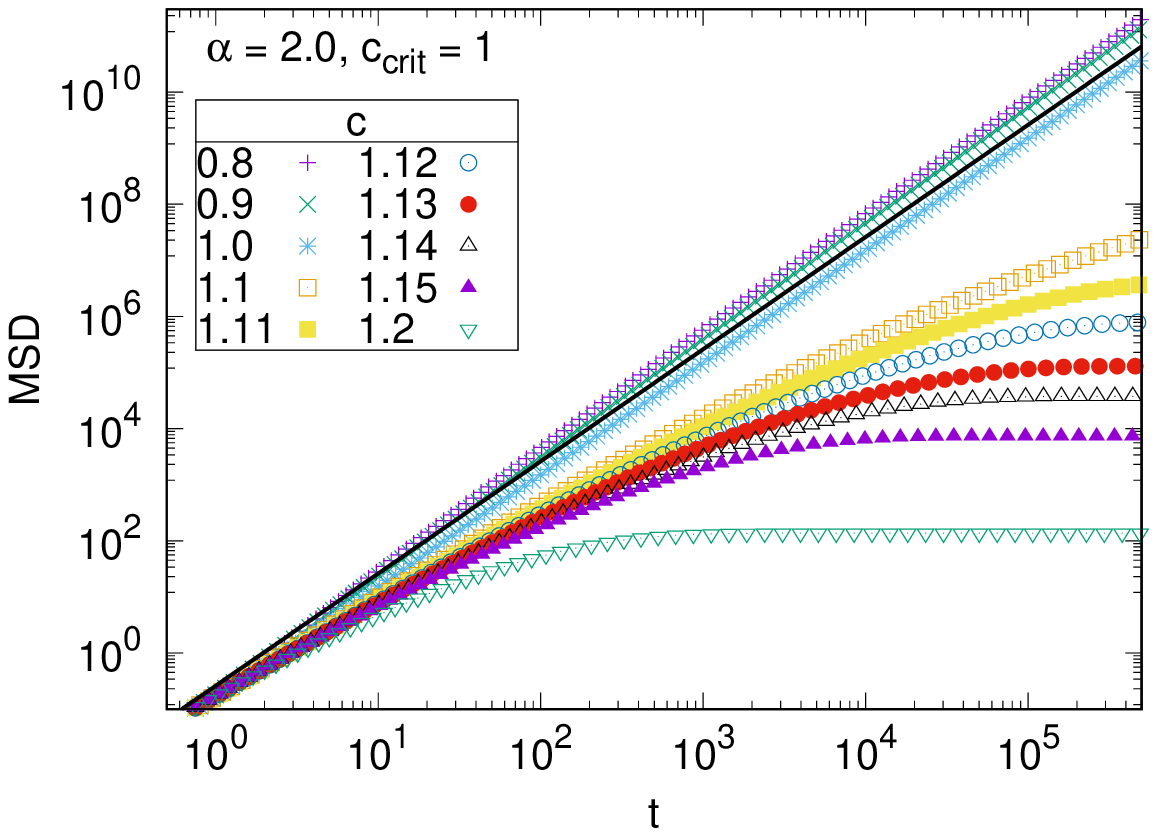}
\caption{MSD for $\alpha=0.6$, $1.0$, $1.4$, and $2.0$, each for different
$c$. The solid black lines show the growth of the MSD of the corresponding
free FBM ($\propto t^\alpha$ with arbitrary prefactor).}
\label{fig:emsd-diffC}
\end{figure}

Figure \ref{fig:emsd-diffC} shows the MSD for fixed $\alpha$ and different
values of the scaling exponent $c$ of the external potential. For $\alpha
\leq1$ stationary states should exist for all $c>0$, while for $\alpha>1$,
they should exist only for $c>c_\mathrm{crit}=2(1-1/\alpha)$. As can be seen
in the figure our simulation results are in agreement with this conjecture,
despite the fact that for $c$ close to the critical value $c_\mathrm{
crit}$ the convergence to stationarity becomes increasingly slow. We emphasise
particularly the clear corroboration of our conjecture in the ballistic limit
$\alpha=2$, for which the critical value is $c_\mathrm{crit} = 1$ (see bottom
right panel in figure \ref{fig:emsd-diffC}).

On top of our discussion of the MSD with regards to the conjecture on the
existence of stationary states, we address some additional properties of
the MSD. First we note that the time to reach stationarity increases with
$\alpha$ (as seen in figures \ref{fig:emsd-c>1} and \ref{fig:emsd-c<=1})
and decreases with $c$ (see figure \ref{fig:emsd-diffC}). For instance, for
$c=1.25$ stationarity is reached at around $t=5$ for $\alpha=1$, while for
$\alpha=1.6$ it is reached at around $t=20$ (see figure \ref{fig:emsd-c>1}).
Likewise, for $\alpha=0.6$ stationarity is reached at around $t=2$ for $c
=2$, while for $c=0.5$ it is reached at around $t=10$ (see figure
\ref{fig:emsd-diffC}). With respect to the dependence on $\alpha$ ($c$),
this effect is more pronounced for smaller $c$ (larger $\alpha$).

\begin{figure}
\includegraphics[width=0.48\textwidth]{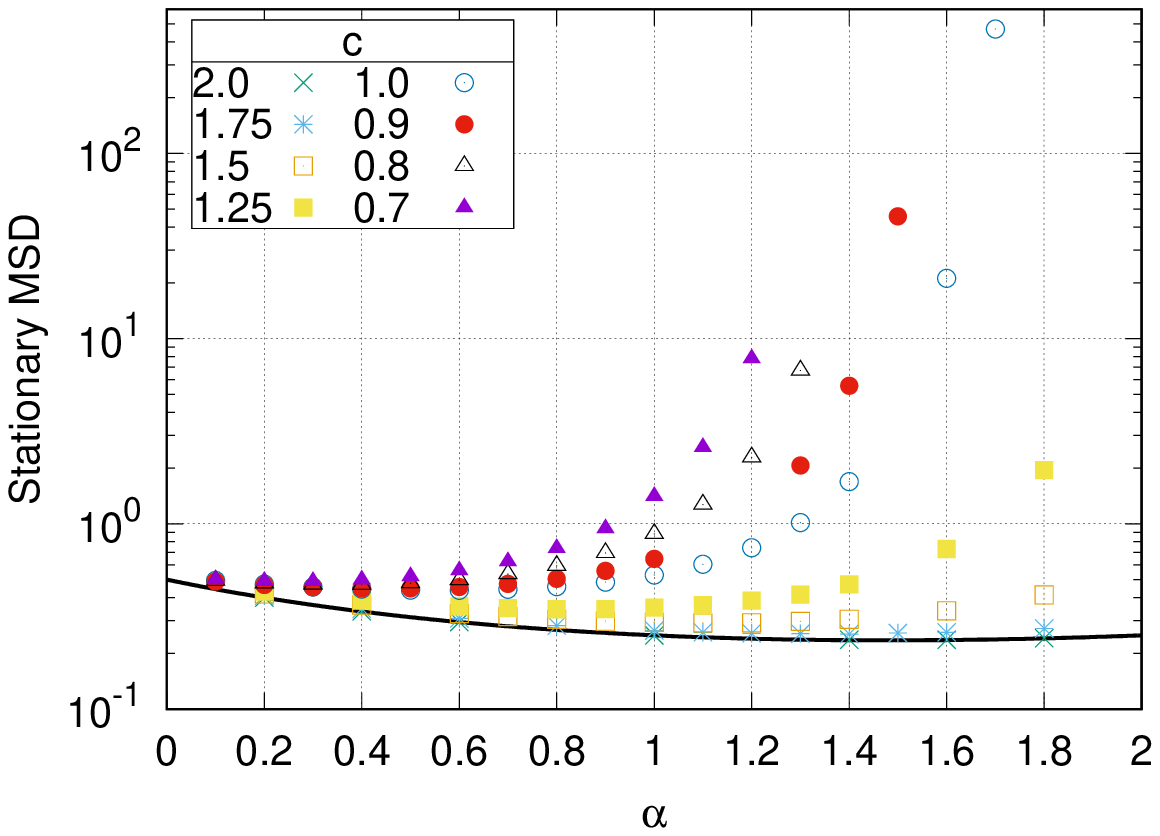}
\includegraphics[width=0.48\textwidth]{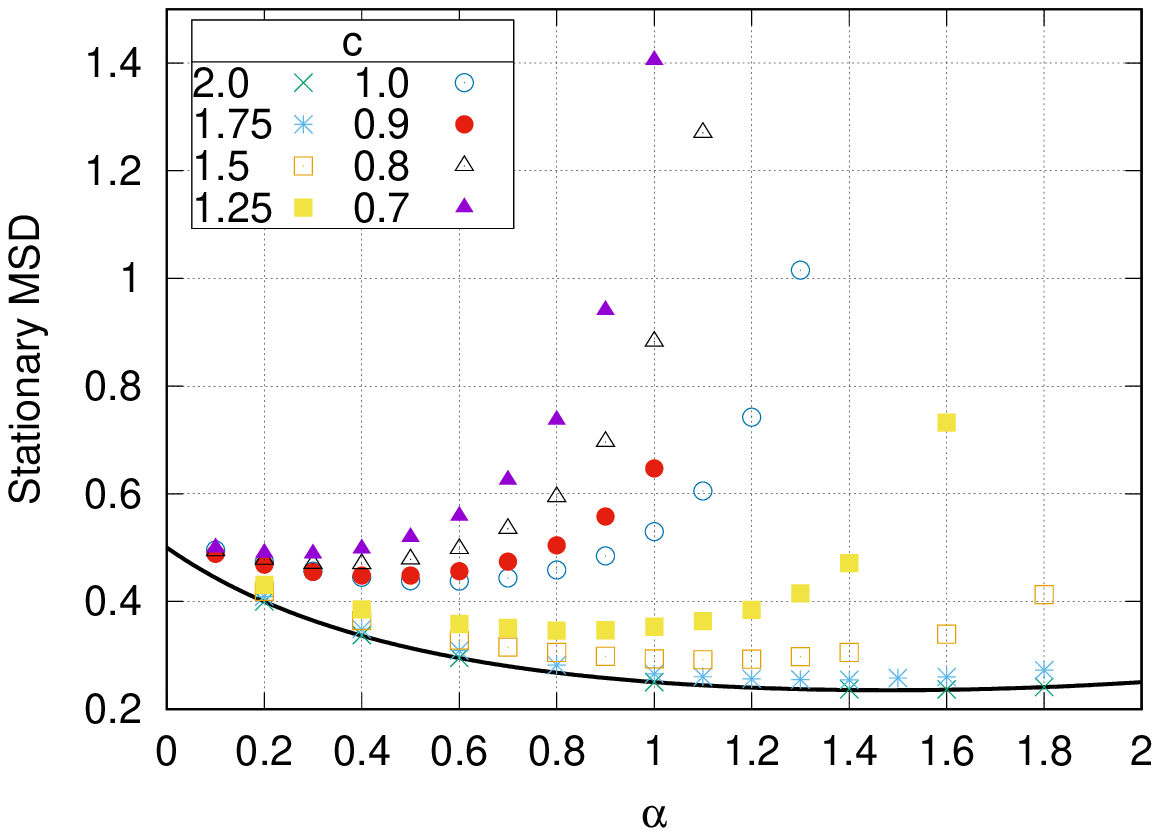}
\caption{Stationary MSD as function of $\alpha$. The values were determined
from the time-dependent MSD by averaging over the plateau regime. The black
line shows the theoretical prediction \eref{eq:statEmsdHarPot} in the
harmonic case. Left: log-lin scales, Right: lin-lin scales (not all data
points shown).}
\label{fig:statEmsdVsAlpha}
\end{figure}

The values of the MSD at stationarity as functions of the exponents $\alpha$
and $c$ are determined from averaging over the plateau regime of the time
dependent MSD. Figure \ref{fig:statEmsdVsAlpha} shows the stationary MSD as
function of $\alpha$. As can be seen the stationary MSD is not monotonic
in $\alpha$: for $\alpha\leq\alpha_0$ it decreases with $\alpha$, while for
$\alpha\geq\alpha_0$ it increases with $\alpha$. Here $\alpha_0$ is the
value, which separates these two regimes. The value $\alpha_0$ increases
with $c$, for instance, we have $\alpha_0(c=0.8)\approx0.4$ and $\alpha_0
(c=1.25)\approx0.9$ (see the right panel of figure \ref{fig:statEmsdVsAlpha}).
We note that this non-monotonic behaviour is already present in the harmonic
case and is in agreement with the theoretical prediction \eref{eq:statEmsdHarPot}.
Conversely, the stationary MSD is monotonically decreasing with $c$, as one
would intuitively expect (see figure \ref{fig:emsd-diffC}). This property can
also be seen from figure \ref{fig:statEmsdVsC} in the appendix which shows
the stationary MSD as function of $c$.

We finally mention the "overshooting" of the MSD before reaching stationarity
for smaller $\alpha$ values ($\alpha<1$). This phenomenon is already present
in the harmonic case (see figure \ref{fig:emsd-c>1}) and is also encoded in
the analytical result (\ref{eq:FirAndSecMomLeqHarPot}), see also the discussion
in \cite{pre12,jochen}. For $\alpha\geq1$ and small $c$ (see figure
\ref{fig:emsd-diffC}) this effect is not observed.

\subsubsection{\label{sec:pdf}PDF}

\begin{figure}
\includegraphics[width=0.48\textwidth]{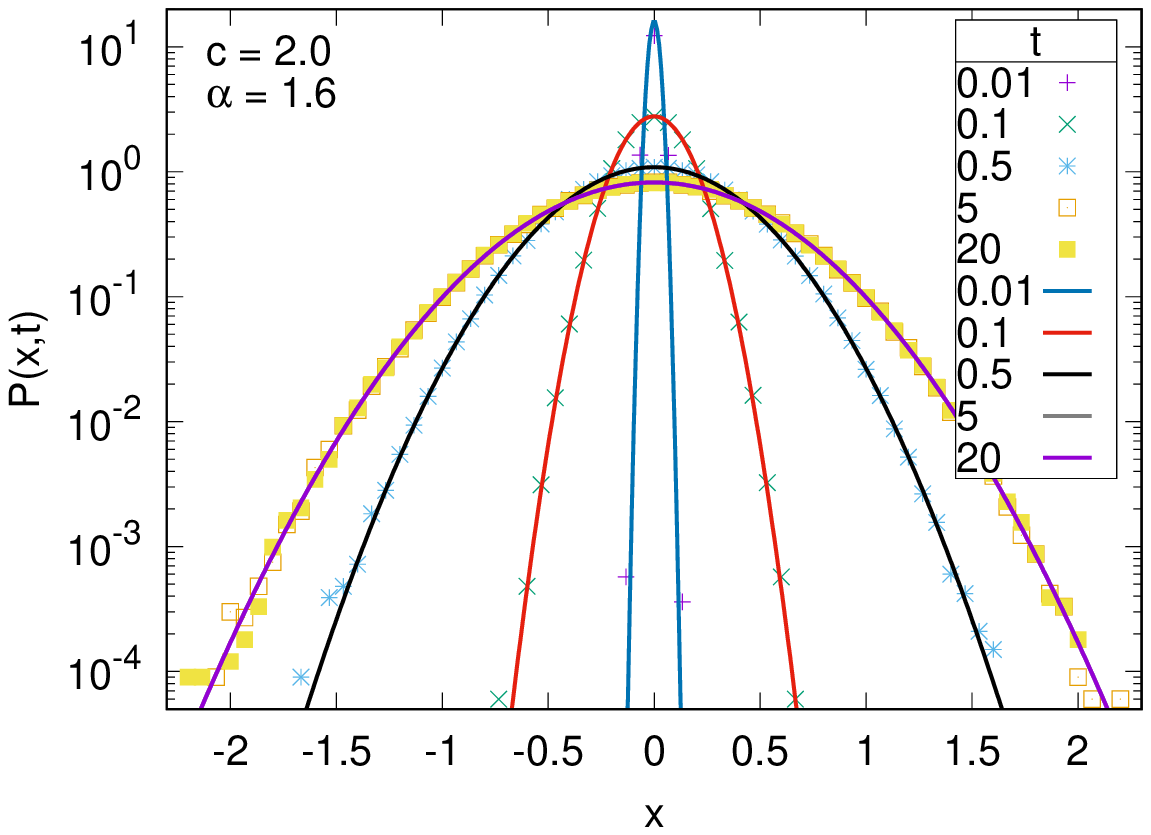}
\includegraphics[width=0.48\textwidth]{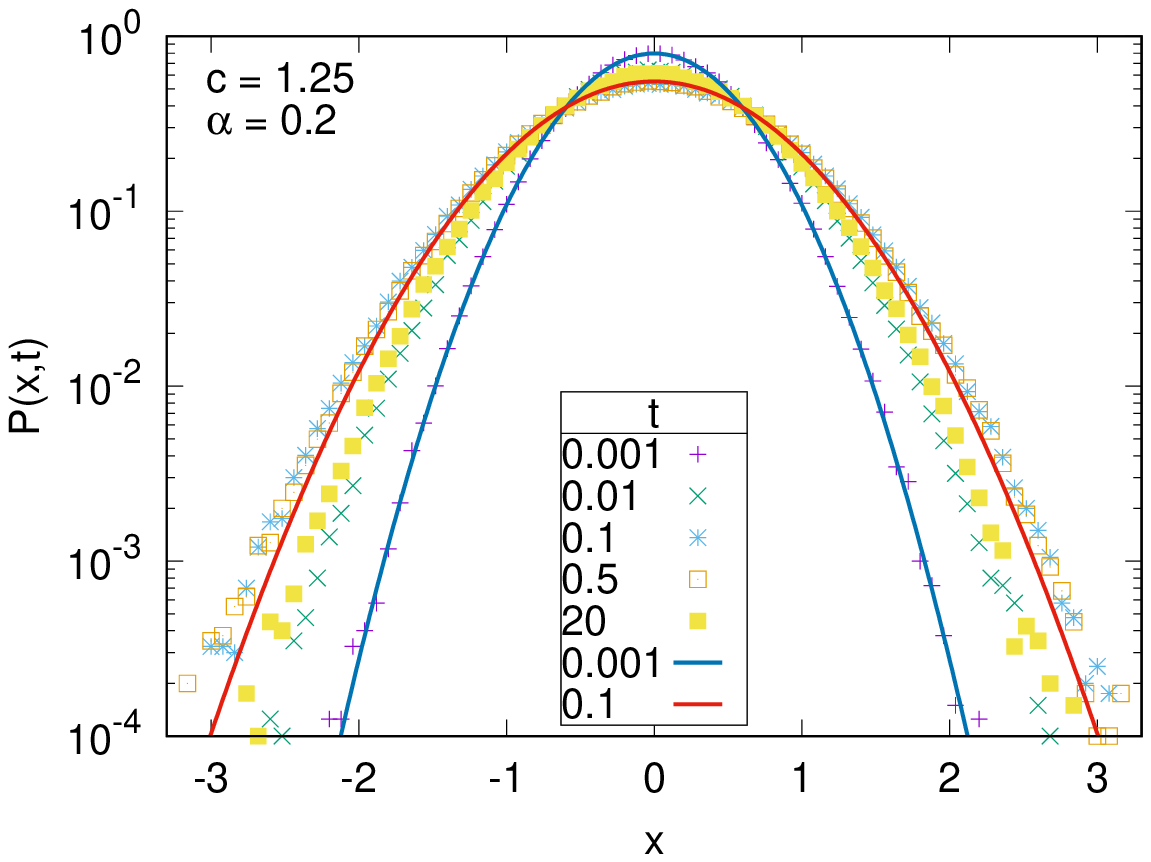}\\
\includegraphics[width=0.48\textwidth]{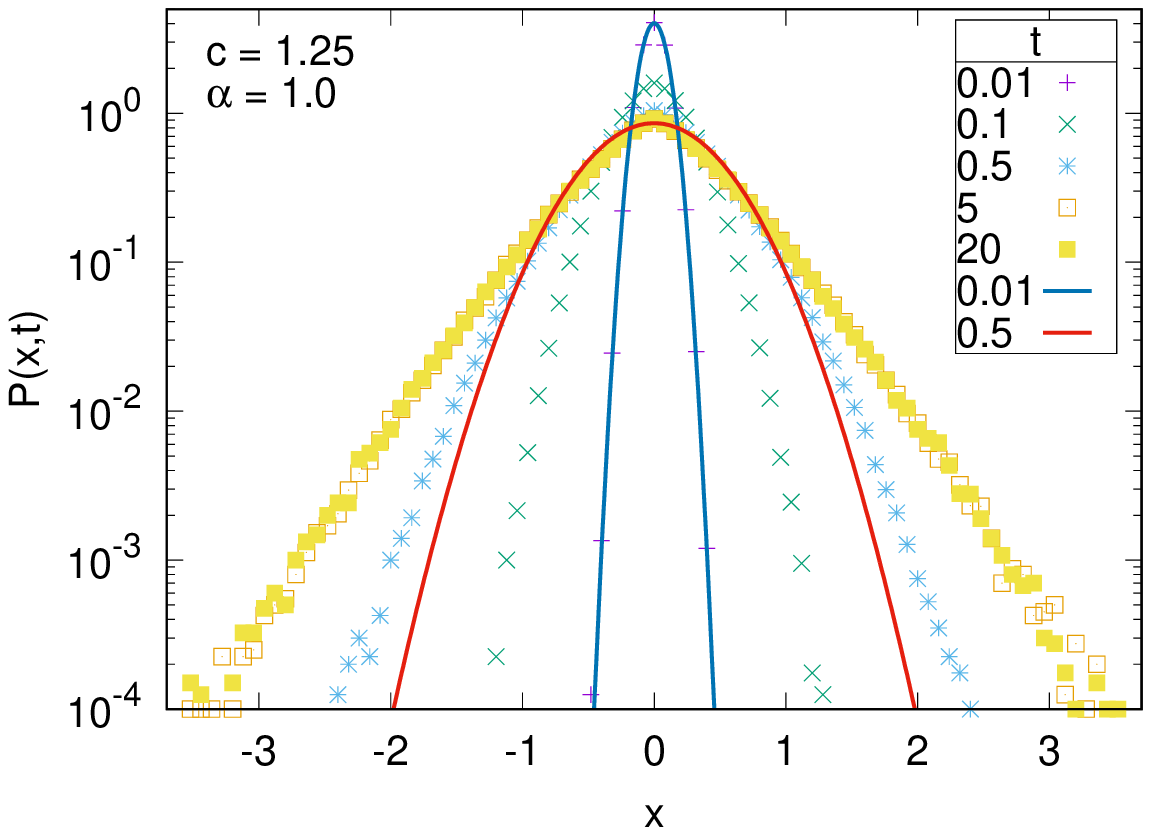}
\includegraphics[width=0.48\textwidth]{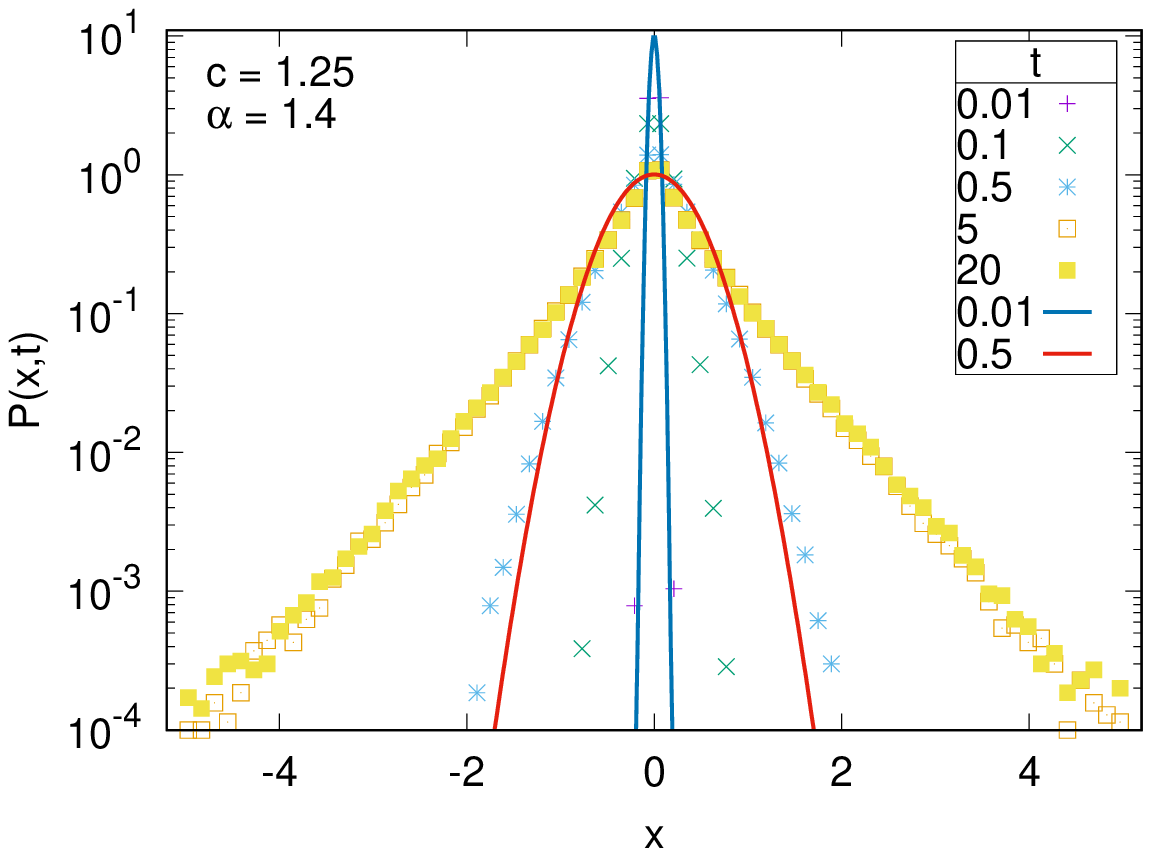}
\caption{Time-dependent PDF for the cases $c=2$ (harmonic potential, top
left panel) and $c=1.25$ for different $\alpha$ (remaining panels). The
solid lines show the corresponding theoretical PDF \eref{eq:solFpeHar}
in the harmonic case. Note that the width of the PDF for $c=1.25$ and
$\alpha=0.2$ initially increases (until approximately the curves for
$t=0.1$ and $0.5$) and then decreases ($t=20$). This corresponds to the
above-mentioned "overshooting" of the MSD (compare with the corresponding
MSD in figure \ref{fig:emsd-c>1}).}
\label{fig:timDepPdf}
\end{figure}

We now turn to the analysis of the PDF. Before addressing the stationary
PDF, figure \ref{fig:timDepPdf} shows as example the time-dependent PDF for
the harmonic case $c=2$ and $c=1.25$. The simulation results agree well with
the theoretical Gaussian PDF \eref{eq:solFpeHar}. For the non-harmonic
potentials with $c>1$ the PDF agrees with the solution in the harmonic
case at short times, an expected behaviour as long as the particle does not
yet fully engage with the external potential. After this initial behaviour
the PDF starts to deviate, and for persistent noise ($\alpha>1$) the PDF
clearly assumes pronouncedly non-Gaussian shapes at long times.

Before analysing the stationary PDF in detail, some words about the
convergence to stationarity are in order. In our numerical analysis we
approximate the stationary PDF by the PDF taken at the longest simulated
time $t_\mathrm{max}$, i.e., we take $P_{\mathrm{st}}(x)\approx P(x,t_
\mathrm{max})$. For this approximation to be meaningful we determined the
time $t_\mathrm{st}$ to reach stationarity as the earliest time when the
MSD reaches stationarity and ensured that $t_{\mathrm{max}}\geq t_\mathrm{
st}$. Following this procedure, in our analysis of the stationary PDF we
limit ourselves to those parameter values of $\alpha$ and $c$ for which
stationarity is fully reached in the simulations.

\begin{figure}
\includegraphics[width=0.48\textwidth]{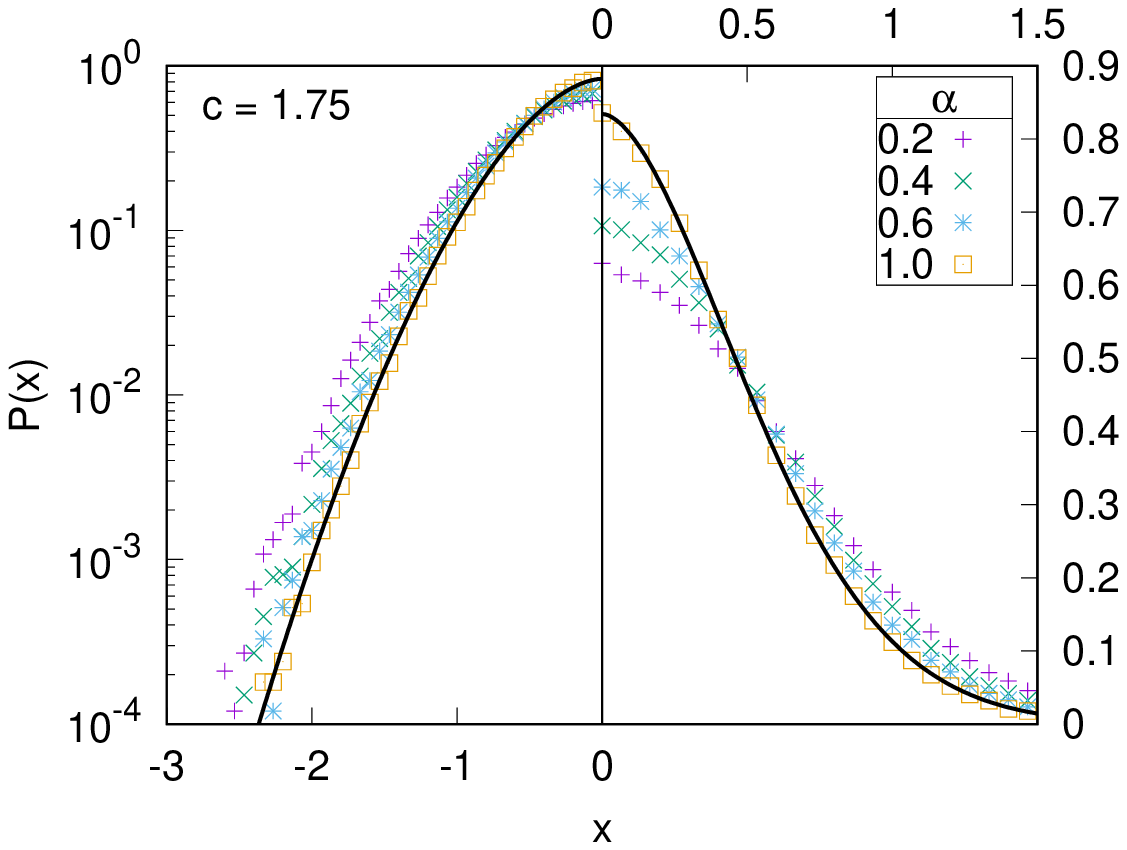}
\includegraphics[width=0.48\textwidth]{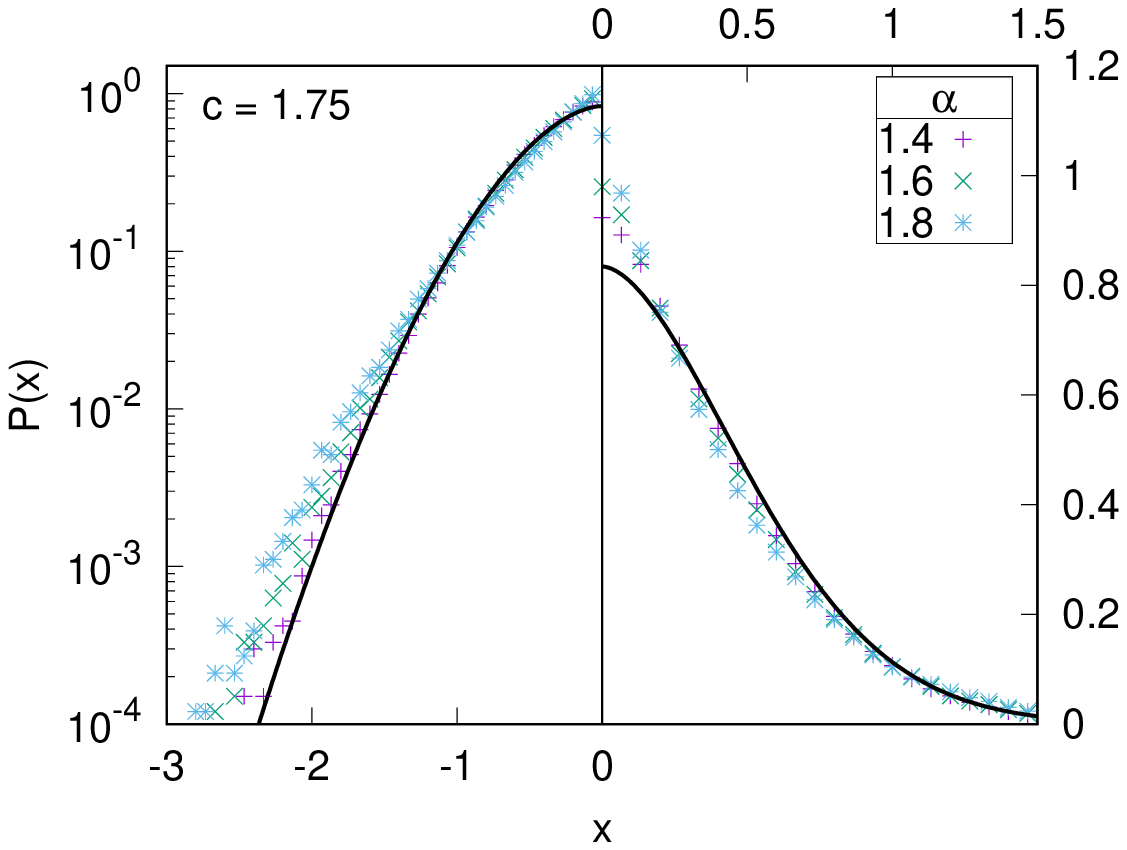}\\
\includegraphics[width=0.48\textwidth]{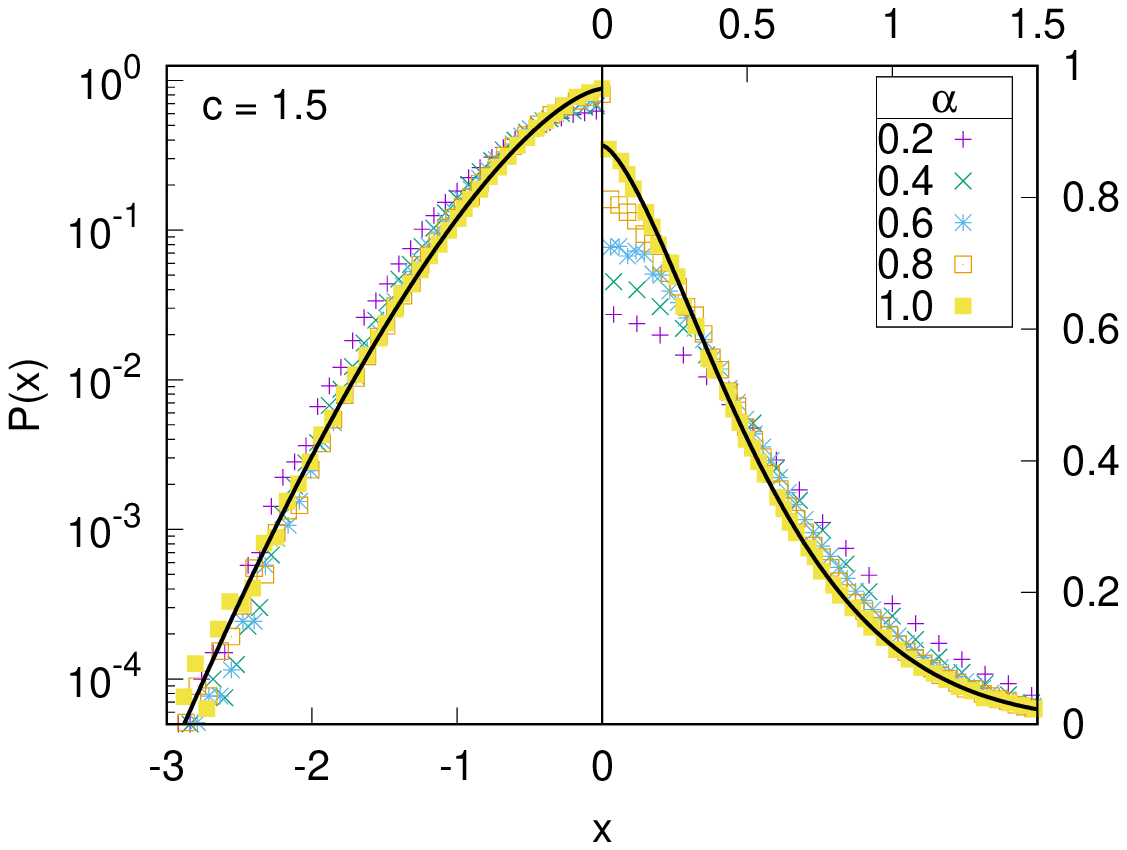}
\includegraphics[width=0.48\textwidth]{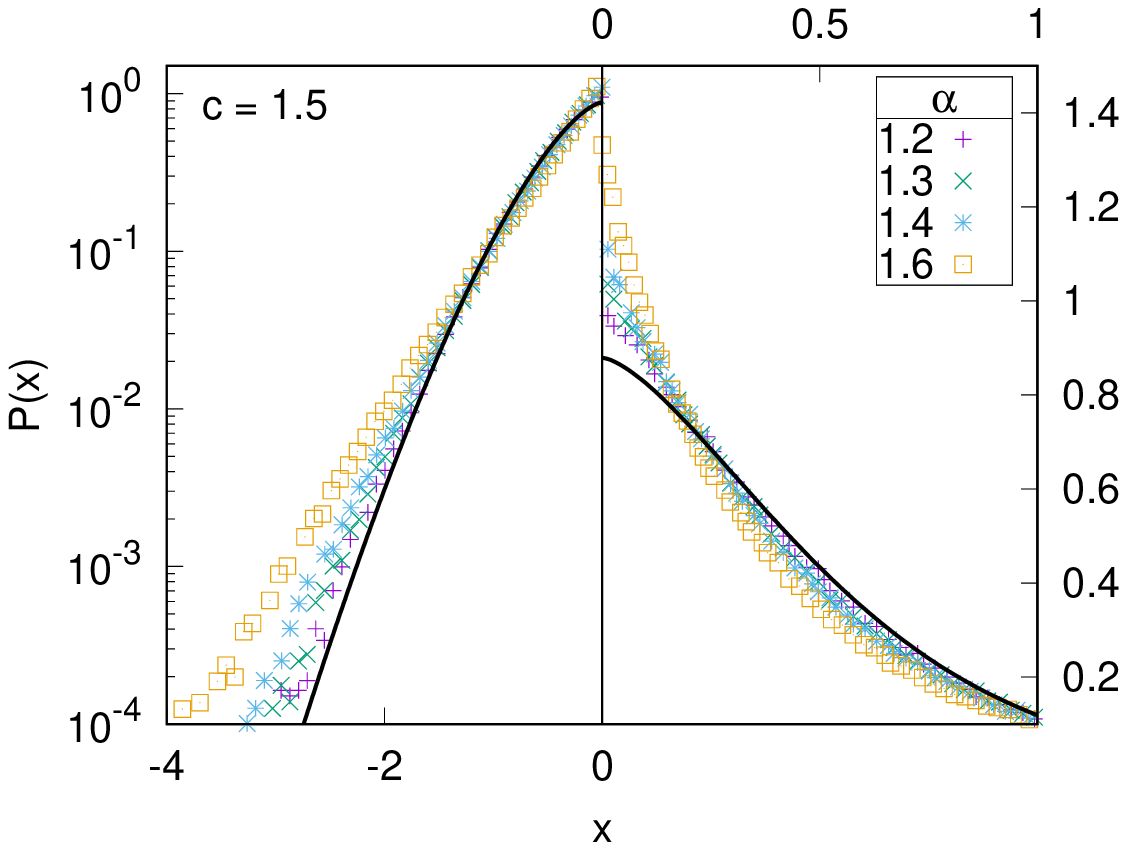}\\
\includegraphics[width=0.48\textwidth]{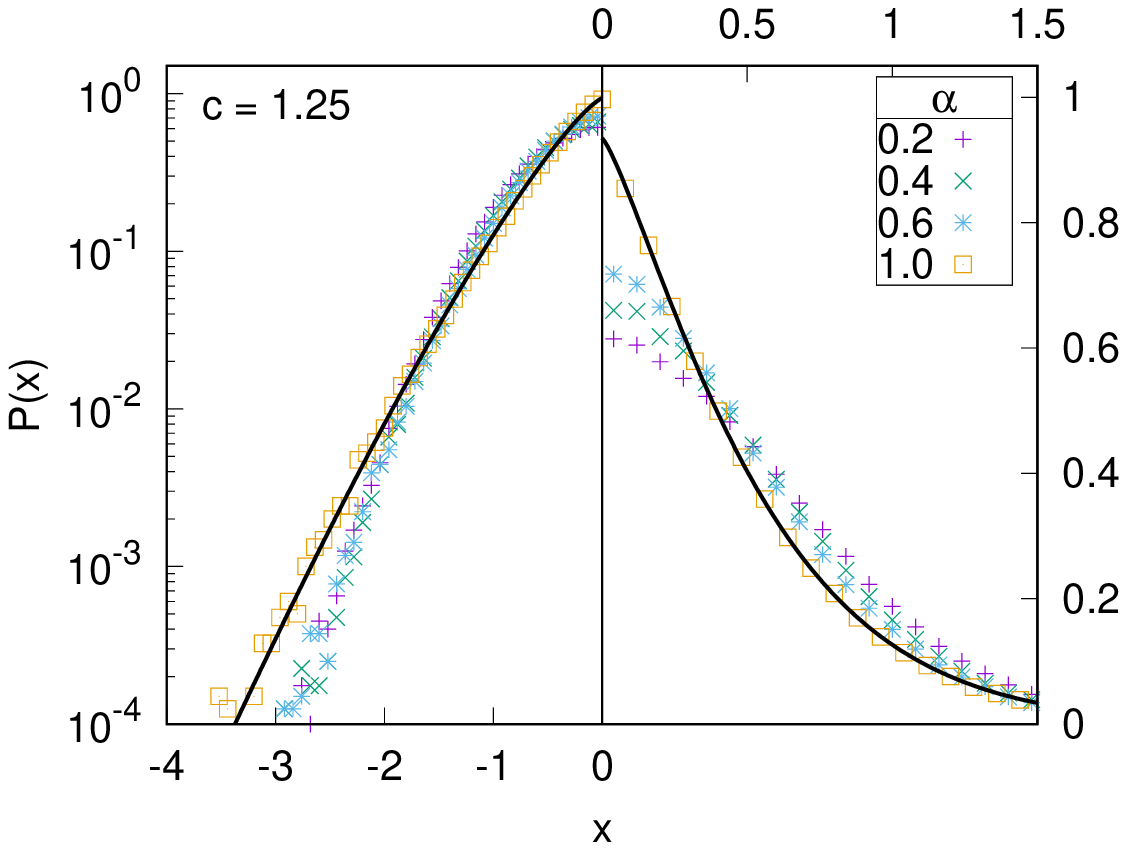}
\includegraphics[width=0.48\textwidth]{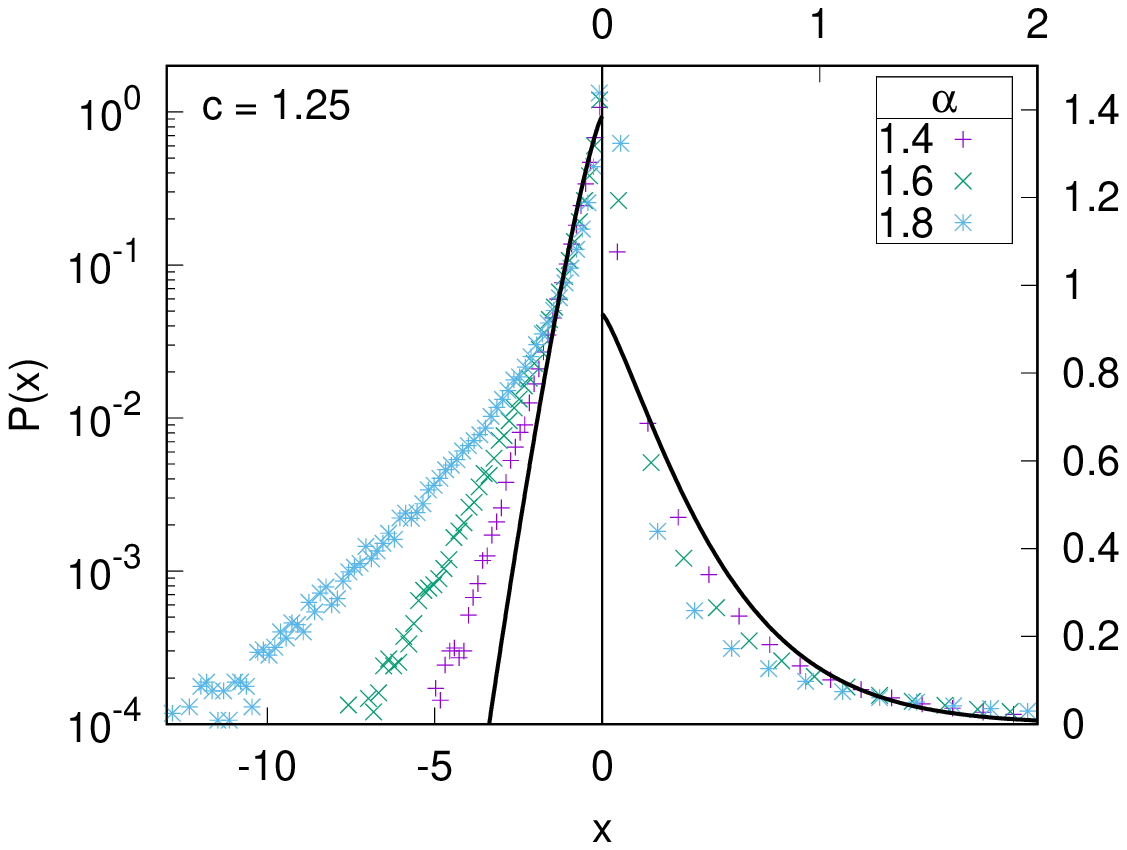}
\caption{Stationary PDF for $c=1.75$, $1.5$, and $1.25$, each shown for
different $\alpha$. For comparison, the black lines show the corresponding
theoretical stationary PDF \eref{eq:statPdfBrown} in the Brownian case.
Since the stationary PDF is symmetric about the $y$-axis, for $x<0$ the
data (including the theoretical PDF) are plotted logarithmically (left
and bottom axes in each panel) and for $x>0$ linearly (right and top
axes).}
\label{fig:statPdf-c>1-aDiff}
\end{figure}

\begin{figure}
\centering
\includegraphics[width=0.48\textwidth]{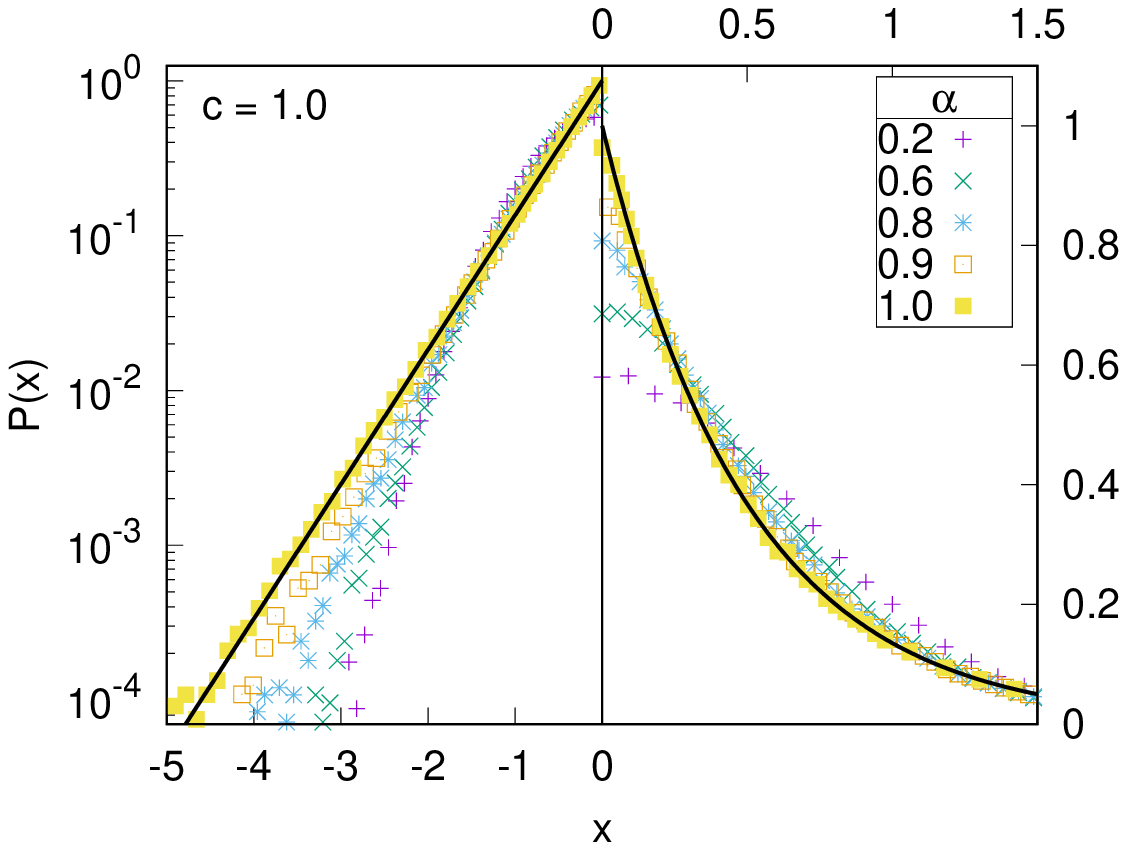}
\includegraphics[width=0.48\textwidth]{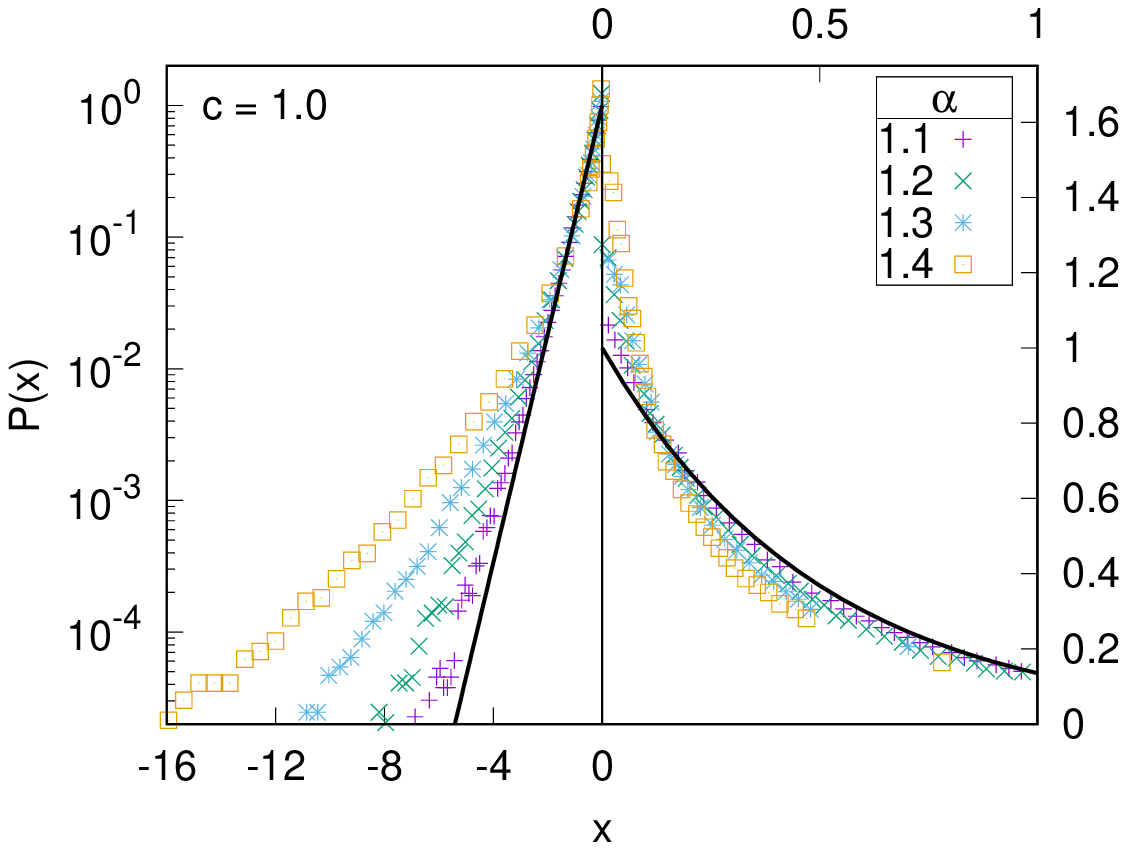}\\
\includegraphics[width=0.48\textwidth]{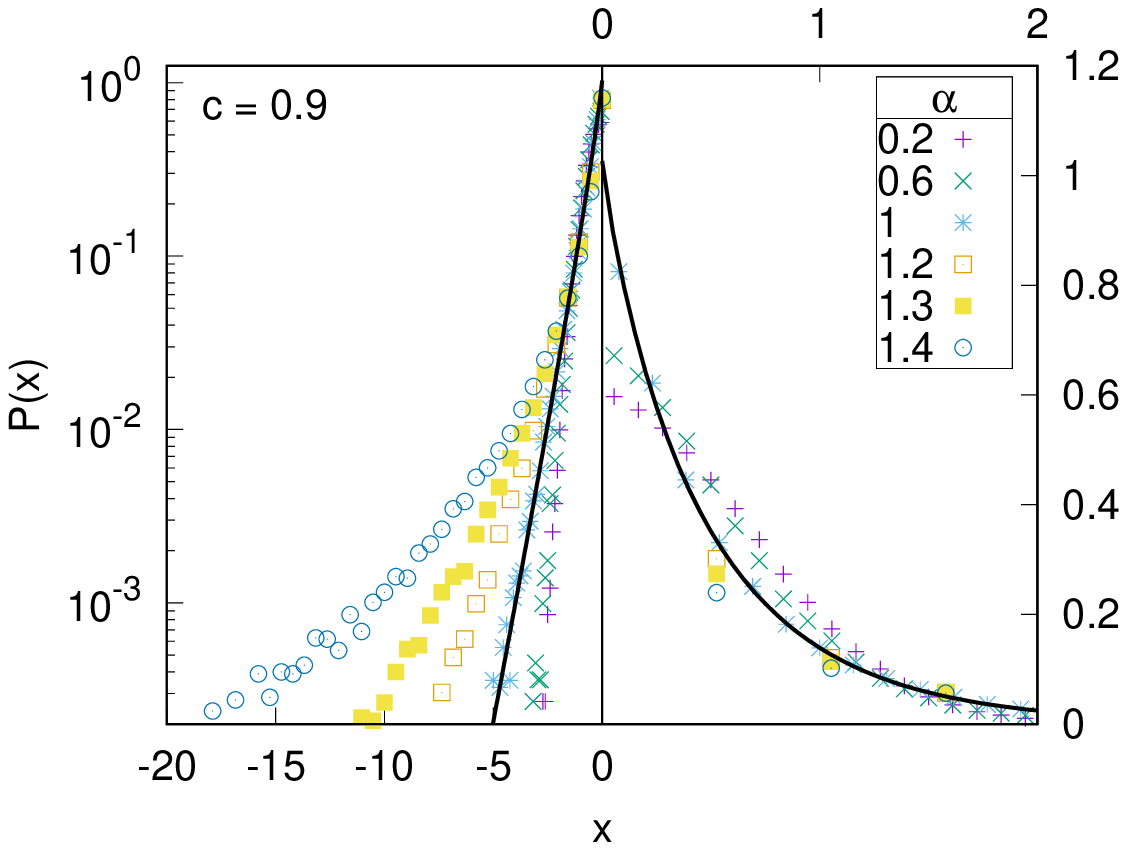}
\includegraphics[width=0.48\textwidth]{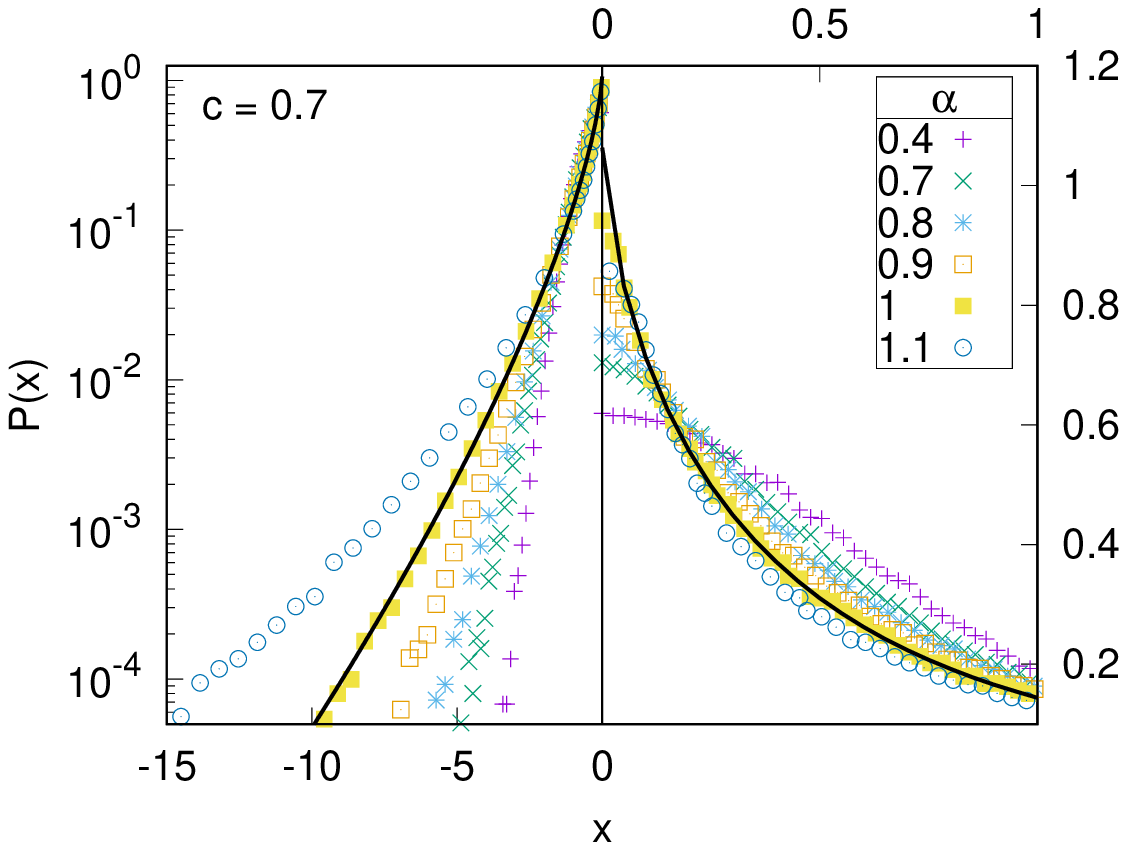}
\caption{Stationary PDF for $c=1.0$, $0.9$, and $0.7$, each for different
$\alpha$. For comparison, the black lines show the corresponding theoretical
stationary PDF \eref{eq:statPdfBrown} in the Brownian case. For $x<0$ the
data (including the theoretical PDF) are plotted logarithmically (left and
bottom axes) and for $x>0$ linearly (right and top axes).}
\label{fig:statPdf-c<1}
\end{figure}

\begin{figure}
\centering
\includegraphics[width=0.48\textwidth]{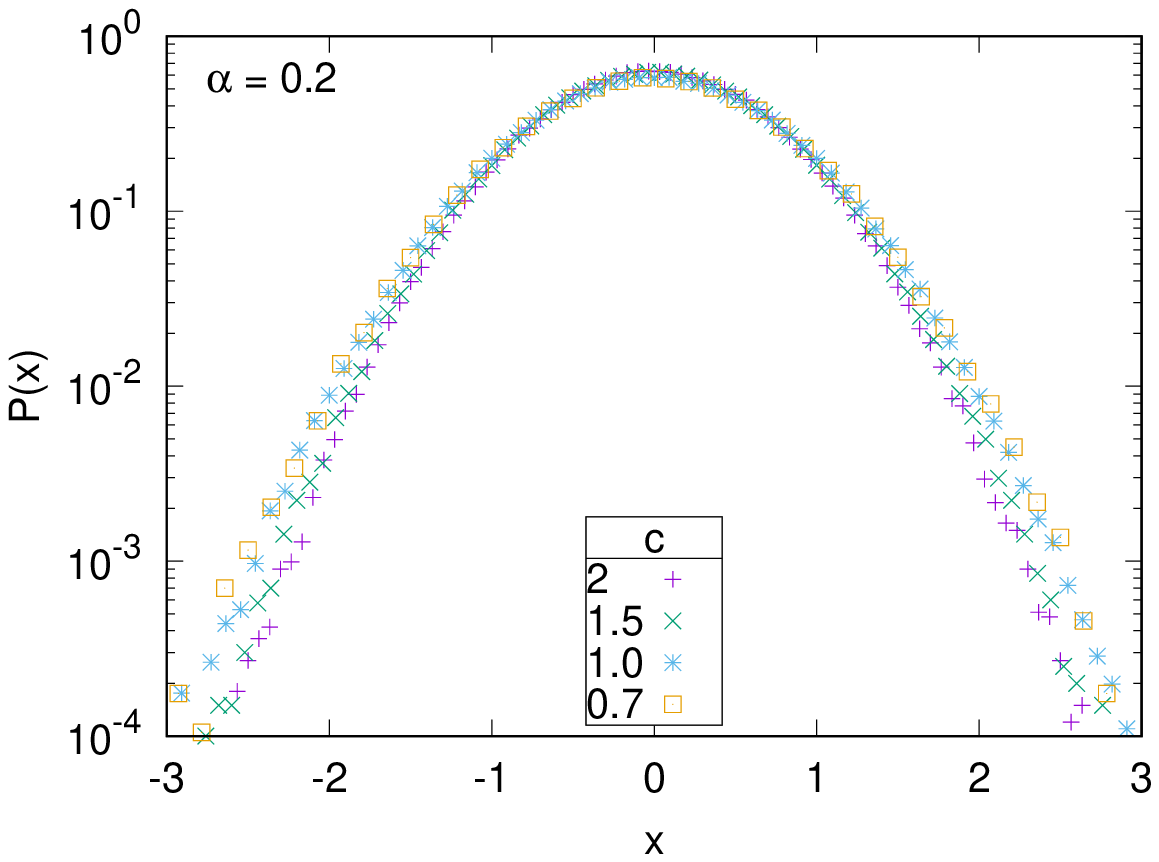}
\includegraphics[width=0.48\textwidth]{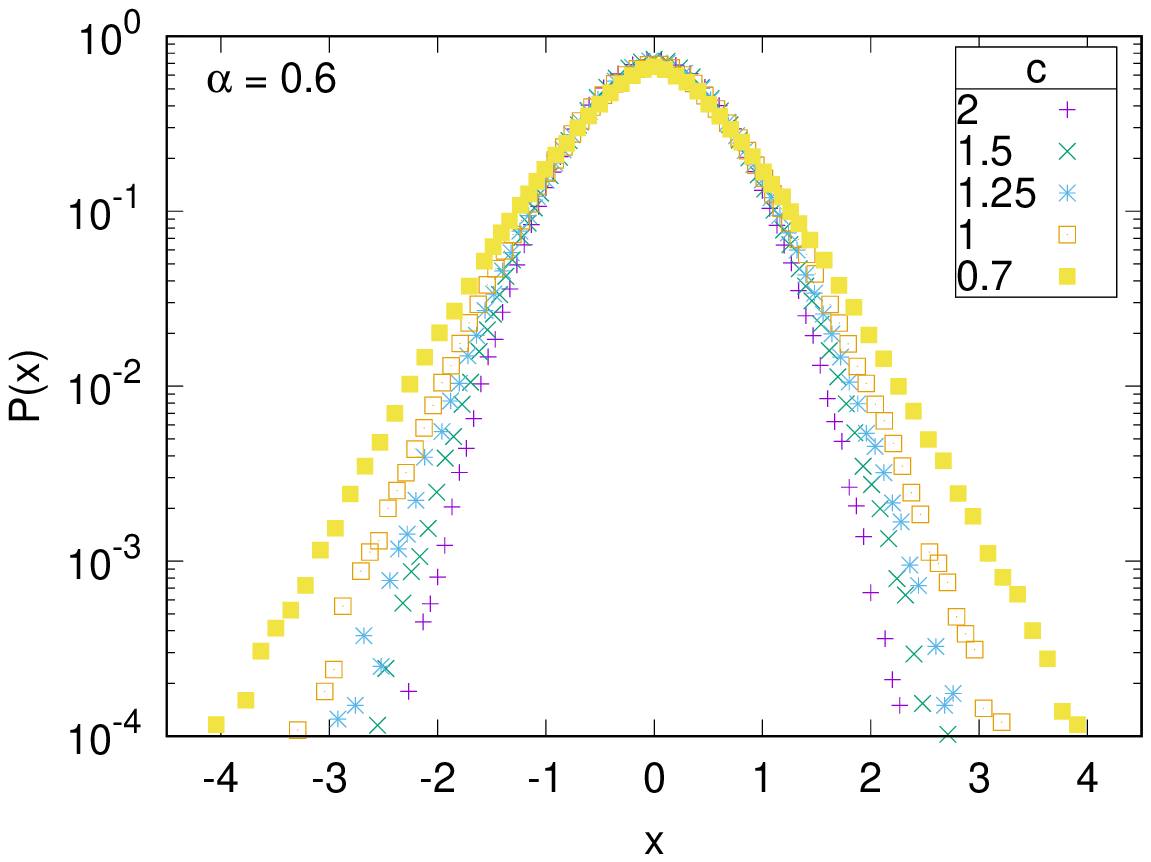}\\
\includegraphics[width=0.48\textwidth]{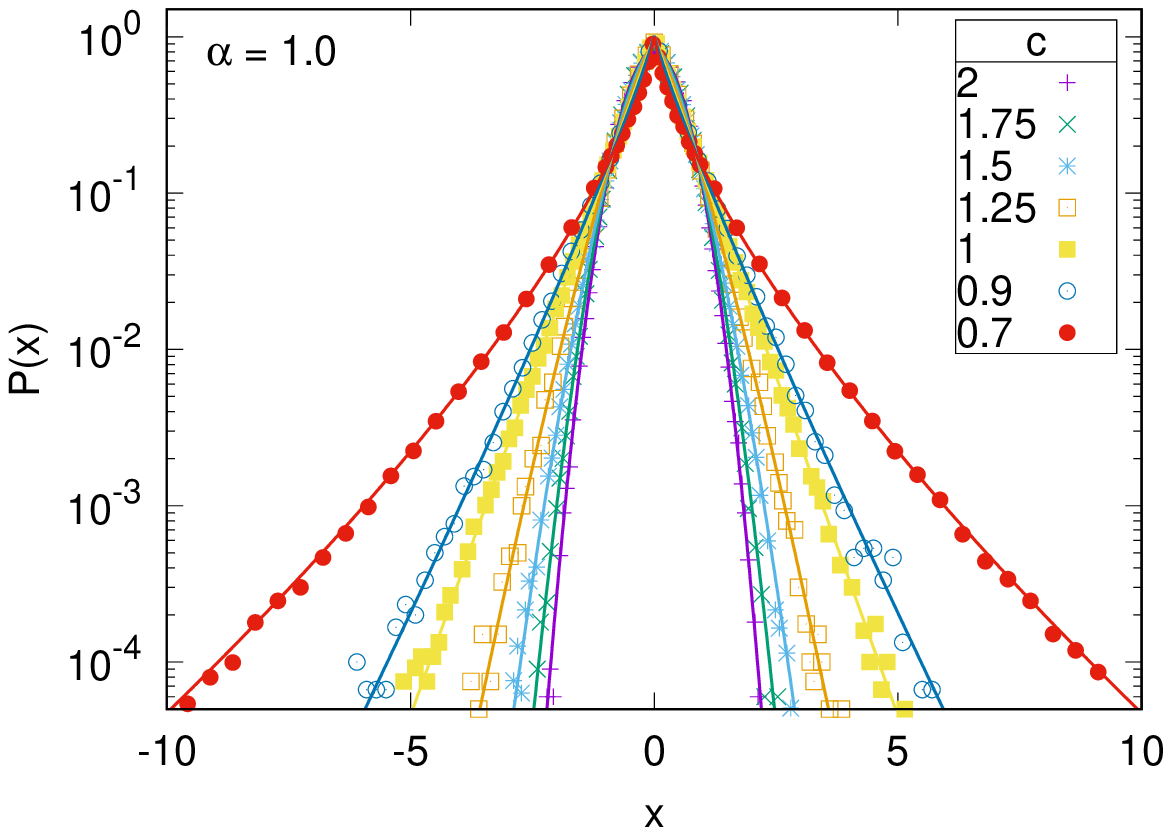}
\includegraphics[width=0.48\textwidth]{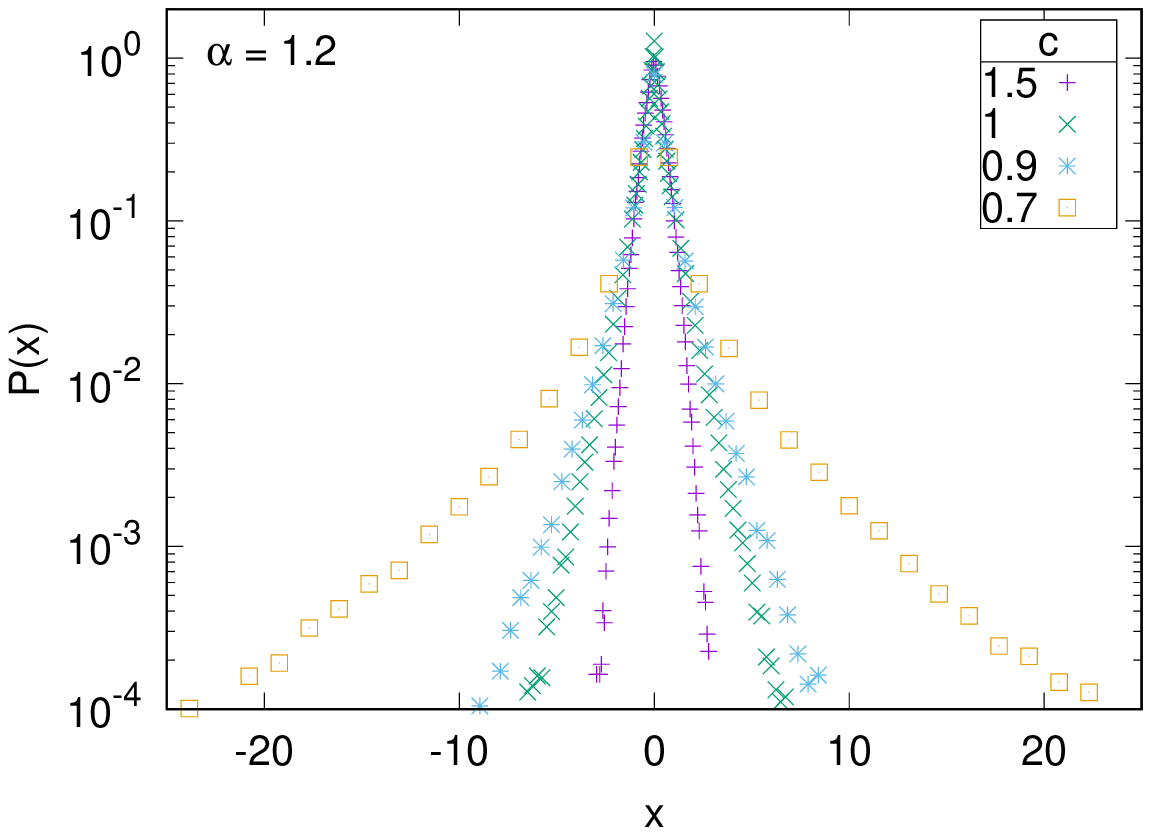}\\
\includegraphics[width=0.48\textwidth]{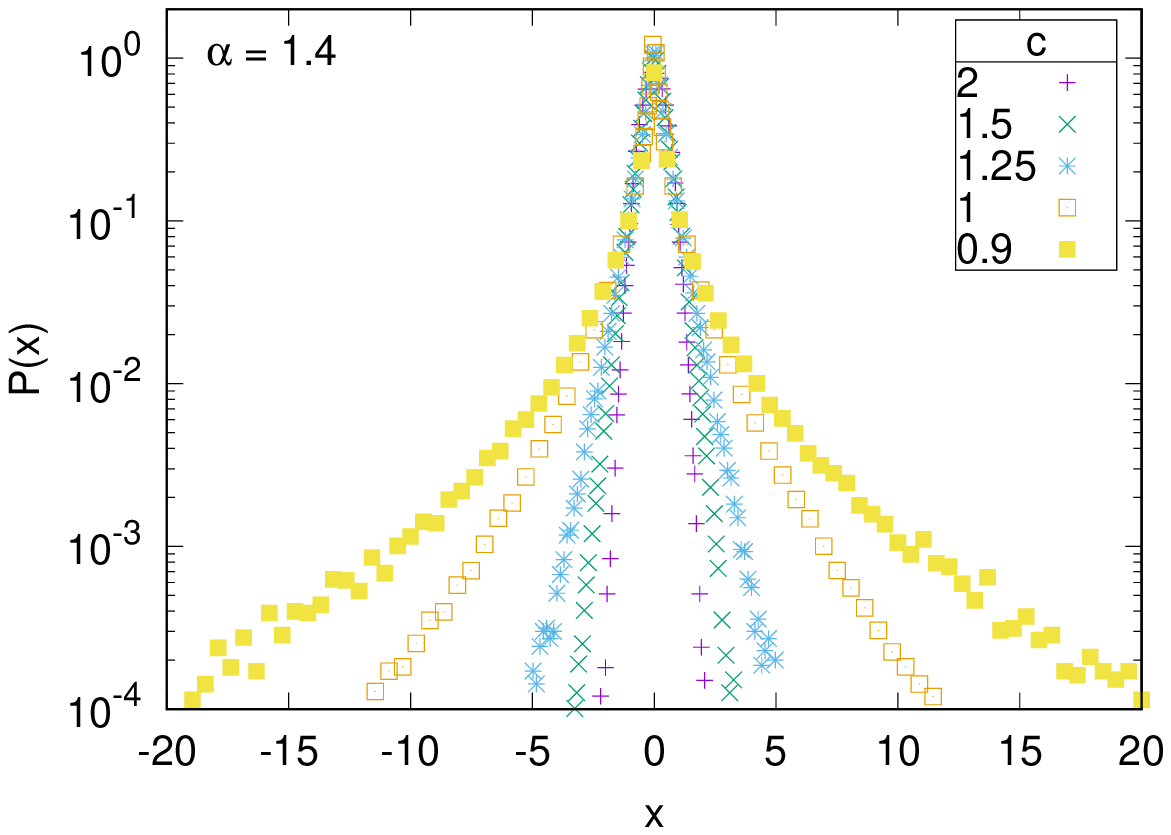}
\caption{Stationary PDF for $\alpha=0.6$, $1.0$, $1.2$, and $1.4$, each for
different $c$ values. The solid lines in the plot for $\alpha=1$ show the
theoretical stationary PDF \eref{eq:statPdfBrown} and are in good agreement
with the simulation results.}
\label{fig:statPdf-aConst-cDiff}
\end{figure}

Figures \ref{fig:statPdf-c>1-aDiff} and \ref{fig:statPdf-c<1} show the
stationary PDF for fixed $c>1$ and $c\leq1$, respectively, each for
different values of $\alpha$. Figure \ref{fig:statPdf-aConst-cDiff}
shows the stationary PDF for fixed $\alpha$ and different $c$.
First we note that the discussed non-monotonicity of the stationary MSD on
$\alpha$ (section \ref{sec:emsd}) is reflected in the width of the
stationary PDF, although this effect is only slightly visible in the plots for
$c=1.75$ and $1.5$, if one takes the full width at half of the maximum value
of the PDF as a measure for the MSD (see figure \ref{fig:statPdf-c>1-aDiff}
for the PDF and figure \ref{fig:emsd-c>1} for the MSD).

Next let us examine the tails of the stationary PDF. As can be seen in figures
\ref{fig:statPdf-c>1-aDiff} and \ref{fig:statPdf-c<1}, for the case of
persistent noise ($\alpha>1$) the tails decay slower than in the Brownian
case, and for anti-persistent noise ($\alpha<1$), although less distinct
at larger $c$ values, they decay faster than in the Brownian case. Generally,
the decay becomes slower with increasing $\alpha$. With respect to $c$ the
tails decay faster with increasing $c$, as one would expect, see figure
\ref{fig:statPdf-aConst-cDiff}.

\begin{figure}
\includegraphics[width=0.48\textwidth]{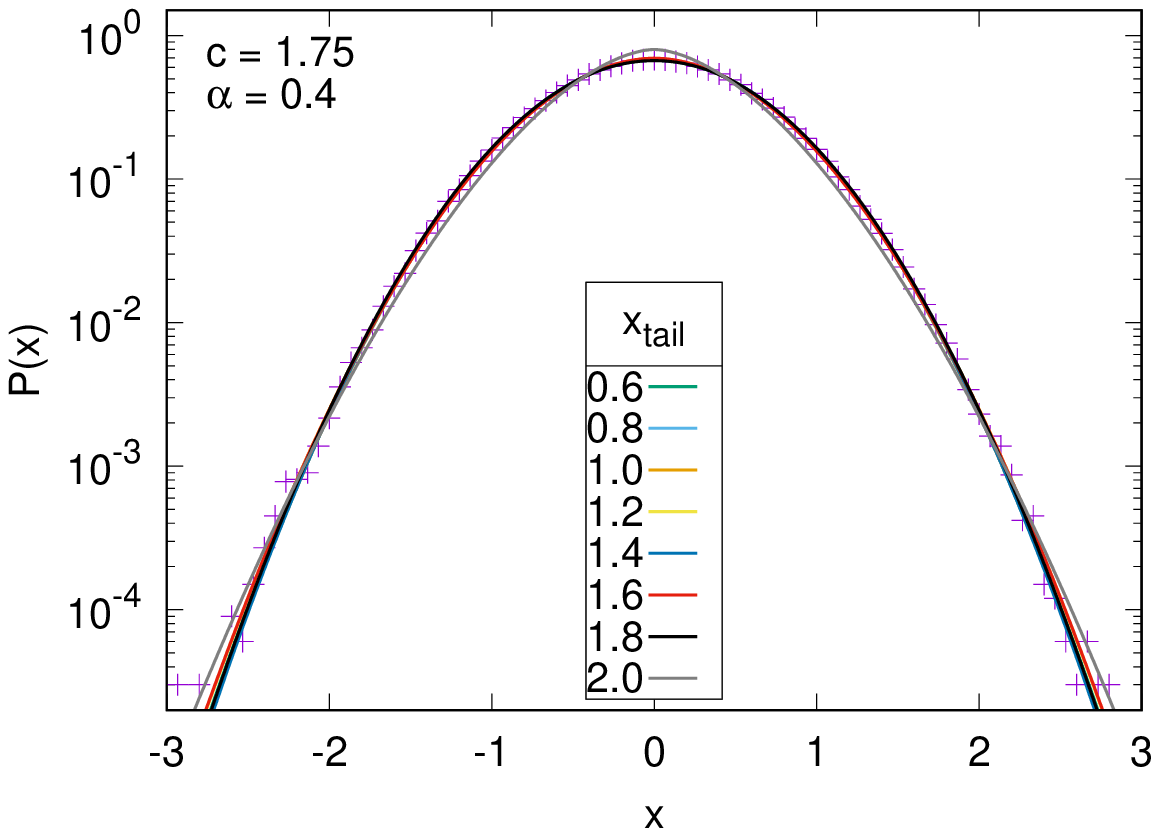}
\includegraphics[width=0.48\textwidth]{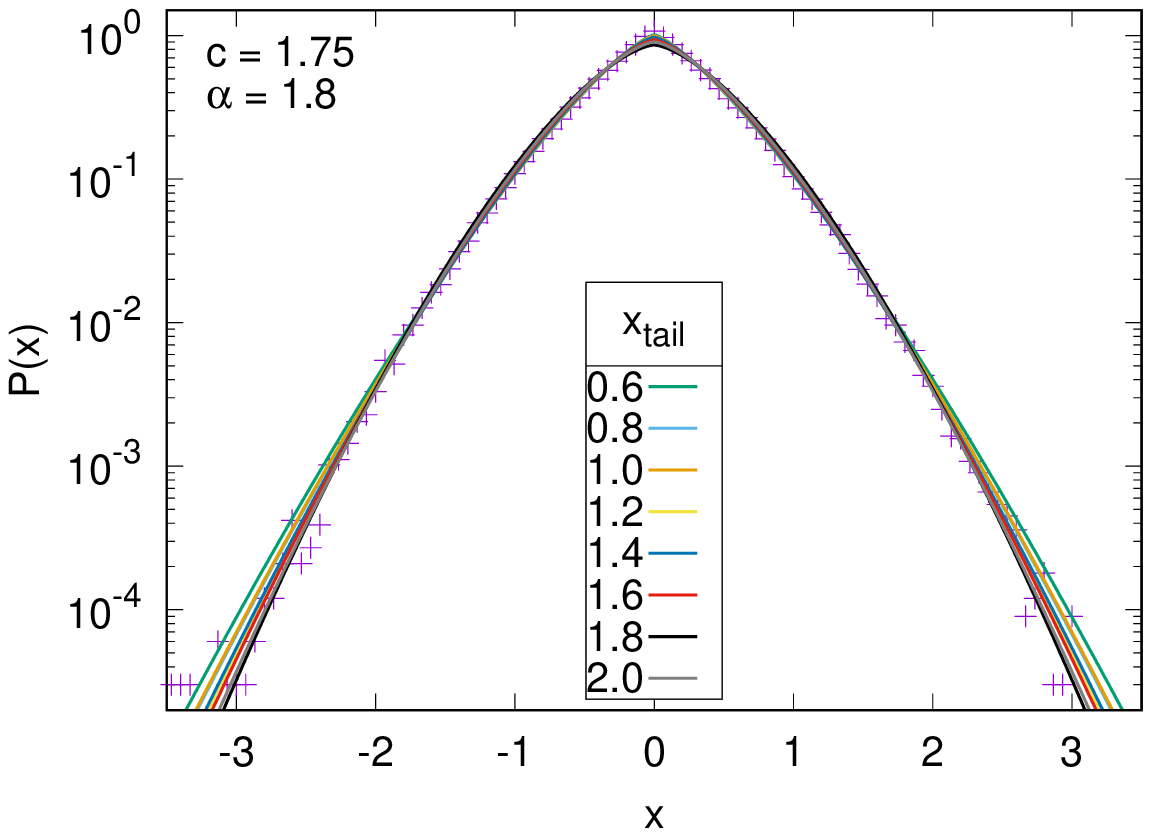}\\
\includegraphics[width=0.48\textwidth]{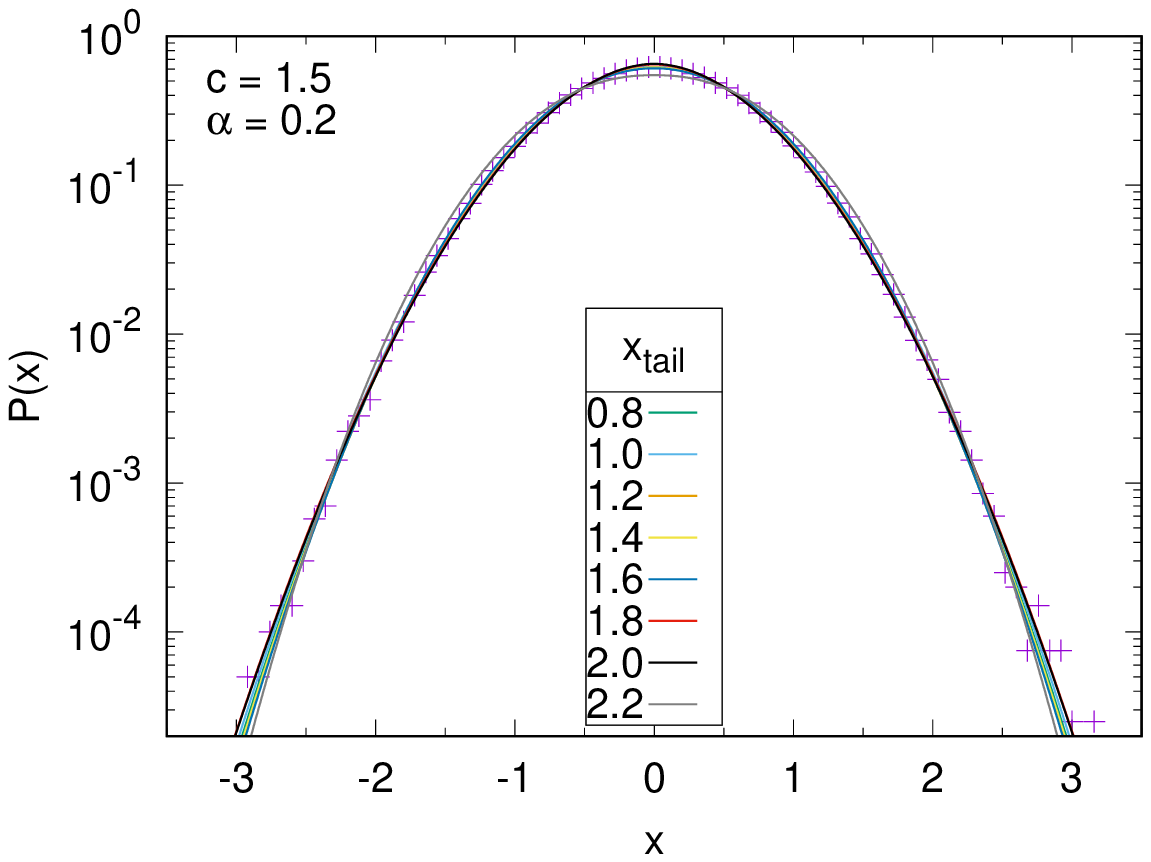}
\includegraphics[width=0.48\textwidth]{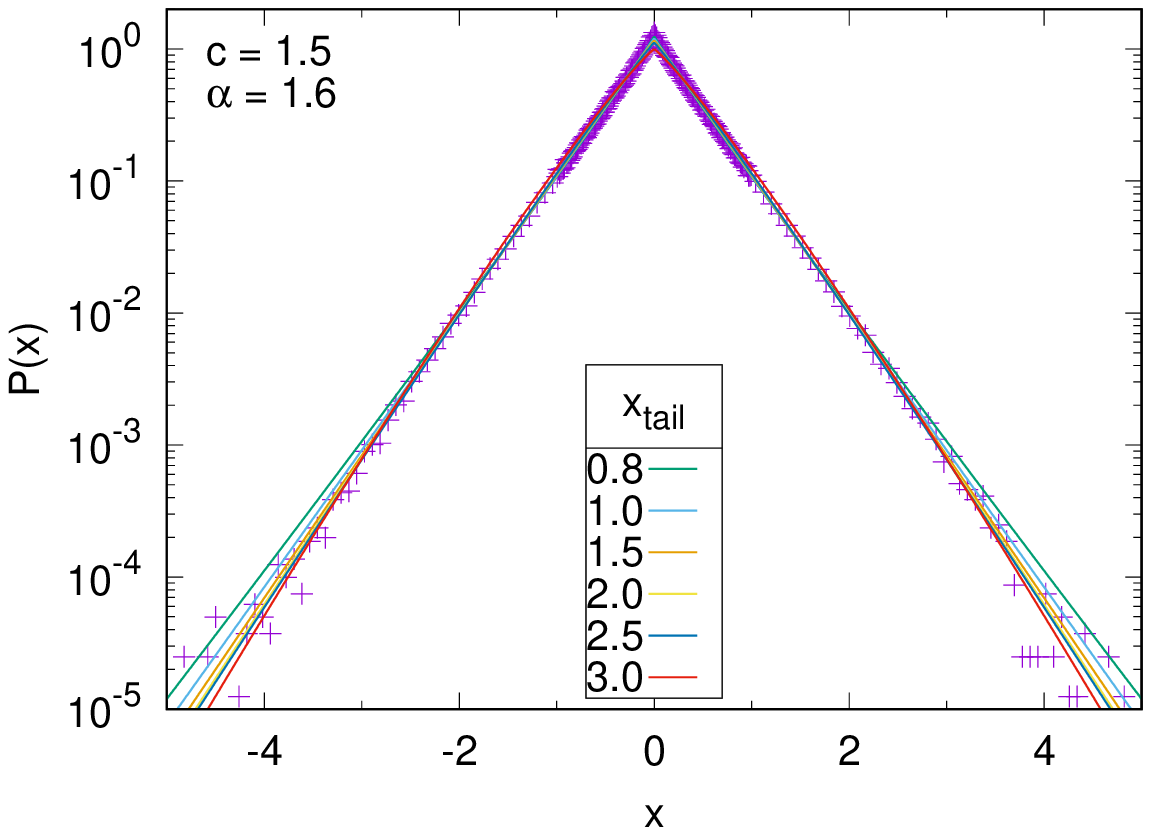}\\
\includegraphics[width=0.48\textwidth]{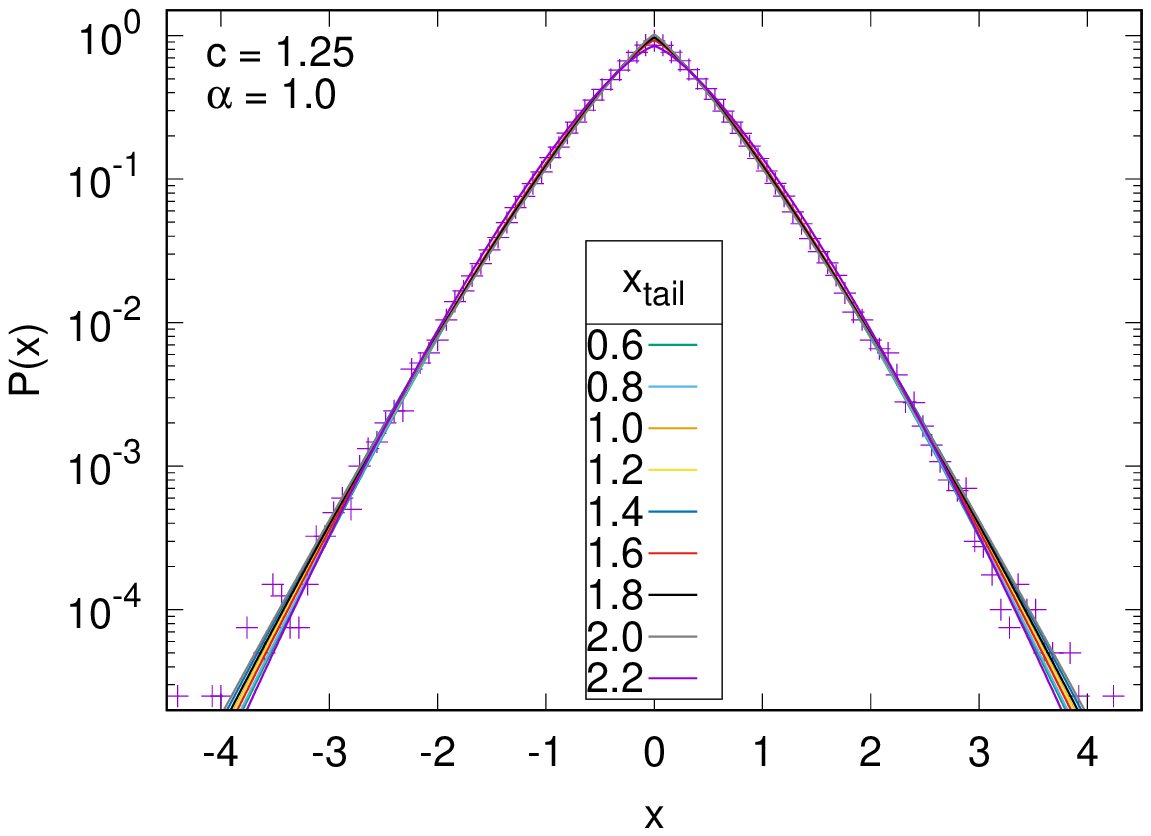}
\includegraphics[width=0.48\textwidth]{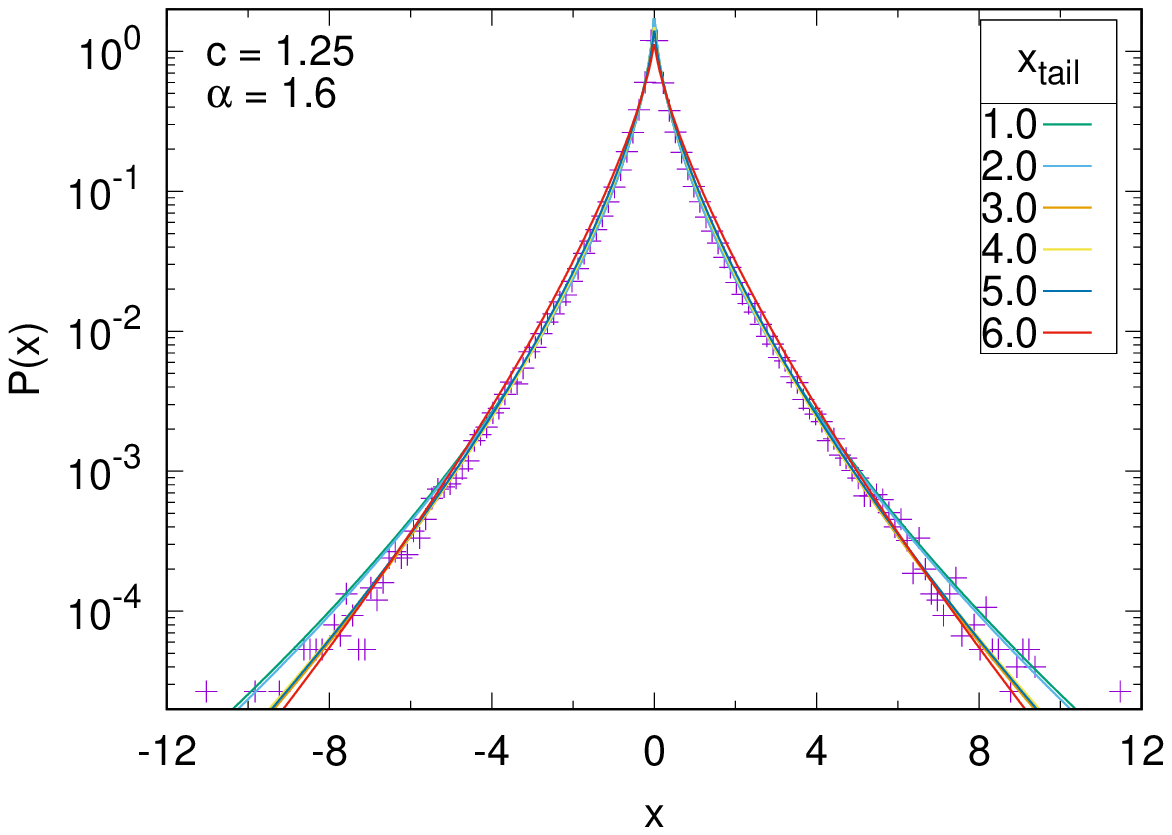}
\caption{Fit of the tails ($|x|>x_{\mathrm{tail}}$) of the stationary PDF
with the generalised exponential function \eref{eq:genPdf} with fit
parameters $a_1$ and $a_2$, for potential scaling exponents $c=1.75$, $1.5$,
and $1.25$, and different $\alpha$.}
\label{fig:tailFits}
\end{figure}

\begin{figure}
\includegraphics[width=0.48\textwidth]{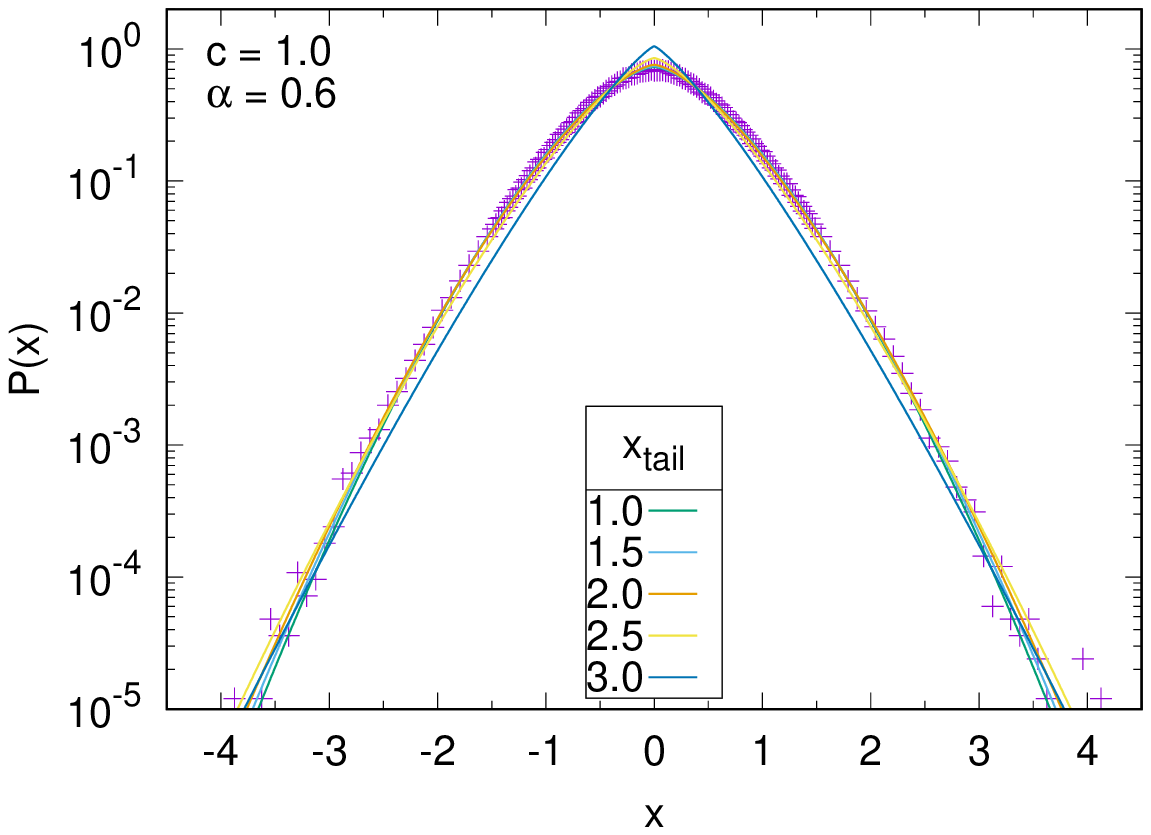}
\includegraphics[width=0.48\textwidth]{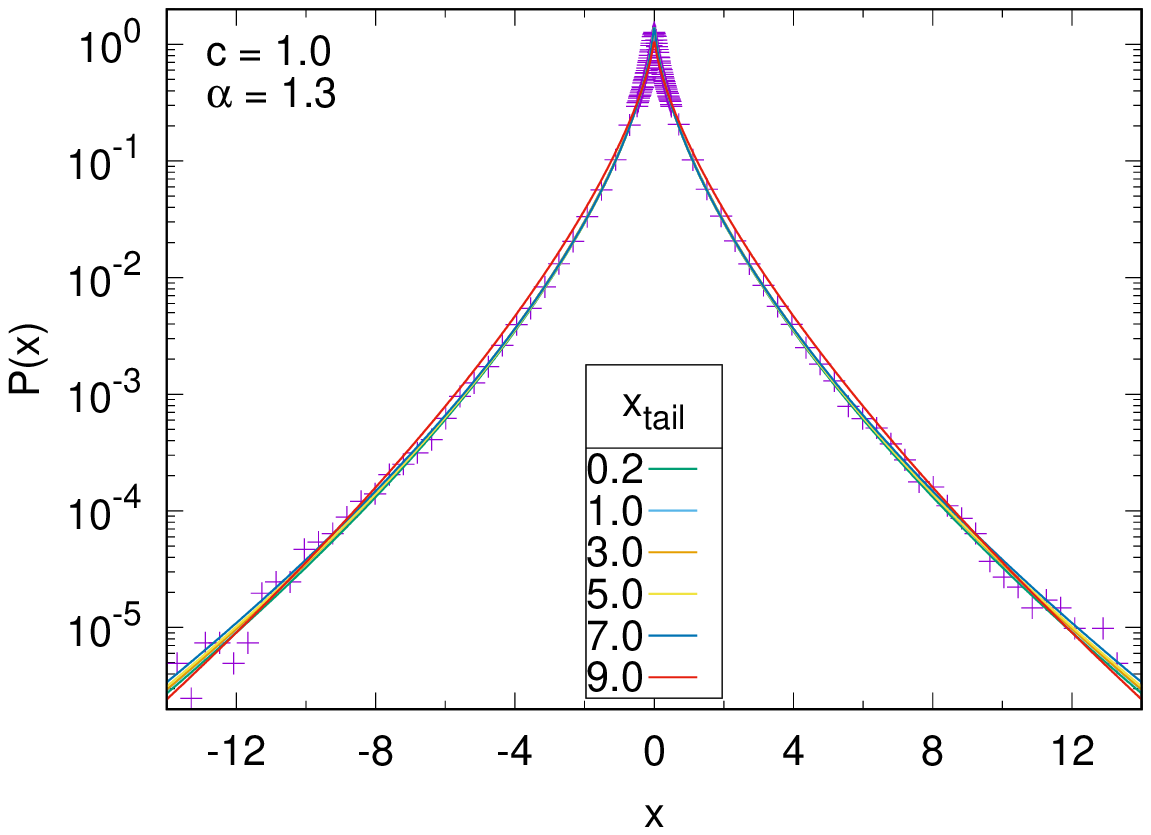}\\
\includegraphics[width=0.48\textwidth]{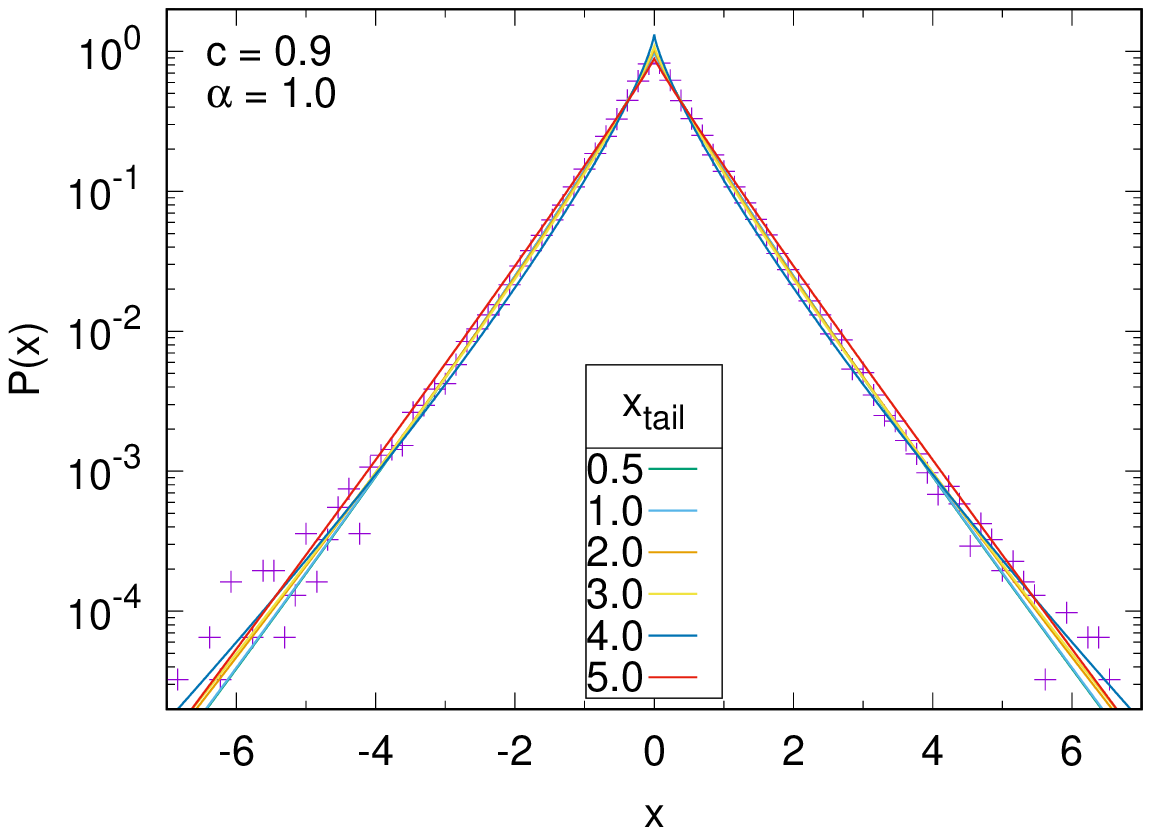}
\includegraphics[width=0.48\textwidth]{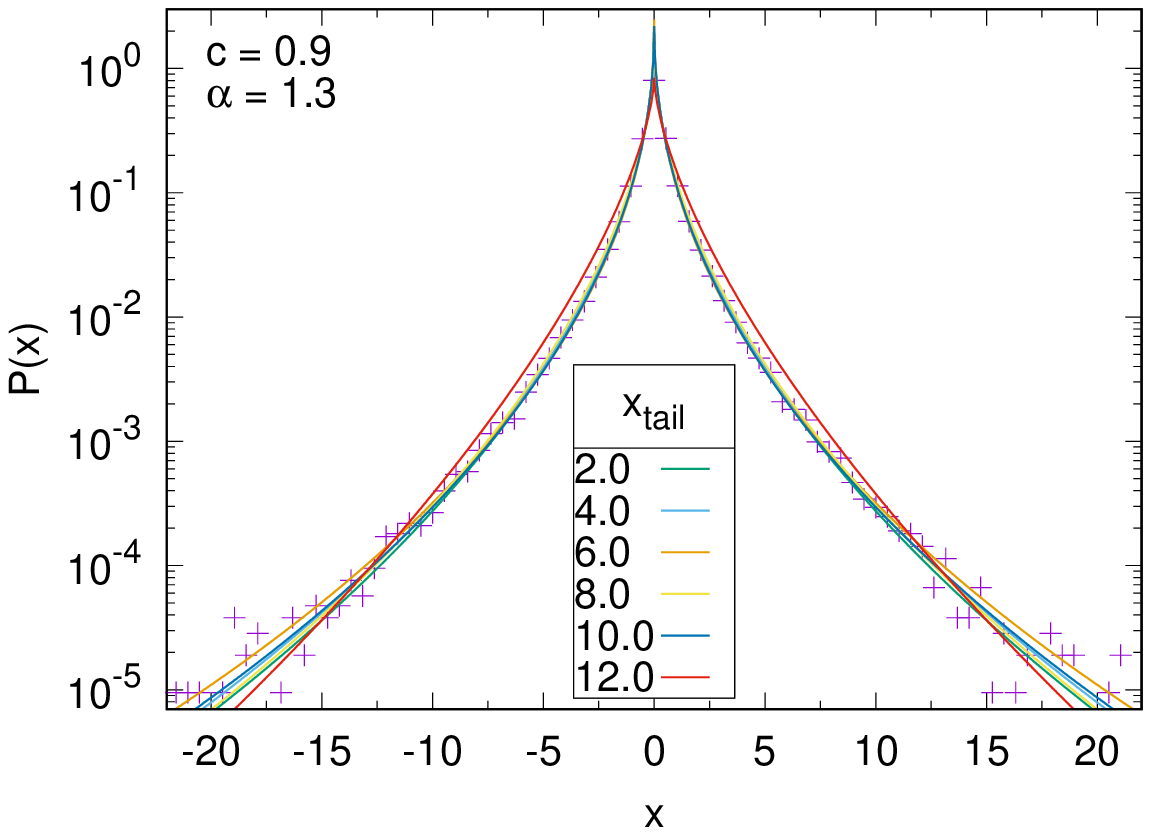}\\
\includegraphics[width=0.48\textwidth]{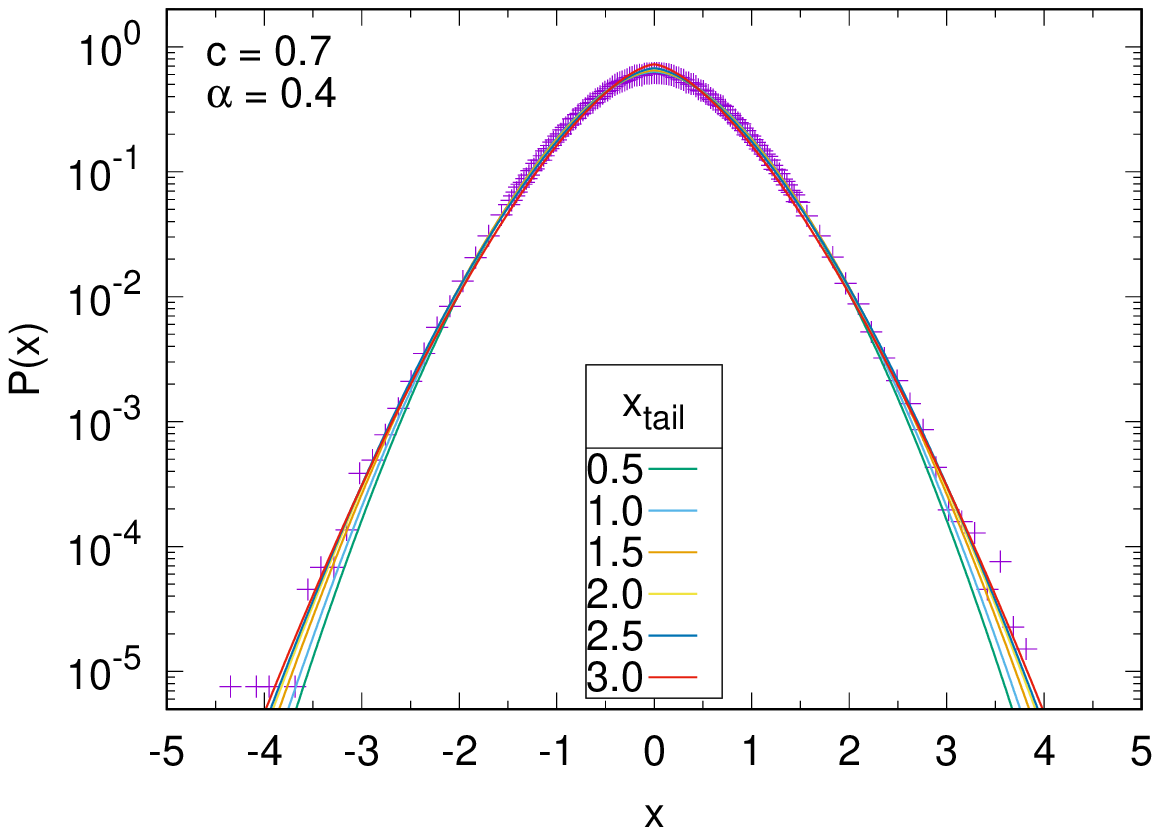}
\includegraphics[width=0.48\textwidth]{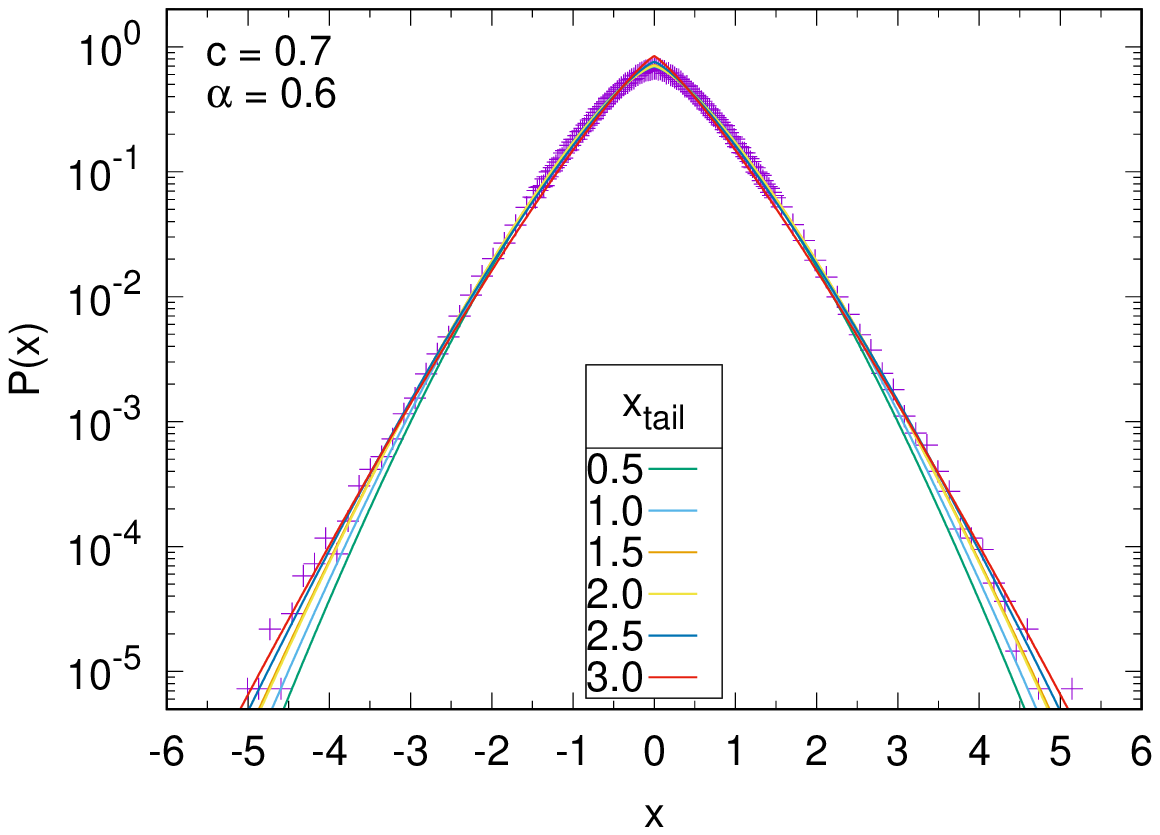}\\
\includegraphics[width=0.48\textwidth]{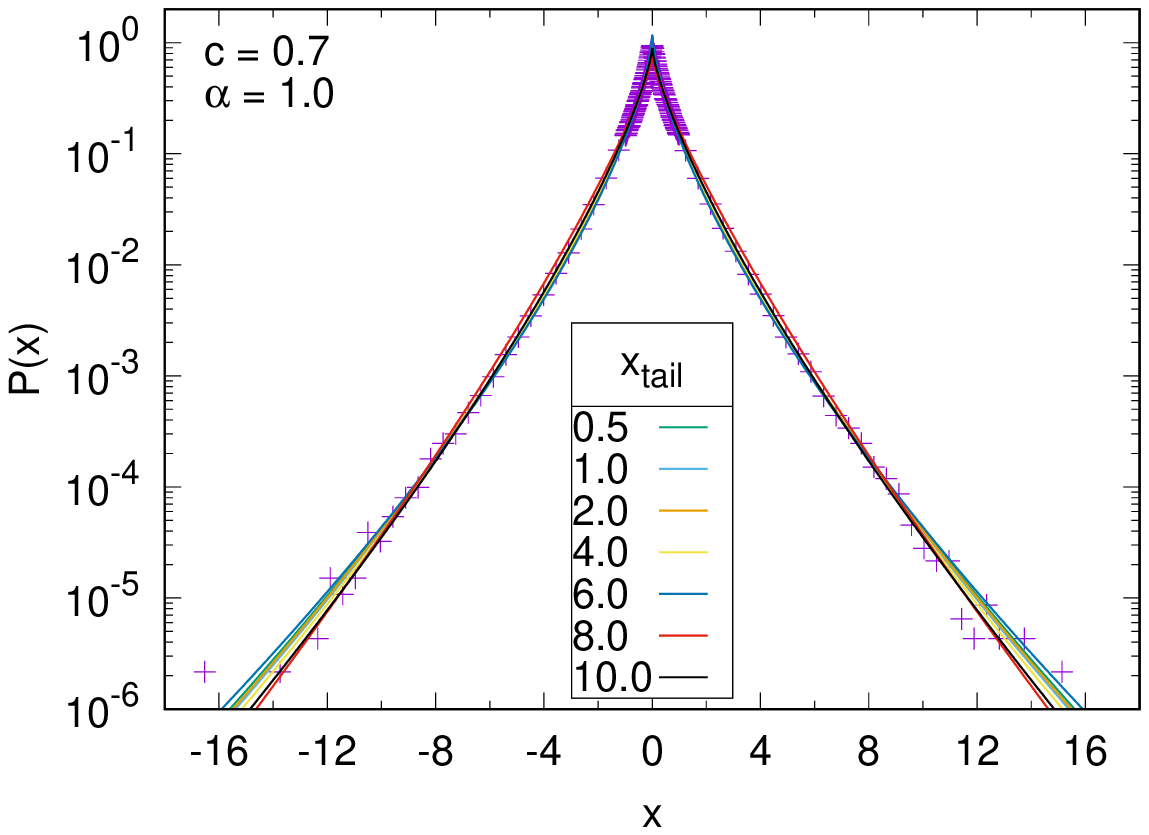}
\includegraphics[width=0.48\textwidth]{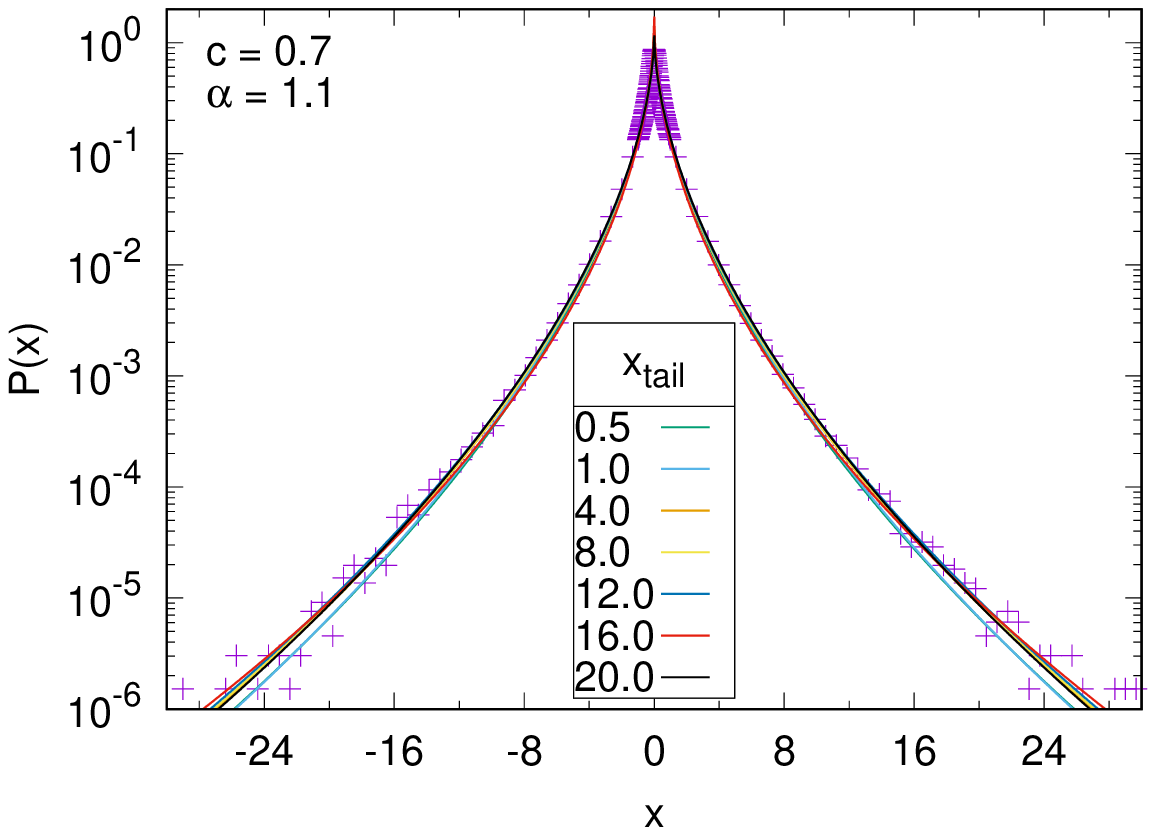}\\
\caption{Fits of the tails ($|x|>x_{\mathrm{tail}}$) of the stationary PDF
with the generalised exponential function \eref{eq:genPdf} with fit
parameters $a_1$ and $a_2$, for potential scaling exponents $c=1.0$, $0.9$,
and $0.7$, and different $\alpha$.}
\label{fig:tailFits2}
\end{figure}

Before we discuss these results further, we introduce the two-sided
generalised exponential PDF
\begin{equation}
\label{eq:genPdf}
f(x)=\frac{1}{\mathscr{N}}e^{-a_1|x|^{a_2}},\qquad\mathscr{N}=\frac{
2\Gamma(1/a_2)}{{a_1}^{1/a_2}a_2},
\end{equation}
with the parameters $a_1,a_2>0$. It encompasses the stationary PDF
in the Brownian (expression \eref{eq:statPdfBrown}) and harmonic (expression
\eref{eq:solStatFpeHarmonic}) cases with $a_1=1/K$ and $a_1=2^{\alpha-1}/
[K\Gamma(1+\alpha)]$, respectively, and $a_2=c$ is given by
the potential shape. Figures \ref{fig:tailFits} and \ref{fig:tailFits2} show
the fits of the tails ($|x|\ge x_{\mathrm{tail}}$) of the stationary PDF with
the generalised exponential fit function \eref{eq:genPdf} and fit-parameters
$a_1$ and $a_2$. Our analysis shows that the fit parameters are quite robust
with respect to the precise choice for $x_{\mathrm{tail}}$. As can be seen,
the agreement with the fit function is quite nice for larger potential scaling
exponents $c$ and smaller FGN exponent $\alpha$.

Due to the symmetry of the PDF (\ref{eq:genPdf}), the first moment
is zero, and for the second and fourth moments we find
\begin{eqnarray}
\label{eq:secAndFourMomGenPdf}
\langle X^2\rangle&={a_1}^{-2/a_2}\frac{\Gamma(3/a_2)}{\Gamma(1/a_2)},\\
\langle X^4\rangle&={a_1}^{-4/a_2}\frac{\Gamma(5/a_2)}{\Gamma(1/a_2)}.
\end{eqnarray}
Hence, the kurtosis becomes
\begin{equation}
\label{eq:kurtGenPdf}
\kappa=\frac{\langle(X-\langle X \rangle)^4\rangle}{\langle(X-\langle X
\rangle)^2\rangle^2}=\frac{\Gamma(5/a_2)\Gamma(1/a_2)}{\Gamma^2(3/a_2)}.
\end{equation}
Note that $\kappa$ is independent of the parameter $a_1$, and in the
Brownian and harmonic cases $a_2=c$.

\begin{figure}
\centering
\includegraphics[width=0.48\textwidth]{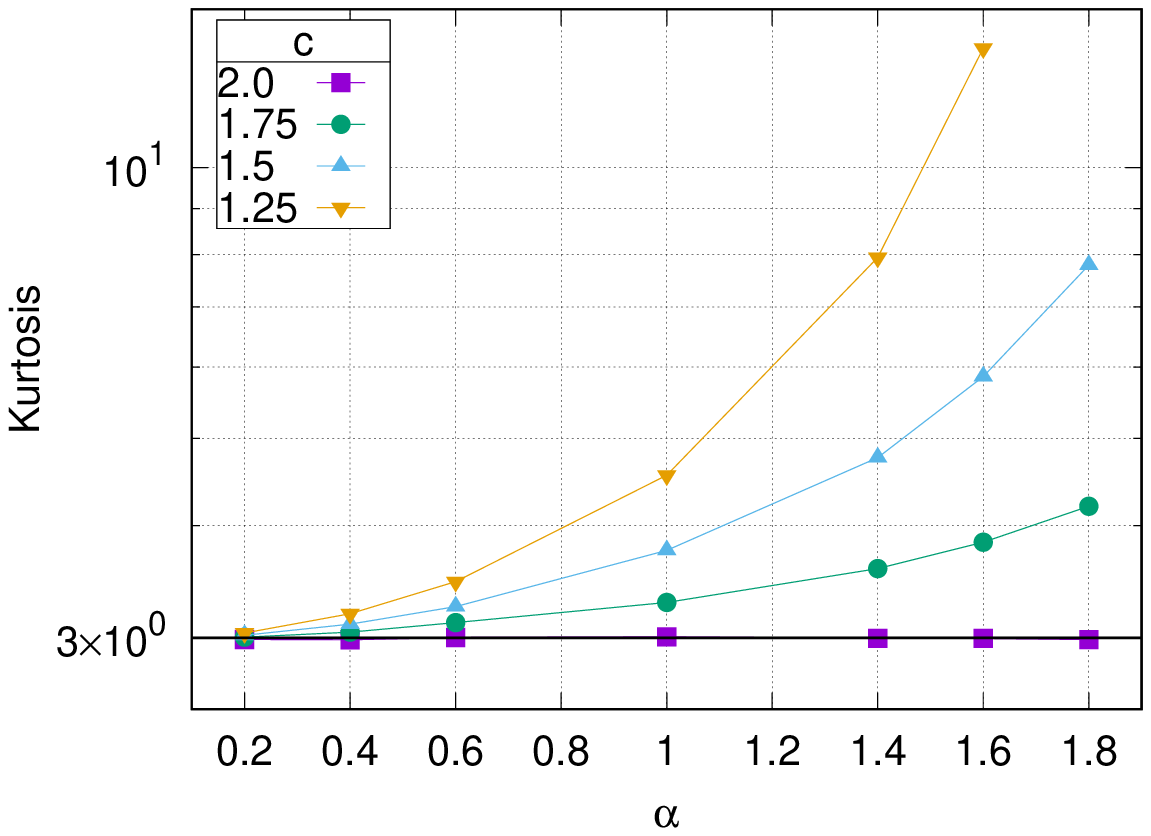}
\includegraphics[width=0.48\textwidth]{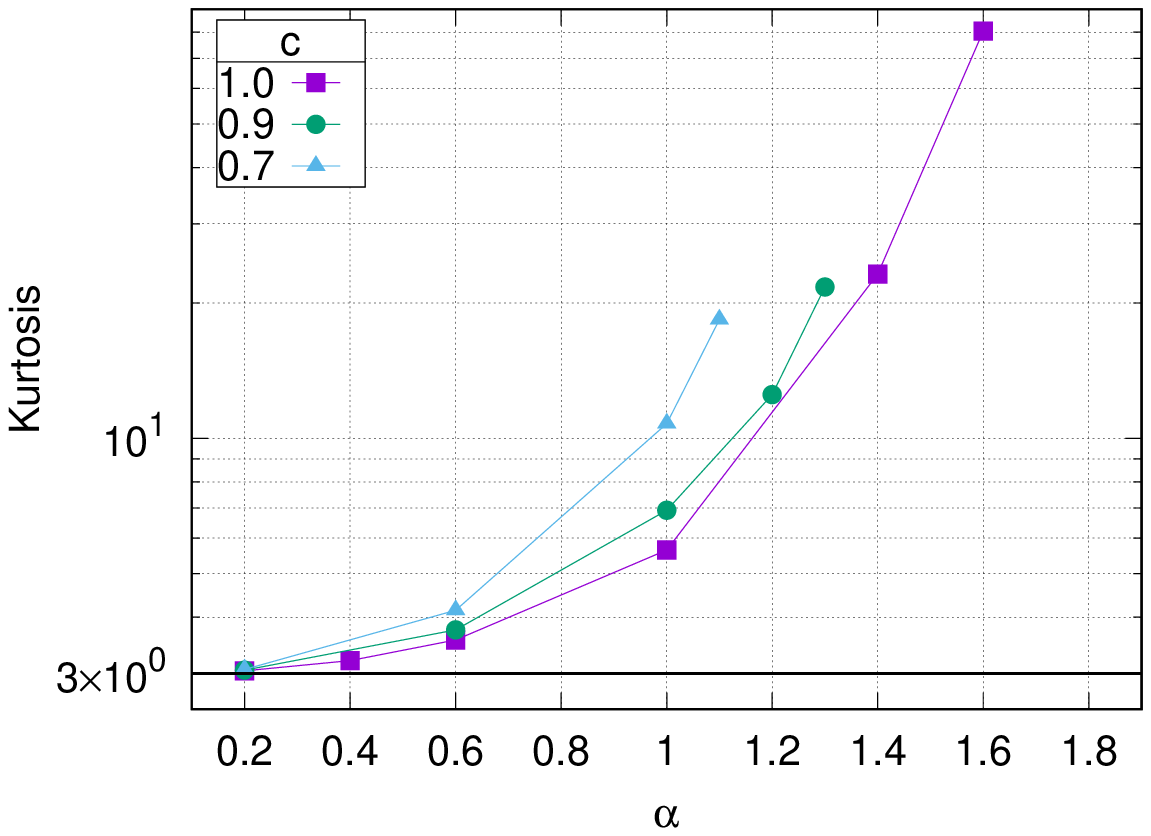}
\includegraphics[width=0.48\textwidth]{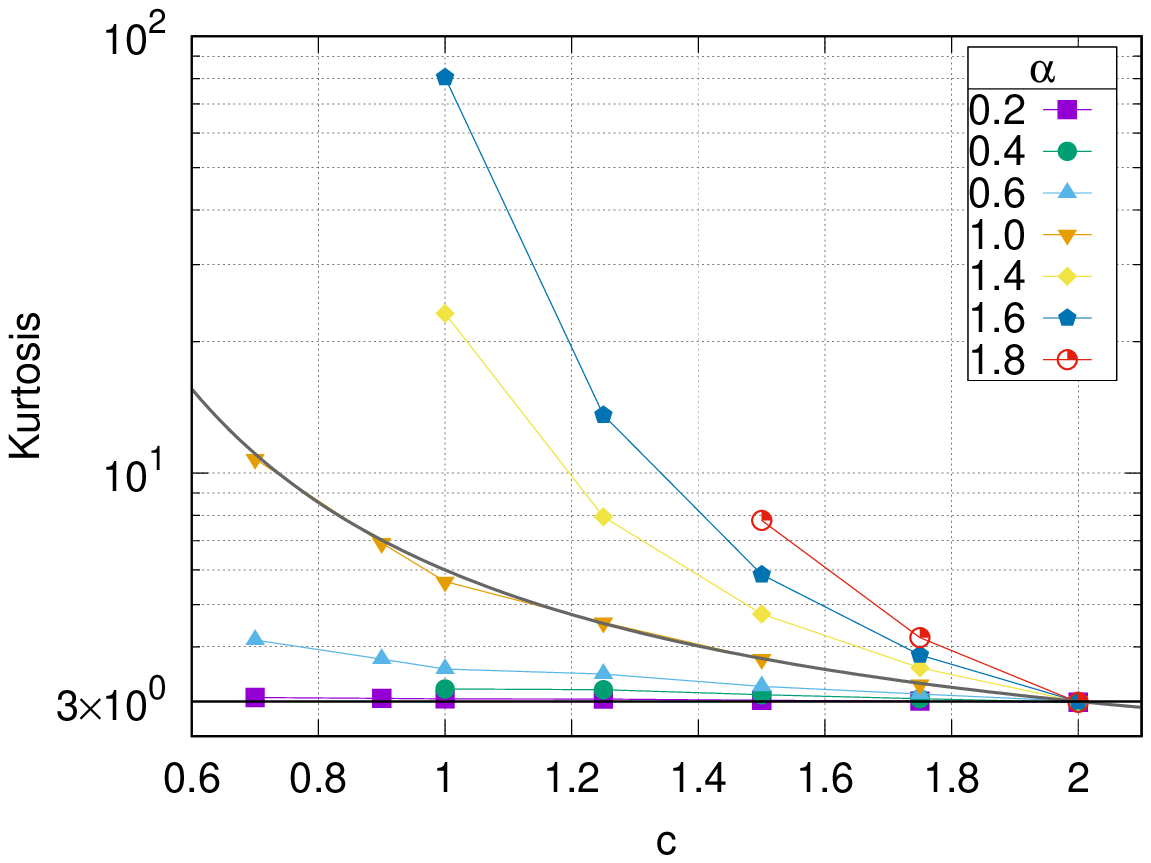}
\caption{Kurtosis calculated from the simulated stationary PDF as function
of $\alpha$ (top) and $c$ (bottom). The horizontal lines show the kurtosis
value of 3 for a Gaussian, the grey line shows the theoretical value of the
kurtosis in the Brownian case for different $c$ values, expression
\eref{eq:kurtGenPdf} with $a_2=c$.}
\label{fig:kurt}
\end{figure}

Figure \ref{fig:kurt} shows the kurtosis, determined from the numerical
simulations, as function of $\alpha$ (top panels) and $c$ (bottom panel).
This measured kurtosis agrees well with the theoretical prediction in
the Brownian and harmonic cases (equation (\ref{eq:kurtGenPdf})).
The kurtosis $\kappa$ monotonically increases
with $\alpha$ and decreases with $c$, which corresponds to the fact that the
tails of the stationary PDF fall off slower in $|x|$ with increasing $\alpha$
(increasing persistence) and faster with increasing $c$. Moreover, compared to
the Brownian case ($\alpha=1$) the kurtosis is larger for persistent noise
($\alpha>1$) and smaller for anti-persistent noise ($\alpha<1$), which is
consistent with the slower decay in $|x|$ of the tails for $\alpha>1$ (and
faster for $\alpha<1$), as compared to the Brownian case.

We note that for all $c\neq2$ the stationary PDF is leptokurtic, i.e., has
"fatter" tails with $\kappa>3$, and approaches the Gaussian value of 3 for
$c\to2$. Interestingly, for small $\alpha$ values the kurtosis stays close
to the Gaussian value of 3, and in fact converges to it for $\alpha\to0$,
independent of $c$ (see top panels in figure \ref{fig:kurt}). This result
is consistent with figure \ref{fig:statPdf-aConst-cDiff}, where larger
$\alpha$-values produce strongly leptokurtic PDFs and smaller $\alpha$
values lead to more Gaussian shapes, compare also \ref{sec:shape}.

\section{Conclusion}
\label{sec:conclusion}

FBM is a strongly non-Markovian stochastic process. Despite the stationary
increments, the long-ranged, power-law noise auto-correlation leads to
distinct effects of (anti-)per\-sistence, which, in turn, lead to a number
of properties for which FBM defies analytical approaches. A long-standing
example is the lack of direct analytical methods to calculate the
first-passage dynamics of FBM, for which only asymptotic \cite{molchan},
numerical \cite{jaeepl,oleksii}, or perturbative \cite{wiese} approaches
exist. This
is related to the fact that, for instance, the seemingly simple Fokker-Planck
equation (\ref{eq:fpeHarmonic}) in the harmonic case or in absence of an
external potential, cannot be used to formally derive the boundary value
solution for a semi-infinite or finite domain with reflecting boundaries
\cite{jaeepl,oleksii}. Even more so, numerical studies show that the PDF
of FBM next to reflecting boundaries is not flat but shows accretion or
depletion next to the boundaries for persistent or anti-persistent cases
\cite{tobias,vojta0,vojta1,vojta2}, with potential implications to the
growth density of serotonergic brain fibres \cite{skirmantas}.
Another remarkable phenomenon was observed for FGN-driven motion subject
to a fluctuation-dissipation relation governed by the fractional Langevin
equation. In this case a critical exponent was found at which a harmonically
bound particle switches between a non-monotonic underdamped phase and a
"resonance" phase, in the presence of an external sinusoidal driving
\cite{stas1}. In many cases, therefore, to explore the detailed properties
of FBM one needs to resort to numerical analyses.

Based on the overdamped Langevin equation driven by FGN, we here
studied in detail the stochastic motion of FBM in a subharmonic potential
by examining the MSD and PDF. The most striking result we obtained is the
conjecture that there exists a long-time stationary state if
the relation $c>2(1-1/\alpha)$ is satisfied. We corroborated this
conjecture via numerical analysis of FBM for a wide range of potential
scaling exponents $c$ and FGN-exponents $\alpha$. In particular, this
implies that while for anti-persistent or uncorrelated FGN ($\alpha\leq1$)
there always exists a long-time stationary state for any $c>0$. For
persistent FGN ($\alpha>1$) the
competition between the confining tendency of the potential and the
persistence of the motion turns out to become a delicate balance. This
behaviour is analogous to what was found for the overdamped Langevin
equation driven by white L{\'e}vy-stable noise \cite{capala}. In the
L{\'e}vy-stable case, however, the confining tendency of the potential
was in competition with the occasional, extremely long jumps due
to the diverging second moment of the driving L{\'e}vy noise. Despite
this fundamental difference in the dynamics of the two processes, in
both cases the condition for the existence of stationarity can be written
as $c>2-1/H$ where $H$ is the self-similarity index of the unconfined
process. We note that the similarity between both FBM and L{\'e}vy
flights also extends to superharmonic potentials, e.g., in the existence
of multimodal states, see the discussion in \cite{tobias1}. We also note
that superdiffusive FBM may explain similar features in the observed
motion of searching and migrating birds as L{\'e}vy flights
\cite{ohad}.

We also demonstrated that the time to reach stationarity increases with
growing $\alpha$ and decreases with growing $c$. Moreover, the stationary
MSD monotonically decreases with growing $c$, as intuitively expected. In
dependence on $\alpha$, the behaviour of the stationary MSD is more
complicated in that it is non-monotonic in $\alpha$. Namely for $\alpha\leq
\alpha_\mathrm{crit}(c)$ it decreases with growing $\alpha$, while for $\alpha
\geq\alpha_\mathrm{crit}(c)$ it increases with growing $\alpha$. The critical
value $\alpha_\mathrm{crit}(c)$ increases monotonically with growing $c$.

In the analysis of the PDF we showed that at short times the behaviour is
close to free motion or motion in an harmonic potential, before the particle
engages with the confining potential. At stationarity the tails
of the PDF decay faster with decreasing $\alpha$ and growing $c$. Particularly,
for $\alpha>1$ ($\alpha<1$) the tails decay slower (faster) in $|x|$ than in
the Brownian case. This is contrary to the case of FBM in a superharmonic
potential ($c>2$), as detailed in \cite{tobias1}. We also showed that the
two-sided generalised exponential PDF (\ref{eq:genPdf})
provides a good description for the stationary PDF
as long as $c$ is not too small and $\alpha$ not too large. Finally we
showed that the stationary PDF is leptokurtic ("fat-tailed") for $c\neq2$
and hence non-Gaussian. For the fully anti-persistent case $\alpha\to0$ the
kurtosis approaches the Gaussian value 3.

It will be interesting to see how this picture extends once the driving FGN
is tempered in terms of an exponential or power-law cutoff \cite{daniel}. Of
course, in this case the long-term PDF beyond the cutoff time always has the
Boltzmannian shape (\ref{eq:statPdfBrown}), however, the transient behaviour
is expected to be quite rich. Such a scenario may be relevant for various
processes in which cutoffs become relevant, e.g., finite system sizes or
systems with finite correlation times, such as lipid motion in membrane
bilayers \cite{jaeprl}. We also mention the analysis of confinement
effects for FBM with random parameters, see, e.g., \cite{diego,wei},
or for particles with stochastically changing mobilities suspended in
non-equilibrium viscoelastic liquids \cite{eiji,fulvio,mario}.

\ack

RM acknowledges funding from the German Science Foundation (DFG, grant no.
ME 1535/12-1). AC acknowledgments the support of the Polish National Agency
for Academic Exchange (NAWA).

\appendix

\section{Curvature of the stationary PDF and stationary MSD as function
of $c$}
\label{sec:shape}

Here we briefly allude to the classification of the stationary PDFs according
to their shape.  More precisely, we can divide the stationary PDFs into two
distinct groups according to their curvature, by which we mean their second
derivative. First, consider the Brownian case ($\alpha=1$) for which the
stationary PDF is given by expression \eref{eq:statPdfBrown}. A straightforward
calculation shows that for $c\leq c_\mathrm{cr}(\alpha=1)=1$ the curvature is
positive for all $x\neq0$, while for $c>c_\mathrm{cr}(\alpha=1)=1$ the
curvature changes sign at $|x|=x_\mathrm{cr}=((c-1)/(2c))^{1/c}$, such that
the curvature is positive for $|x|>x_\mathrm{cr}$ and negative for $|x|<
x_\mathrm{cr}$. Compare also the plot for $\alpha=1$ in figure
\ref{fig:statPdf-aConst-cDiff}.

In general, we observe that for all $\alpha$ there is a critical value
$c_\mathrm{cr}(\alpha)$ such that for all $c\leq c_\mathrm{cr}(\alpha)$
the stationary PDFs exhibit a positive curvature for all $x\neq0$, while
for all $c>c_\mathrm{cr}(\alpha)$ the curvature has a change of sign at
some $|x|=x_\mathrm{cr}(\alpha,c)>0$ such that the curvature is positive
for $|x|>x_\mathrm{cr}$ and negative for $|x|<x_\mathrm{cr}$.

The critical value $c_\mathrm{cr}(\alpha)$ increases with $\alpha$. For
instance, for $\alpha=1.8$ and $c=1.25$ the stationary PDF exhibits a
positive curvature, while for $\alpha=1$ and $c=1.25$ the curvature of
the stationary PDF changes sign. Also, for $\alpha=0.2$ and $c=0.7$
the curvature of the stationary PDF changes sign, while for $\alpha=1$
and $c=0.7$ the curvature of the stationary PDF is positive.

\begin{figure}
\centering
\includegraphics[width=0.48\textwidth]{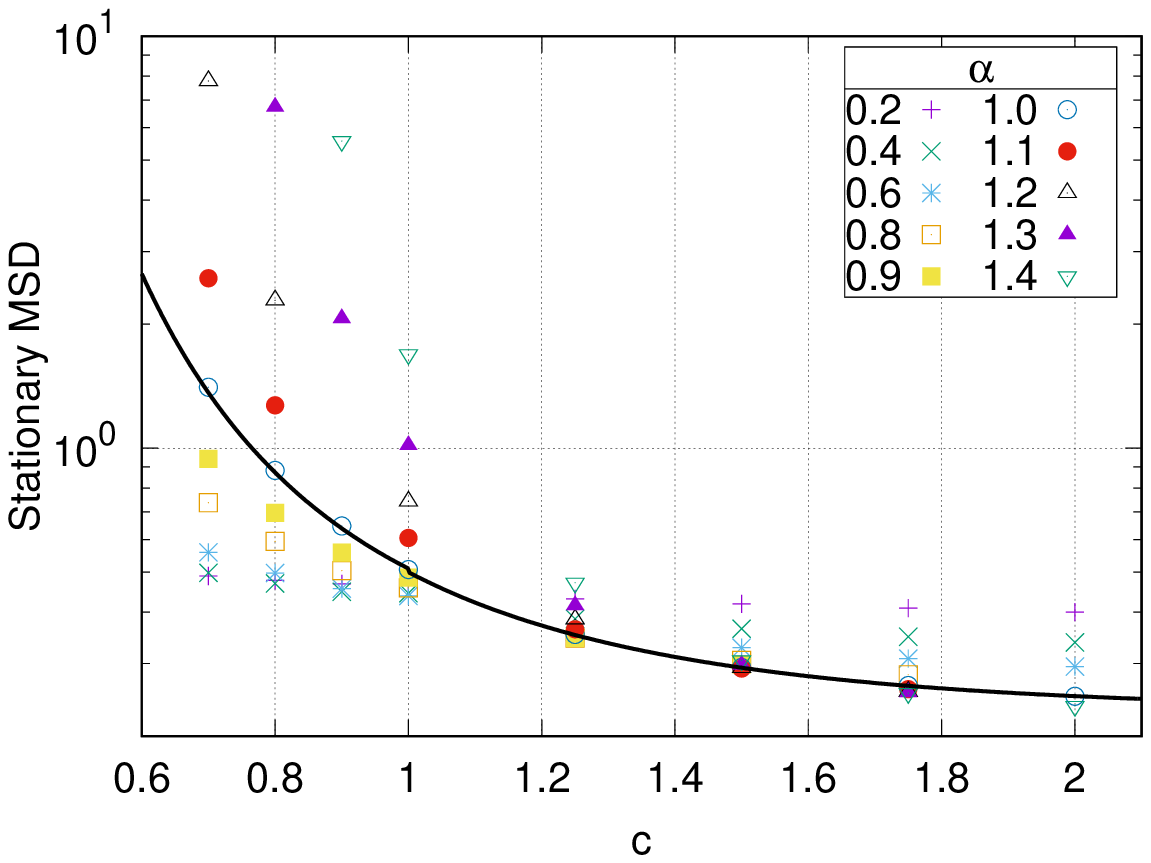}
\includegraphics[width=0.48\textwidth]{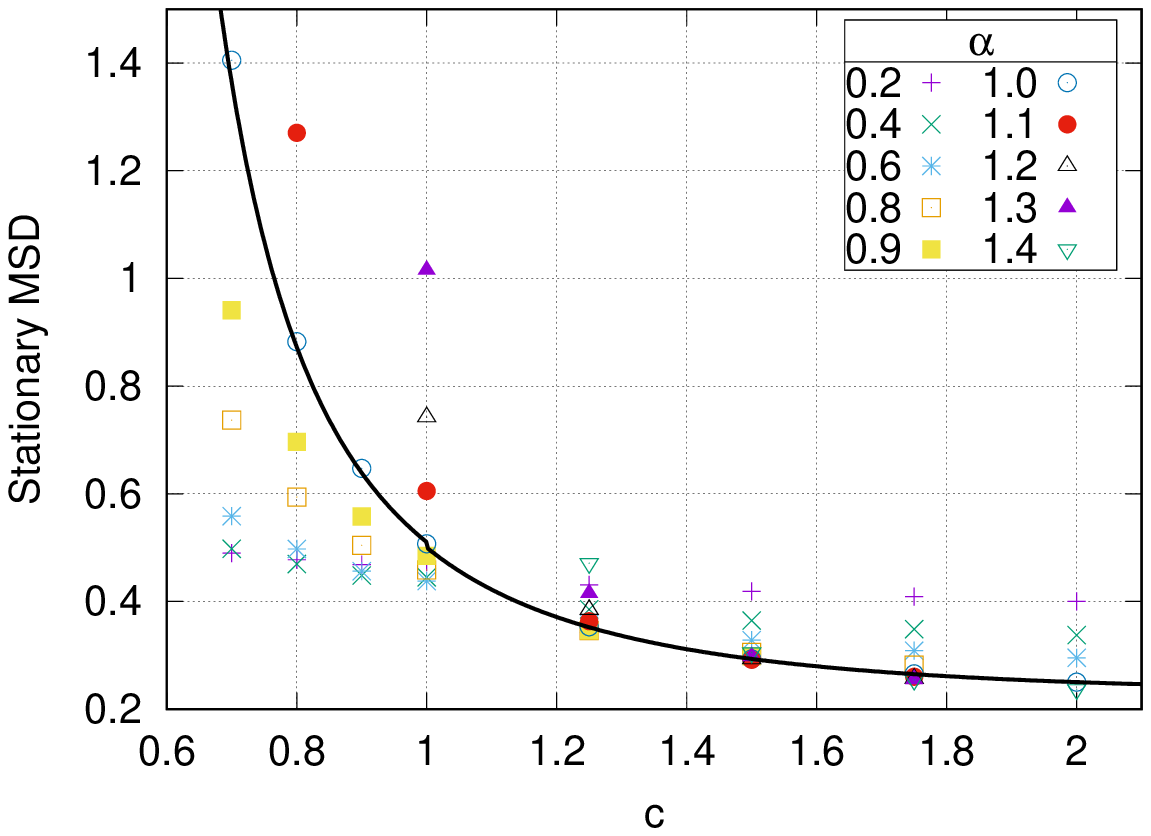}
\caption{Stationary MSD as a function of $c$. The values were determined
from the time-dependent MSD by averaging over the plateau regime. The black
line shows the theoretical prediction \eref{eq:statEmsdBrown} in the
Brownian case. Left: log-lin plot, Right: lin-lin plot (not all all data
points shown).}
\label{fig:statEmsdVsC}
\end{figure}

Finally, in figure \ref{fig:statEmsdVsC} we show the stationary MSD as
function of the potential scaling exponent $c$ for various $\alpha$,
thus complementing figure \ref{fig:emsd-diffC} in the main text.

\section*{References}

\begin{thebibliography}{99}

\bibitem{kappler} E. Kappler, Ann. Phys. (Leipzig) \textbf{11}, 233 (1931).

\bibitem{landau} L. D. Landau and E. M. Lifshitz, Landau and Lifshitz Course
of Theoretical Physics 5: Statistical Physics Part 1 (Butterworth-Heinemann,
Oxford UK, 1980).

\bibitem{schafer} D. A. Schafer, J. Gelles, M. P. Sheetz, and R. Landick,
Nature \textbf{352}, 444 (1991).

\bibitem{simon} S. F. Tolic-N{\o}rrelykke, M. B. Rasmussen, F. S. Savone,
K. Berg-S{\o}rensen, and L. B. Oddershede, Biophys. J. \textbf{90}, 3694
(2006).

\bibitem{franosch} T. Franosch, M. Grimm, M. Belushkin, F. M. Mor, G.
Foffi, L. Forr{\'o}, and S. Jeney, Nature \textbf{478}, 85 (2011).

\bibitem{landau1} E. M. Lifshitz and L. P. Pitaevski, Landau and Lifshitz
Course of Theoretical Physics 10: Physical Kinetics (Butterworth-Heinemann,
Oxford UK, 1981).

\bibitem{vankampen} N. van Kampen, Stochastic processes in physics and
chemistry (North Holland, Amsterdam, 1981).

\bibitem{coffey} W. T. Coffey and Y. P. Kalmykov, The Langevin equation
(World Scientific, Singapore, 2012).

\bibitem{lene1} J.-H. Jeon, N. Leijnse, L. B. Oddershede, and R. Metzler,
New J. Phys. \textbf{15}, 045011 (2013).

\bibitem{pre12} J.-H. Jeon and R. Metzler, Phys. Rev. E {\bf 85},
021147 (2012).

\bibitem{jochen} J. Kursawe, J. Schulz, and R. Metzler, Phys. Rev. E
\textbf{88}, 062124 (2013).

\bibitem{mebakla} R. Metzler, E. Barkai, and J. Klafter, Phys. Rev. Lett.
\textbf{82}, 3563 (1999).

\bibitem{report} R. Metzler and J. Klafter, Phys. Rep. \textbf{339}, 1
(2000).

\bibitem{stas} S. Burov, R. Metzler, and E. Barkai, Proc. Natl. Acad. Sci.
USA \textbf{107}, 13228 (2010).

\bibitem{staspccp} S. Burov, J.-H. Jeon, R. Metzler, and E. Barkai, Phys.
Chem. Chem. Phys. \textbf{13}, 1800 (2011).

\bibitem{lene} J.-H. Jeon, V. Tejedor, S. Burov, E. Barkai, C. Selhuber-Unkel,
K. Berg-S{\o}rensen, L. Oddershede, and R. Metzler, Phys. Rev. Lett. \textbf{
106}, 048103 (2011).

\bibitem{xie} H. Yang, G. Luo, P. Karnchanaphanurach, T.-M. Louie, I. Reich,
S. Cova, L. Xun, and X. S. Xie, Science \textbf{302}, 262 (2003).

\bibitem{jeremy} X. Hu, L. Hong, M. D. Smith, T. Neusius, X. Cheng, and
J. C. Smith, Nature Phys. \textbf{12}, 171 (2016). 

\bibitem{chechkin} A. Chechkin, V. Gonchar, J. Klafter, R. Metzler, and L.
Tanatarov, Chem. Phys. \textbf{284}, 233 (2002).

\bibitem{chechkin1} A. V. Chechkin, J. Klafter, V. Yu. Gonchar, R. Metzler,
and L. V. Tanatarov, Phys. Rev. E \textbf{67}, 010102(R) (2003).

\bibitem{spagno} A. A. Dubkov, B. Spagnolo, and V. V. Uchaikin, Int. J.
Bifurc. Chaos \textbf{18}, 2649 (2008).

\bibitem{capala} K. Capa{\l}a, A. Padash, A. V. Chechkin, B. Shokri, R.
Metzler, and B. Dybiec, Chaos \textbf{30}, 123103 (2020).

\bibitem{tobias1} T. Guggenberger, A. Chechkin, and R. Metzler,
J. Phys. A \textbf{54}, 29LT01 (2021).

\bibitem{tobias} T. Guggenberger, G. Pagnini, T. Vojta, and R. Metzler,
New J. Phys. \textbf{21}, 022002 (2019).

\bibitem{igorejp} I. M. Sokolov, Euro. J. Phys. \textbf{31}, 1353 (2010).

\bibitem{haenggi} P. S. Burada, G. Schmid, D. Reguera, J. M. Rubi, and
P. H{\"a}nggi, Phys. Rev. E \textbf{75}, 051111 (2007).

\bibitem{risken} H. Risken, The Fokker-Planck equation (Springer,
Heidelberg, 1989).

\bibitem{eli} D. A. Kessler and E. Barkai, Phys. Rev. Lett. \textbf{105},
120602 (2010).

\bibitem{dybiec2010} B. Dybiec, I. M. Sokolov, and A. V. Chechkin,
J. Stat. Mech. {\bf 2010}, P07008 (2010).

\bibitem{kolmo} A. N. Kolmogorov, C. R. (Doklady) Acad. Sci. URSS (N.S.)
\textbf{26}, 115 (1940).

\bibitem{mandelbrot1968} B. B. Mandelbrot and J.Van Ness,
SIAM Rev. {\bf 10}, 422 (1968).

\bibitem{qian2003} H. Qian, in Processes with Long-Range Correlations,
edited by G. Rangajaran and M. Z. Ding (Springer, Heidelberg, 2003).

\bibitem{kloeden2011} P. E. Kloeden and E. Platen, Numerical Solution
of Stochastic Differential Equations (Springer, Heidelberg, 1992).

\bibitem{dieker2004} T. Dieker, Simulation of fractional Brownian motion,
MSc thesis, Vrije Universiteit Amsterdam, revised version (2004).

\bibitem{sune} S. Jespersen, R. Metzler, and H. C. Fogedby, Phys. Rev. E
\textbf{59}, 2736 (1999).

\bibitem{fogedby} H. C. Fogedby, Phys. Rev. Lett. \textbf{73}, 2517 (1994).

\bibitem{fogedby1} H. C. Fogedby, Phys. Rev. E \textbf{58}, 1690 (1998).

\bibitem{chechkinjsp} A. V. Chechkin, V. Yu. Gonchar, J. Klafter, R. Metzler,
and L. V. Tanatarov, J. Stat. Phys. \textbf{115}, 1505 (2004).

\bibitem{epl} R. Metzler, E. Barkai, and J. Klafter, Europhys. Lett.
\textbf{46}, 431 (1999).

\bibitem{weron} A. Weron and M. Magdziarz, Europhys. Lett. \textbf{86},
60010 (2009).

\bibitem{abramowitz72} M. Abramowitz and I. A. Stegun, Handbook of
Mathematical Functions (National Bureau of Standards, Bethesda, 1972)

\bibitem{oleksii} O. Sliusarenko, V. Yu. Gonchar, A. V. Chechkin, I. M.
Sokolov, and R. Metzler, Phys. Rev. E \textbf{81}, 041119 (2010).

\bibitem{zwanzig} R. Zwanzig, Nonequilibrium statistical mechanics
(Oxford University Press, Oxford, UK, 2001).

\bibitem{klimo} Yu. L. Klimontovich, Turbulent motion and the structure
of chaos (Kluwer, Dordrecht, 1991).

\bibitem{adelman1976} S. A. Adelman, J. Chem. Phys. {\bf 64}, 124 (1976).

\bibitem{haenggi1994} P. H{\"a}nggi and P. Jung, Colored Noise in
Dynamical Systems (John Wiley, New York, 1994)

\bibitem{molchan} G. M. Molchan, Commun. Math. Phys. \textbf{205}, 97 (1999).

\bibitem{jaeepl} J.-H. Jeon, A. V. Chechkin, and R. Metzler, Europhys. Lett.
\textbf{94}, 20008 (2011).

\bibitem{wiese} K. J. Wiese, S. N. Majumdar, and A. Rosso, Phys. Rev. E
\textbf{83}, 061141 (2011).

\bibitem{vojta0} A. H. O. Wada and T. Vojta, Phys. Rev. E \textbf{97},
020102(R) (2018).

\bibitem{vojta1} T. Vojta, S. Skinner, and R. Metzler, Phys. Rev. E
\textbf{100}, 042142 (2019).

\bibitem{vojta2} T. Vojta, S. Halladay, S. Skinner, S. Janu\v{s}onis, T.
Guggenberger, and R. Metzler, Phys. Rev. E \textbf{102}, 032108 (2020).

\bibitem{skirmantas} S. Janu\v{s}onis, N. Detering, R. Metzler, and T. Vojta,
Frontiers Comp. Neurosci. \textbf{14}, 56 (2020).

\bibitem{stas1} S. Burov and E. Barkai, Phys. Rev. Lett. \textbf{100}, 070601
(2008).

\bibitem{ohad} O. Vilk, E. Aghion, T. Avgar, C. Beta, O. Nagel, A. Sabri,
R. Sarfati, D. K. Schwartz, M. Weiss, D. Krapf, R. Nathan, R. Metzler, and
M. Assaf, E-print arXiv:2109.04309.

\bibitem{daniel} D. Molina-Garcia, T. Sandev, H. Safdari, G. Pagnini, A.
Chechkin, and R. Metzler, New J. Phys. \textbf{20}, 103027 (2018).

\bibitem{jaeprl} J.-H. Jeon, H. Martinez-Seara Monne, M. Javanainen, and R.
Metzler, Phys. Rev. Lett. \textbf{109}, 188103 (2012).

\bibitem{wei} W. Wang, F. Seno, I. M. Sokolov, A. V. Chechkin, and R.
Metzler, New J. Phys. \textbf{22}, 083041 (2020).

\bibitem{diego} A. Sabri, X. Xu, D. Krapf, and M. Weiss, Phys. Rev. Lett.
\textbf{125}, 058101 (2020).

\bibitem{eiji} E. Yamamoto, T. Akimoto, A. Mitsutake, and R. Metzler,
Phys. Rev. Lett. \textbf{126}, 128101 (2021).

\bibitem{fulvio} F. Baldovin, E. Orlandini, and F. Seno,
Frontiers Phys. \textbf{7}, 124 (2019).

\bibitem{mario} M. Hidalgo-Soria and E. Barkai,
Phys. Rev. E \textbf{102}, 012109 (2020).

\end{thebibliography}
\end{document}